\documentclass[reprint, superscriptaddress, nofootinbib, prd]{revtex4-1}

\usepackage{amsmath}
\usepackage{amssymb}
\usepackage{graphicx}
\usepackage[utf8]{inputenc}
\usepackage{verbatim}
\usepackage[T1]{fontenc}
\usepackage[english]{babel}
\usepackage{dcolumn}
\usepackage{bm}
\usepackage{mathrsfs}  
\usepackage{mathtools}
\usepackage{cancel}
\usepackage{esvect}
\usepackage{subfigure}
\usepackage{booktabs}
\usepackage{soul}
\usepackage{bbold}
\usepackage[dvipsnames]{xcolor}

\definecolor{custompurple}{HTML}{9300d3}
\definecolor{customgreen}{HTML}{019d73}
\definecolor{customblue}{HTML}{57b5e8}

\usepackage[colorlinks=true,linkcolor=customblue, citecolor=custompurple,urlcolor=customgreen]{hyperref}

\renewcommand{\d}{\mathrm{d}}
\newcommand{\tr}[1]{\mathrm{Tr}\left[#1\right]}
\renewcommand{\Re}{\mathrm{Re}}

\begin{document}

\preprint{}
\title{Simulating jets and heavy quarks in the Glasma\\using the colored particle-in-cell method}
\author{Dana Avramescu}
\email[Corresponding author: ]{dana.d.avramescu@jyu.fi}
\affiliation{Department of Physics, P.O. Box 35, 40014 University of Jyv\"{a}skyl\"{a}, Finland}
\affiliation{Helsinki Institute of Physics, P.O. Box 64, 00014 University of Helsinki, Finland}
\author{Virgil Băran}
\email{virgil.baran@unibuc.ro}
\affiliation{Faculty of Physics, University of Bucharest, Atomiștilor 405, Măgurele, Romania}
\author{Vincenzo Greco}
\email{greco@lns.infn.it}
\affiliation{Department of Physics and Astronomy, University of Catania, Via S. Sofia 64, I-95123 Catania}
\affiliation{INFN-Laboratori Nazionali del Sud, Via S. Sofia 62, I-95123 Catania, Italy}
\author{Andreas Ipp}
\email{ipp@hep.itp.tuwien.ac.at}
\affiliation{Institute for Theoretical Physics, TU Wien, Wiedner Hauptstraße 8, A-1040 Vienna, Austria}
\author{David M\"{u}ller}
\email{dmueller@hep.itp.tuwien.ac.at}
\affiliation{Institute for Theoretical Physics, TU Wien, Wiedner Hauptstraße 8, A-1040 Vienna, Austria}
\author{Marco Ruggieri}
\email{marco.ruggieri@dfa.unict.it}
\affiliation{Department of Physics and Astronomy, University of Catania, Via S. Sofia 64, I-95123 Catania}

\date{\today}

\begin{abstract}
We explore the impact of strong classical color fields, which occur in the earliest stages of heavy-ion collisions and are known as the Glasma, on the classical transport of hard probes, namely heavy quarks and jets. To achieve this, we simulate SU(3) color fields using classical real-time lattice gauge theory and couple them to an ensemble of test particles whose dynamics are described by Wong's equations. We provide an overview of how classical color algebras are constructed and introduce a method to generate random classical SU(3) color charges. We extensively test our numerical particle solver in the limits of infinitely massive heavy quarks and ultra-relativistic light-like jets and obtain excellent quantitative agreement with previous studies. Going towards realistic masses and initial momenta, we extract longitudinal and transverse momentum broadening for heavy quarks and jets. The resulting accumulated momenta and the anisotropy of these dynamical hard probes exhibit deviations from limiting scenarios, showing that the full dynamics have a significant effect. 
\end{abstract}

\maketitle


\section{Introduction}

Relativistic heavy-ion collision experiments, as conducted at the Large Hadron Collider (LHC) or the Relativistic Heavy Ion Collider (RHIC), provide the remarkable opportunity to study hadronic matter under extreme conditions with increasing statistics and precision. Immediately after the collision, the medium is characterized by large gluon occupation numbers and a highly non-linear regime, known as the Glasma \cite{Lappi:2006fp, Lappi:2008eq, Fujii:2008km, Fukushima:2011nq}. Particularly sensitive probes of the very early stage of the collision are heavy quarks and jets. Due to their short formation time, they experience the initial stage of the collision. By understanding the imprint of the Glasma fields on these probes, one can disentangle important information about the structure of initially produced matter, in both proton-nucleus and nucleus-nucleus collisions.

The Glasma is described using a wider framework entitled Color Glass Condensate (CGC) \cite{Iancu:2003xm, Gelis:2010nm, Gelis:2012ri} which is formulated at the high-energy limit of Quantum Chromodynamics (QCD). The field equations for the color fields of the gluons are solved numerically using methods from lattice QCD \cite{Lappi:2003bi, Lappi:2006hq, Muller:2019bwd}. To describe the properties of hard probes from high-energy nuclear collisions, numerous approaches based on perturbative QCD (pQCD) techniques \cite{Liu:2006ug, Arnold:2008vd, Caron-Huot:2008zna, Majumder:2009cf}, lattice computations \cite{Schenke:2008gg, Banerjee:2011ra, Panero:2013pla, Boguslavski:2018beu, Altenkort:2020fgs}, or non-Abelian Yang Mills transport theories \cite{Litim:1999id, Litim:1999ns, Litim:2001db} have been used. These probes are produced immediately after the collision and may be affected by the entire evolution of the resulting Quark Gluon Plasma (QGP).  

Previous approaches that focus on the effect of the Glasma on hard probes include a study on jets in the Glasma \cite{Ipp:2020mjc, Ipp:2020nfu} based on a lattice discretization of the Yang-Mills equations, where the transport properties of jets are evaluated by treating them as ultra-relativistic light-like partons. More precisely, the jet momentum broadening is extracted from Glasma field correlators computed on the lattice, without explicitly solving the dynamical particle equations of motion. Another lattice study \cite{Boguslavski:2018beu, Boguslavski:2020tqz, Boguslavski:2020mzh} with over-occupied Yang-Mills plasma instead of Glasma, evaluates the heavy quark transport coefficient from electric field correlators (assuming the heavy quarks to be infinitely massive and static) and emphasizes the emergence of plasmon mass induced oscillations. In another series \cite{Das:2015aga, Das:2017dsh, Ruggieri:2018rzi, Sun:2019fud, Liu:2019lac, Liu:2020cpj, Khowal:2021zoo, Ruggieri:2022kxv}, the effect of the Glasma phase on the diffusion of heavy quarks is extensively studied and compared to the standard Langevin description of heavy quark dynamics, with a recent focus on memory effects. A different approach is taken in \cite{Carrington:2020sww, Carrington:2021dvw, Carrington:2022bnv}, where both the Glasma fields and particle transport equations are derived using analytical frameworks. The Glasma fields are obtained in the proper time expansion and the transport of the hard probes is treated using the Fokker-Planck equations adapted to the Glasma. Complementary, it was shown that the initial stage, implemented in different frameworks, has an effect on jet quenching \cite{Andres:2019eus, Andres:2022bql}. Even though these approaches vary with respect to the approximations which are used, they all converge to the same key result: the Glasma phase has a considerable effect on the transport of hard probes. Nevertheless, very few of these studies have a built-in way to describe the very early stage consistently and in many cases they are constructed on approximations applicable at later stages. 

In this work, we present a novel framework that simulates the full dynamics of hard particles right after the collision on top of an evolving boost-invariant SU(3) Glasma background field. This is practically achieved by developing a numerical solver for the equations of motion of particles propagating in these fields. The particles are initialized with finite masses, formation times and initial momenta. The solver is used to extract relevant quantities such as the momentum accumulated as the partons propagate in the background fields. The novelty consists in the numerical methods developed for the particle solver and the techniques used to efficiently solve both the Glasma and particle equations concurrently. In particular, we introduce a novel way to generate SU(3) classical color charges using the Haar measure. The code runs on GPUs and allows for the systematic study of the full dynamics of particles and the dependence on many parameters used for particle initialization. 

There exist two relevant limiting cases in which the accumulated momentum of hard probes in Glasma may be evaluated only from Glasma lattice field correlators, without solving the particle equations of motion. These correspond to infinitely massive heavy quarks and highly energetic jets. When we consider such quarks in our particle solver, we reproduce the limiting results. The limiting case of extremely fast light-like jets is extracted using two setups, namely the classical transport framework using Wong's equations and a quantum pQCD computation. By comparing the resulting momentum broadening, we notice a discrepancy between the classical computation and the quantum one, and propose a way to resolve it. Going beyond these limiting cases, towards realistic dynamical results, we quantitatively study whether the full dynamics has a considerable effect. We extract the instantaneous transport coefficients, namely $\kappa$ for heavy quarks and $\hat{q}$ for jets, and check if the large transport coefficients of hard probes in the Glasma obtained by previous studies are an artifact of the approximations used or still persists with our full numerical setup. Most remarkably, we observe that that momentum broadening along rapidity oscillates as a function of proper time, which could indicate plasmon modes in the Glasma \cite{Boguslavski:2020tqz}. Preliminary results obtained using our solver have been presented previously in \cite{Avramescu:2022vkd}.

This study is structured as follows. Section \ref{sec:glasma} contains an overview of the classical description of the early stage  in terms of Glasma initial conditions and classical boost-invariant Yang-Mills equations. In Section \ref{sec:particles} we present Wong's equations. In Section \ref{sec:classiccharges} we describe how SU(2) and SU(3) color charges are sampled correctly. Section \ref{sec:limcases} describes the limiting cases for infinitely massive heavy quarks and for extremely fast light-like jets. In Section \ref{sec:divisionbydr} we show how to reconcile classical particle simulations with the calculation of momentum broadening within pQCD. The results obtained with our particle solver are showcased in Section \ref{sec:results}. Finally, Section \ref{summary} includes a summary of all the results, along with viable future extensions of our study. Detailed calculations including the correct sampling of color charges can be found in Appendices \ref{appen:wong_details} through \ref{appen:numchecks}.


\section{Glasma in a nutshell}
\label{sec:glasma}

Within the Color Glass Condensate framework \cite{Iancu:2003xm, Gelis:2010nm, Gelis:2012ri}, the medium produced after the collision of relativistic nuclei is a state dominated by strong classical color fields known as the Glasma \cite{Lappi:2003bi, Lappi:2006fp,Lappi:2006hq, Lappi:2008eq, Fujii:2008km, Fukushima:2011nq}. The CGC is an effective theory for high energy nuclei and relies on the separation of scales between degrees of freedom with small and large longitudinal momentum fraction $x$. Hard ({large-$x$}) partons behave as highly Lorentz-contracted, static color sources $J^\mu$ for the gauge fields $A_\mu$ described by the soft (small-$x$) partons. At leading order in the coupling constant $g$, the hard and soft sectors are coupled via the Yang-Mills equations
\begin{equation}
    \label{eq:yang_mills}
    \mathscr{D}_\mu F^{\mu\nu}=J^\nu,
\end{equation}
with $\mathscr{D}_\mu (\,\dots) \equiv \partial_\mu(\,\dots)-\mathrm{i}g \big[A_\mu, \,\dots \big]$ denoting the gauge-covariant derivative, $F^{\mu\nu}=\partial_\mu A_\nu-\partial_\nu A_\mu -\mathrm{i}g\big[A_\mu,A_\nu\big]$ the field strength tensor and $J^\mu$ the color current.
At sufficiently high energies, we can approximate the nuclei to be propagating along the light-cone directions $x^\pm\equiv(x^0\pm x^3)/\sqrt{2}$. Their color currents are given by
\begin{equation}
    \label{eq:lc_current}
    J^\mu_{A,B}=\delta^{\mu\pm}\rho_{A,B}(x^\mp,\vec{x}_\perp),
\end{equation}
where $\rho_{A,B}$ represent classical color charge densities and the subscripts $A$ and $B$ denote the two colliding nuclei. The color charge densities are treated as stochastic variables whose statistics are determined by the probability functional $W[\rho]$. We take it to be given by the McLerran-Venugopalan (MV) model \cite{McLerran:1993ni, McLerran:1993ka, McLerran:1994vd}. It considers the color charges $\rho$ to follow Gaussian statistics, which are determined by the one- and two-point correlators
\begin{gather}
    \begin{aligned}
        \langle\rho^a(x^\mp, \vec{x}_\perp)\rangle_{A,B} & =0,\notag\\
        \langle\rho^a(x^\mp, \vec{x}_\perp)\rho^b(y^\mp, \vec{y}_\perp)\rangle_{A,B} &= g^2\lambda_{A,B}(x^\mp)\delta^{ab}\notag \\
& \times \delta(x^\mp-y^\mp)\delta^{(2)}(\vec{x}_\perp-\vec{y}_\perp),
    \end{aligned}\\
\end{gather}
where $\lambda_{A,B}$ is the average color charge per unit volume. One may extract $\mu^2_{A,B}=\int \mathrm{d}x^\pm \lambda_{A,B}(x^\mp)$, which denotes the MV model parameter (in units of energy squared) and represents the variance of the color charge density fluctuations of each nucleus. 

We can solve the Yang-Mills equations from Eq.~\eqref{eq:yang_mills} for the special choice of color current in Eq.~\eqref{eq:lc_current} in the covariant gauge $\partial_\mu A^\mu_\mathrm{cov}=0$. The only non-zero components of the gauge field are given by $A^\pm_\mathrm{cov}(x^\mp, \vec{x}_\perp)\equiv\alpha_{A,B}(x^\mp, \vec{x}_\perp)$ where $\alpha_{A,B}$ obeys a Poisson equation restricted to the transverse plane
\begin{equation}
    \label{eq:2dpoisson}
    \Delta_\perp\alpha_{A,B}(x^\mp, \vec{x}_\perp)=-\rho^\mathrm{cov}_{A,B}(x^\mp,\vec{x}_\perp),
\end{equation}
in which $\Delta_\perp$ is the transverse Laplace operator. The Poisson equation can be formally solved via Fourier transformation
\begin{align}
    \alpha_{A,B}(x^\mp, \vec{x}_\perp) = \int d^2 \vec{k}_\perp \frac{\tilde \rho^\mathrm{cov}_{A,B}(x^\mp,\vec{k}_\perp)}{\vec{k}_\perp^2 + \lambda^2} \exp\big({-\mathrm{i} \vec{k}_\perp \cdot \vec{x}_\perp}\big),
\end{align}
where $\lambda$ is an infrared regulator and $\tilde{\rho}^\mathrm{cov}_{A,B}$ are the Fourier transformed charge densities in the covariant gauge. 
By performing a gauge transformation to the light-cone gauge $A^+_\mathrm{lc}=0$, the gauge field only has transverse components given by
\begin{equation}
    A^i_{A,B}(x^\mp,\vec{x}_\perp)=\frac{\mathrm{i}}{g} V (x^\mp,\vec{x}_\perp)\partial^i V ^\dagger(x^\mp,\vec{x}_\perp),
\end{equation}
with the light-like Wilson line
\begin{equation}
     V ^\dag_{A,B}(x^\mp,\vec{x}_\perp)=\mathscr{P}\,\exp\Bigg(\mathrm{i}g\int\limits_{-\infty}^{x^\mp}\mathrm{d}y^\mp\alpha_{A,B}(y^\mp,\vec{x}_\perp)\Bigg).
\end{equation}
where $\mathscr{P}(\dots\,)$ denotes the path-ordering operation.

In the ultrarelativistic limit, the nuclei are contracted to infinitesimally thin sheets. This can be expressed via $J^\mu_{A,B}=\delta^{\mu\pm}\delta(x^\mp)\rho_{A,B}(x_\perp)$ where the two-dimensional charge densities obey the correlator
\begin{equation}
    \label{eq:MVchargecorr}
    \langle\rho^a(\vec{x}_\perp)\rho^b(\vec{y}_\perp)\rangle_{A,B}=g^2\mu^2_{A,B}\delta^{ab}\delta^{(2)}(\vec{x}_\perp-\vec{y}_\perp).
\end{equation}
The transverse gauge fields are given by
\begin{equation}
    \label{eq:puregaugefields}
    A^i_{A,B}(x^\mp,x_\perp)=\theta(x^\mp)\alpha^i_{A,B}(\vec{x}_\perp),
\end{equation}
in which $\theta$ represents the Heaviside function and
\begin{equation}
    \alpha^i_{A,B}(\vec{x}_\perp)=\dfrac{\mathrm{i}}{g} V_{A,B} (\vec{x}_\perp)\partial^i V_{A,B} ^\dagger(\vec{x}_\perp),
\end{equation}
involves a Wilson line depending on the transverse coordinate, obtainable as $ V_{A,B} (\vec{x}_\perp)=\lim\limits_{x^\mp\rightarrow\infty} V_{A,B} (x^\mp,\vec{x}_\perp)$. 

We now consider the classical collision problem
\begin{equation}
    \mathscr{D}_\mu F^{\mu\nu}=J^\nu_A + J^\nu_B,
\end{equation}
where the initial conditions in the asymptotic past are provided by the color fields of the nuclei. The Glasma is described by the gauge field in the future light-cone of the collision. In the ultra-relativistic limit, the total color current generated by the two nuclei $J^\mu=J^\mu_A +J^\mu_B$ possesses invariance under longitudinal Lorentz boosts, which implies that any observables of the Glasma must be invariant under boosts as well. An appropriate choice of coordinates is given by the Milne coordinates $(\tau,\eta)$ defined as
\begin{align}
    \label{eq:taueta}
    \tau=\sqrt{2x^+x^-},\quad \eta=\frac{1}{2}\ln{\left(\frac{x^+}{x^-}\right)},
\end{align}
with proper time $\tau$ and space-time rapidity $\eta$. By fixing the residual gauge freedom by imposing the temporal gauge condition $A^\tau=0$ and requiring boost invariance of the gauge fields as $A^\mu(\tau, \eta, \vec{x}_\perp) = A^\mu(\tau, \vec{x}_\perp)$, one may formulate initial conditions for the Glasma fields along the boundary of the future light-cone as \cite{Kovner:1995ja}
\begin{equation}
    \label{eq:Glasmainitcond}
        \begin{aligned}
        &A^{i}\left(\tau, \vec{x}_{\perp}\right)\Big|_{\tau=0}=\alpha_A^{i}\left(\vec{x}_{\perp}\right)+\alpha_B^{i}\left(\vec{x}_{\perp}\right),\\ &A^{\eta}\left(\tau, \vec{x}_{\perp}\right)\Big|_{\tau=0}=\frac{\mathrm{i}g}{2}\left[\alpha_A^{i}\left(\vec{x}_{\perp}\right), \alpha_B^{i}\left(\vec{x}_{\perp}\right)\right],
        \end{aligned}
\end{equation}
accompanied by
\begin{equation}
    \partial_{\tau} A^{i}\left(\tau, \vec{x}_{\perp}\right)\Big|_{\tau=0}=\partial_{\tau} A^{\eta}\left(\tau, \vec{x}_{\perp}\right)\Big|_{\tau=0}=0.
\end{equation}

The conjugate momenta associated with the gauge fields are
\begin{equation}
    \label{eq:glasmapipeta}
    P^i=\tau\partial_\tau A_i,\quad P^\eta=\frac{1}{\tau}\partial_\tau A_\eta.
\end{equation}
The Yang-Mills action expressed in Milne coordinates, together with boost-invariance, yields the field equations
\begin{equation}
    \label{eq:ymfieldeqs}
    \begin{aligned}
    &\partial_\tau P^i=\tau\mathscr{D}_jF_{ji}-\dfrac{\mathrm{i}g}{\tau}\Big[A_\eta,\mathscr{D}_iA_\eta\Big],\\
    &\partial_\tau P^\eta=\dfrac{1}{\tau}\mathscr{D}_i\left(\mathscr{D}_iA_\eta\right),
    \end{aligned}
\end{equation}
along with the Gauss constraint $\mathscr{D}_iP^i+\mathrm{i}g\left[A_\eta, P^\eta\right]=0$ which is fulfilled throughout the evolution. In order for these equations to preserve gauge invariance upon discretization, they need to be recast in a lattice QCD formulation. 

\subsection{Numerical implementation}
\label{subsec:glasmanumimpl}
The Yang-Mills equations of the Glasma may be solved numerically. The work presented here is based on an approach that employs classical real-time lattice gauge theory \cite{Krasnitz:1998ns,Lappi:2003bi}.
In order to assure gauge invariance of the field equations from Eqs.~\eqref{eq:ymfieldeqs} upon discretization, one may proceed as follows: the Minkowski space is discretized on a hypercubic lattice replacing the gauge fields with gauge links, which are Wilson lines connecting neighboring points on this lattice. A particularity of the boost-invariant collision scenario is that one needs to employ this procedure only in the transverse plane. This is due to the fact that $A_\eta(\tau,\vec{x}_\perp)$ acts as a scalar under $\eta$-independent gauge transformations, and thus the $\eta$ direction is left continuous with respect to a lattice discretization. 

The transverse plane, taken as a square of length $L$ and accompanied by periodic boundary conditions for the fields, is discretized in $N^2$ points in which the fields are assigned values at various proper times. In the continuum limit, assuming small lattice spacings $a=L/N$, a gauge link connecting the lattice point located at $\vec{x}_\perp$ and the neighboring point along a direction $\hat{i}$, where $\hat{i}$ is the unit vector along $x^i$, is given by
\begin{equation}
     U _{\hat{i}}(\tau, \vec{x}_\perp)\approx\exp\Bigg(\mathrm{i}ga A_i\Big(\tau, \vec{x}_\perp+\dfrac{a}{2}\hat{i}\Big)\Bigg).
\end{equation}
Links in opposite directions can be expressed through the Hermitian operation $ U _{-\hat{i}}(\tau,\vec{x}_\perp)\equiv U ^\dagger_{\hat{i}}(\tau,\vec{x}_\perp-\hat{i})$. These gauge links are then used to construct a plaquette variable as $ U _{\hat{i}\hat{j}}(\tau, \vec{x}_\perp)\equiv U _{\hat{i}}(\tau, \vec{x}_\perp) U _{\hat{j}}(\tau, \vec{x}_\perp+\hat{i}) U _{-\hat{i}}(\tau, \vec{x}_\perp+\hat{i}+\hat{j}) U _{-\hat{j}}(\tau, \vec{x}_\perp+\hat{j})$. 

The Yang-Mills action can be approximated using link and plaquette variables along with the conjugate momenta
\begin{equation}
    \begin{aligned}
        P^\eta(\tau,\vec{x}_\perp)&=\dfrac{1}{\tau}\partial_\tau A_\eta(\tau,\vec{x}_\perp),\\
        P^i(\tau,\vec{x}_\perp)&=-\mathrm{i}\dfrac{\tau}{ga}\Big[\partial_\tau U _{\hat{i}}(\tau,\vec{x}_\perp)\Big] U _{\hat{i}}^\dagger(\tau,\vec{x}_\perp).
    \end{aligned}
\end{equation}
Varying the discretized action yields discretized equations of motion 
\begin{align}
    \begin{split}
    \partial_\tau P^\eta(\tau,\vec{x}_\perp)=&\dfrac{1}{\tau}\mathsf{D}_i^2A_\eta(\tau,\vec{x}_\perp),\\
    \partial_\tau P^i(\tau,\vec{x}_\perp)=&-\sum_j\dfrac{\tau}{ga^3}\Big[ U _{\hat{i}\hat{j}}(\tau, \vec{x}_\perp)+ U _{\hat{i}\,-\!\hat{j}}(\tau, \vec{x}_\perp)\Big]_\mathrm{ah}\\
    &-\dfrac{\mathrm{i}g}{\tau}\Big[A_\eta^\mathrm{transp}(\tau,\vec{x}_\perp),\mathsf{D}_i^FA_\eta(\tau,\vec{x}_\perp)\Big],
    \end{split}
\end{align}
where $\mathrm{D}_i^2\equiv\mathrm{D}_i^F\mathrm{D}_i^B$ contains the forward $\mathrm{D}_i^F$ and backward $\mathrm{D}_i^B$ gauge-covariant finite differences on the lattice, $\left(\,\dots\right)_\mathrm{ah}$ denotes the anti-Hermitian traceless part of a matrix, and ${A_\eta^\mathrm{transp}(\tau,\vec{x}_\perp)\equiv U _{\hat{i}}(\tau,\vec{x}_\perp)A_\eta(\tau,\vec{x}_\perp+\hat{i}) U _{\hat{i}}^\dagger(\tau,\vec{x}_\perp)}$ represents the parallel transported scalar field. These equations are accompanied by the Gauss constraint and are solved numerically by employing the leapfrog algorithm. In this numerical method, the conjugate momenta are evaluated at half-integer time steps, whereas the rest of the fields are computed at integer proper times. 

Finally, the MV model initial conditions must be discretized as well. The naive use of Eq.~\eqref{eq:MVchargecorr} leads to loss of randomness in the infinitesimal direction $x^\pm$ along which the nucleus propagates, due to the non-trivial path-ordering of the involved Wilson lines. Nevertheless, by sticking together infinitesimally thin sheets of color charge and regularizing the correlator as \cite{Fukushima:2007ki}
\begin{equation}
    \label{eq:MVchargecorrnum}
    \langle\rho^a_m(\vec{x}_\perp)\rho^b_n(\vec{y}_\perp)\rangle_{A,B}=\dfrac{1}{N_sa^2}g^2\mu^2_{A,B}\delta_{mn}\delta^{ab}\delta(\vec{x}_\perp-\vec{y}_\perp),
\end{equation}
where $m,n\in\{1,2, \dots N_s\}$ denotes the index of the sheet, and with $N_s$ the number of such color sheets, this issue is resolved. Numerically, color charges are generated by sampling random numbers distributed according to a Gaussian with zero mean and variance chosen to obey Eq.~\eqref{eq:MVchargecorrnum}. Once the color charges are provided, the solutions of Eq.~\eqref{eq:2dpoisson} now expressed for each color sheet as $\Delta_\perp\alpha^a_n(\vec{x}_\perp)=-\rho^a_n(\vec{x}_\perp)$ may be obtained using Fast Fourier Transformation (FFT), where the infrared and ultraviolet cut-offs are $\lambda$ and $\Lambda$. Furthermore, the Wilson lines are constructed as products computed for each sheet as $V^\dagger(\vec{x}_\perp)=\prod\limits_{n=1}^{N_s}\exp\Big(-\mathrm{i}g\alpha_n(\vec{x}_\perp)\Big)$ with $\alpha_n\equiv\alpha_n^aT^a$. Subsequently, the transverse gauge links are computed from these discretized Wilson lines and the initial Glasma conditions given in Eq.~\eqref{eq:Glasmainitcond} are also numerically discretized. Once all these steps are completed, the Glasma fields are numerically solved using our numerical simulation routines\footnote{The simulation code for the Glasma fields is publicly available at \href{https://gitlab.com/openpixi/curraun}{https://gitlab.com/openpixi/curraun}.}.


\section{Partons immersed in Glasma}
\label{sec:particles}

The dynamics of particles propagating in classical Yang-Mills fields is given by Wong’s equations \cite{Wong:1970fu} which describe how the positions and momenta of the particles evolve in time, while their charges rotate in color space \cite{Boozer_2011}. In the laboratory frame they read as
\begin{equation}
    \label{eq:wonglab}
        \begin{aligned}
            &\dfrac{\d x^i}{\d t}=\frac{p^i}{E},\\
            &\dfrac{\d p^i}{\d t}=gQ^aF^{i\mu,a}\frac{p_\mu}{E},\\
            &\dfrac{\d Q^a}{\d t}=-gf^{abc} A_\mu^bQ^c\frac{p^\mu}{E},
        \end{aligned}
\end{equation}
where $i\in\{x,y,z\}$ and $a=\{ 1,2, \dots, D_A \}$ with the dimension of the adjoint representation $D_A = N_c^2 - 1$. The energy is given by $E=\sqrt{\vec{p}^2+m^2}$ with $m$ being the mass of the particle, $\vec{p}\equiv(p^x,p^y,p^z)$, and $f^{abc}$ the structure constants for the SU($N_c$) group.

The dynamic equation for the energy $p^0 = E$ is given by
\begin{equation}
    \label{eq:dEdt}
    \frac{\mathrm{d}E}{\mathrm{d}t}=gQ^aF^{0i,a}\frac{p_i}{E}=g Q^a\vec{E}^a\cdot\vec{v},
\end{equation}
with $E^i\equiv F^{0i}$ denoting the color-electric field. This relation states that the energy of a moving particle changes due to the work exerted by the color-electric field upon it, where $\vec{p}=\gamma m \vec{v}$, with $\gamma=E/m$ the Lorentz factor and $\vec{v}$ the laboratory frame velocity.

Wong's equations may be recast into a covariant form, with quantities computed along the worldline of the particle
\begin{equation}
    \label{eq:wongcurv}
    \begin{aligned}
        &\frac{\mathrm{d}x^\mu}{\mathrm{d}\boldsymbol{\tau}}=\frac{p^\mu}{m},\\
        &\frac{\mathrm{D}p^\mu}{\mathrm{d}\boldsymbol{\tau}}=gQ^aF^{\mu\nu,a}\frac{p_\nu}{m},\\
        &\frac{\mathrm{d}Q^a}{\mathrm{d}\boldsymbol{\tau}}=-gf^{abc}A_\mu^bQ^c\frac{p^\mu}{m},
    \end{aligned}
\end{equation}
where $(\mathrm{d}\boldsymbol{\tau})^2=g_{\mu\nu}\mathrm{d}x^\mu\mathrm{d}x^\nu$ denotes the relativistic proper time and in which $E\,\mathrm{d}/\mathrm{d}t=m\,\mathrm{d}/\mathrm{d}\boldsymbol{\tau}$ is employed and $\mathrm{D}/\mathrm{d}\boldsymbol{\tau}$ is the covariant derivative taken along the particle worldline. Further, we make use of the Lie-algebra-valued color charges $Q=Q^a T^a$ to write $Q^aF^{\mu\nu,a}=\tr{QF^{\mu\nu}}/T_R$,  where $T_R$ is the representation-dependent Dynkin index defined through $\tr{T^a T^b}=T_R \delta^{ab}$, with $T^a\in\mathrm{SU(N_c)}$. The above equations simplify to
\begin{equation}
    \label{eq:wongq}
    \begin{aligned}
    \frac{\mathrm{D}p^\mu}{\mathrm{d}\boldsymbol{\tau}}&=\frac{g}{T_R}\tr{QF^{\mu\nu}}\frac{p_\nu}{m},\\ \frac{\mathrm{d}Q}{\mathrm{d}\boldsymbol{\tau}}&=-\mathrm{i}g[A_\mu,Q]\frac{p^\mu}{m}.
\end{aligned}
\end{equation}

The equation governing the evolution of the color charge may formally be solved by
\begin{equation}
    \label{eq:chargepropertime}
    Q(\boldsymbol{\tau})=\mathcal{U}(\boldsymbol{\tau},\boldsymbol{\tau}_0)\,Q(\boldsymbol{\tau}_0)\,\mathcal{U}(\boldsymbol{\tau}_0,\boldsymbol{\tau}),
\end{equation}
where $Q$ is rotated with Wilson lines. These Wilson lines involve the path-ordered exponential computed along the trajectory of the particle and are given by
\begin{equation}
    \label{eq:wilsonlinepropertime}
    \mathcal{U}(\boldsymbol{\tau},\boldsymbol{\tau}_0)=\mathscr{P}\exp\Bigg(-\mathrm{i}g\int\limits_{\boldsymbol{\tau}_0}^{\boldsymbol{\tau}}\mathrm{d}\boldsymbol{\tau}^\prime\frac{\mathrm{d}x^\mu}{\mathrm{d}\boldsymbol{\tau}^\prime}A_\mu(x^\mu)\Bigg).
\end{equation}
The last relation may be derived by making use of the parallel transport equation for a Wilson line 
\begin{equation}
    \dfrac{\mathrm{d}}{\mathrm{d}\boldsymbol{\tau}}\mathcal{U}(\boldsymbol{\tau}, \boldsymbol{\tau}_0)=-\mathrm{i}g\dfrac{\mathrm{d}x^\mu}{\mathrm{d}\boldsymbol{\tau}}A_\mu(x^\mu(\boldsymbol{\tau}))\,\mathcal{U}(\boldsymbol{\tau}, \boldsymbol{\tau}_0).
\end{equation}

The use of Wilson lines in the evolution of the color charge automatically conserves the quadratic
\begin{align} 
\label{eq:q2_main}
    Q^aQ^a \equiv q_2(R)
\end{align}
and cubic \emph{classical Casimirs}\footnote{One may also define Casimir invariants for classical Lie algebras. For SU(2) and SU(3), we construct them in analogy with the group-theoretical Casimir invariants. More details are offered in Appendix \ref{appen:clascol}.} 
\begin{align} 
\label{eq:q3_main}
    d_{abc}Q^aQ^bQ^c\equiv q_3(R),
\end{align}
where $d_{abc}$ are the symmetric structure constants and the values $q_2(R)$, $q_3(R)$ depend on the chosen representation $R$ for the color charge $Q$. We go into detail about how these invariants are fixed in Section \ref{sec:classiccharges} and Appendix \ref{appen:clascol}.

In this work we approximate hard partons as test particles, which means that we neglect any back reaction of the partons onto the Glasma.

\subsection{Dynamics of particles in Glasma}
\label{subsec:particlesglasma}
Let us express Eqs.~\eqref{eq:wongcurv} and~\eqref{eq:chargepropertime} in Milne coordinates and choose the background fields to be those of the boost-invariant Glasma. A detailed derivation can be found in Appendix \ref{appen:wong_milne}. 
The coordinate Wong equations are given by
\begin{equation}
    \label{eq:wongpos}
   \frac{\mathrm{d} x^\mu}{\mathrm{d}\tau}=\frac{p^{\mu}}{p^\tau}
\end{equation}
where $x^\mu\in\{x,y,\eta\}$ and $(p^\tau)^2=(p^x)^2+(p^y)^2+\tau^2 (p^\eta)^2+m^2$. The Wong equations for momenta read as
\begin{align}
\label{eq:wonginmilne}
\begin{split}
    &\tau\dfrac{\mathrm{d}p^\eta}{\mathrm{d}\tau}+2 p^\eta \\
    &=\frac{g}{T_R}\left(\tr{Q E_\eta}\! - \! \tr{QB_x}\frac{p^y}{p^\tau} \! + \! \tr{QB_y} \frac{p^x}{p^\tau}\right),\\
    &\frac{dp^x}{d\tau}=\frac{g}{T_R}\left(\tr{QE_x} \! + \! \tr{QB_\eta}\frac{p^y}{p^\tau}\! - \! \tr{QB_y} \frac{\tau p^\eta}{p^\tau}\right),\\
    &\frac{dp^y}{d\tau}=\frac{g}{T_R}\left(\tr{QE_y}\! -\! \tr{QB_\eta} \frac{p^x}{p^\tau} \! + \! \tr{QB_x}\frac{\tau p^\eta}{p^\tau}\right),
\end{split}
\end{align}
and are accompanied by the dynamic equation for the temporal component
\begin{align}
\label{eq:ptaueq}
    \begin{split}
    &\dfrac{\mathrm{d}p^\tau}{\mathrm{d}\tau}+\frac{\tau p^\eta}{p^\tau}p^\eta\\
    &=\frac{g}{T_R}\left(\tr{Q E_\eta}\frac{\tau p^\eta}{p^\tau}+\tr{QE_x} \frac{p^x}{p^\tau}+\tr{QE_y} \frac{p^y}{p^\tau}\right),
    \end{split}
\end{align}
where the color-electric and -magnetic fields are determined from the field strength tensor via
\begin{equation}
    \label{eq:glasmafields_main}
    \begin{aligned}
        &E_i\equiv F_{\tau i},  &&B_i\equiv \epsilon_{ij}\dfrac{1}{\tau}F_{\eta j},\\
        &E_\eta\equiv \dfrac{1}{\tau}F_{\tau\eta}, &&B_\eta\equiv -F_{xy}.
    \end{aligned}
\end{equation}

As for the proper time evolution of the color charge, the Wilson line involved in the color rotation, see Eq.~\eqref{eq:wilsonlinepropertime}, can be expressed as a path-ordered integral along the worldline 
\begin{align} 
\label{eq:wilson_path}
    \mathcal{U}(\tau,\tau_0)=\mathscr{P}\exp\Bigg(-\mathrm{i}g\int\limits_{x^\mu(\tau_0)}^{x^\mu(\tau)}\mathrm{d}x^\mu A_\mu\left(x^\mu(\tau)\right)\Bigg).
\end{align}
As used in the Glasma framework, we employ the temporal gauge $A_\tau = 0$ and the gauge field is taken to be independent of space-time rapidity $\eta$, which simplifies the Wilson lines. 

As colored particles pass through the Glasma, the momentum of the particles $p_\mu$ changes according to Wong's equations. The main observable we focus on, which represents a measure of the accumulated momentum, is the momentum broadening $\delta p_\mu$  defined as 
\begin{align}
    \label{eq:mombroad}
    \delta p_\mu^2(\tau)\equiv p_\mu^2(\tau)-p_\mu^2(\tau_\mathrm{form}).
\end{align}
Here $\tau_\mathrm{form}$ denotes the formation time at which the particle is introduced into the system and $p_\mu(\tau_\mathrm{form})$ is the initial momentum of the particle. The momentum broadening thus reflects how much momentum is accumulated through interactions with the Glasma background field compared to the initial momentum of the particle.

\subsection{Numerical implementation}
\label{subsec:numimplement}

\begin{figure*}[t]
\centering
\subfigure[Trajectories in the transverse plane]{\includegraphics[width=0.9\columnwidth]
{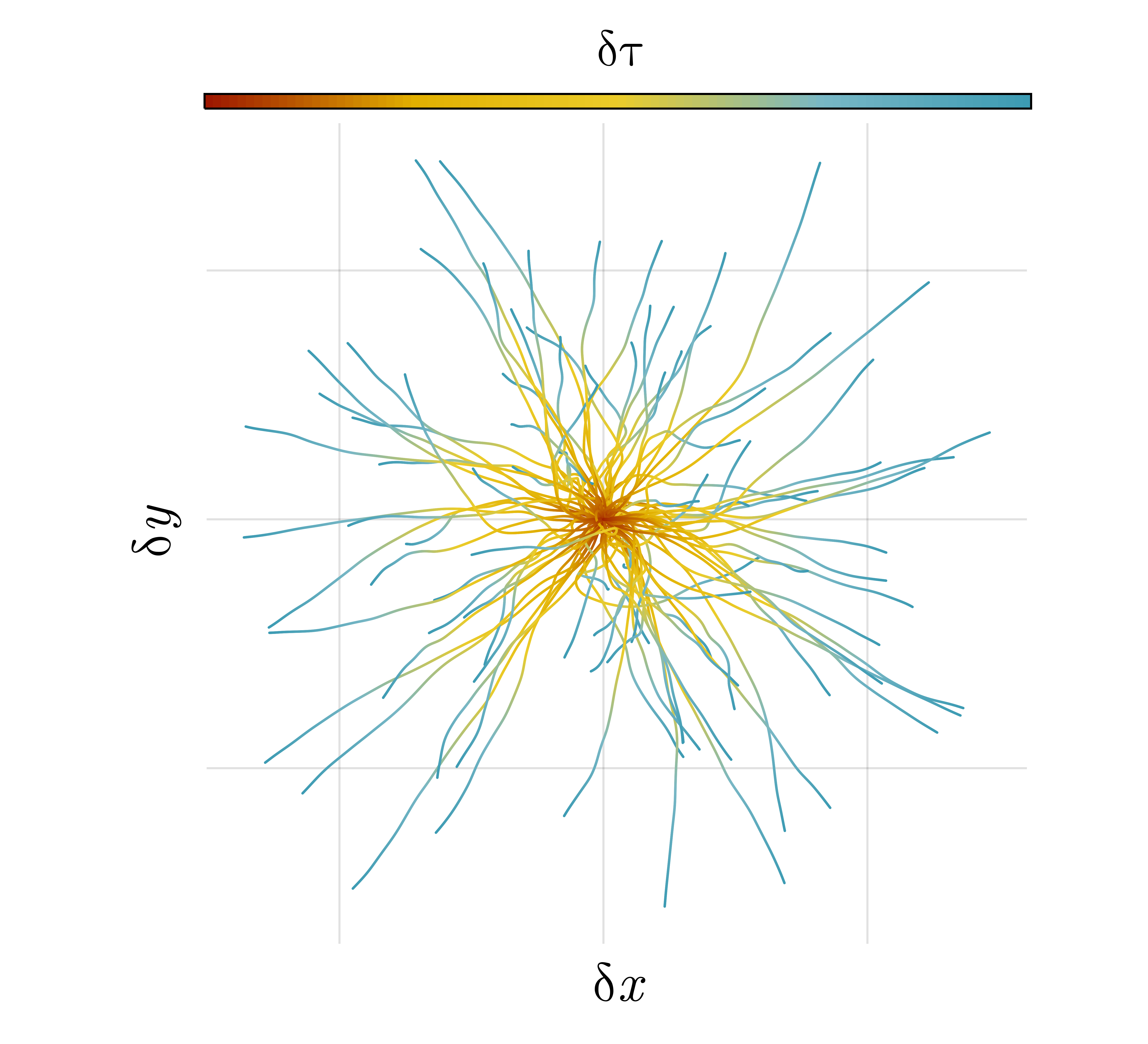}
}
\quad
\subfigure[Trajectories in momentum space]{\includegraphics[width=0.9\columnwidth]
{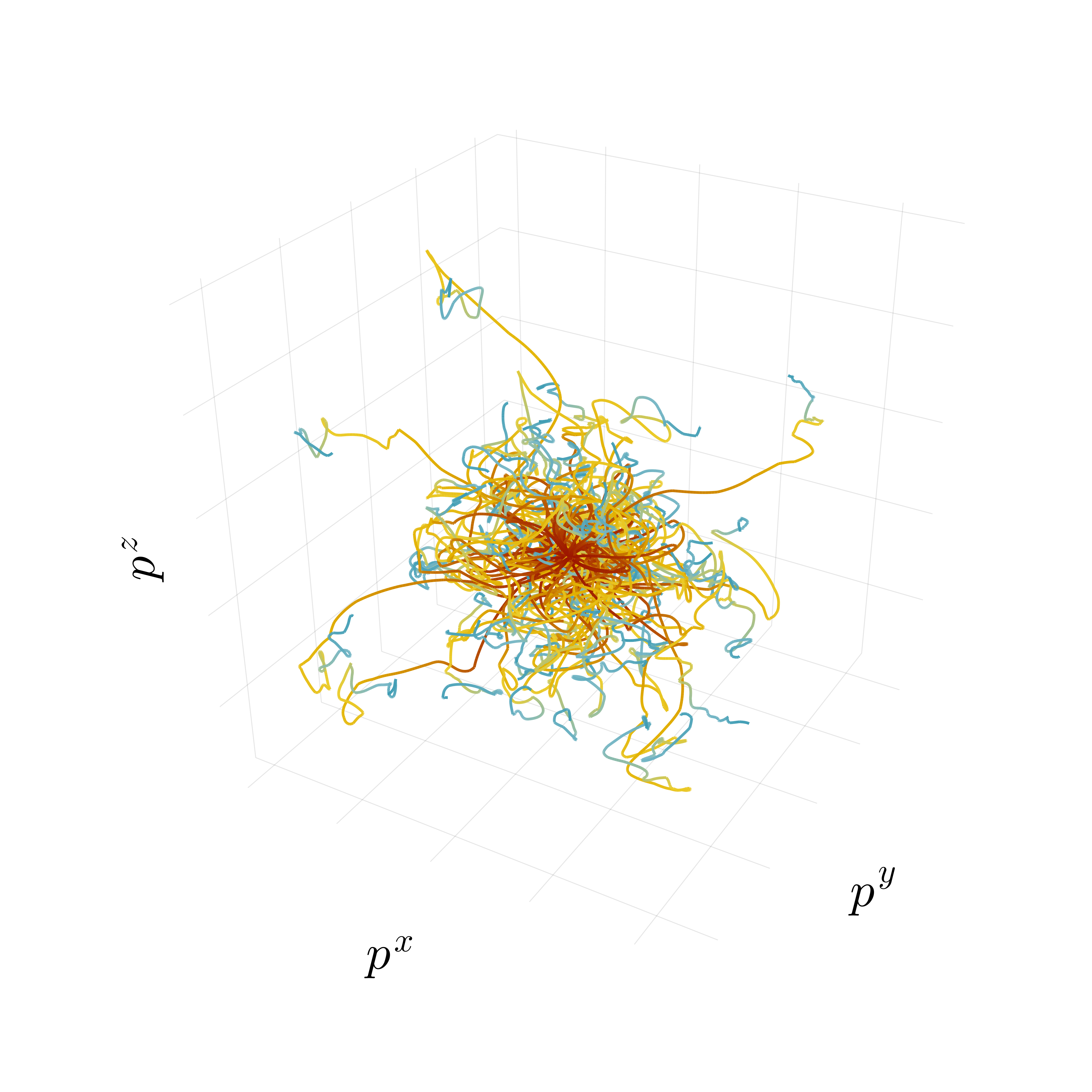}
}
\caption{Trajectories of charm quarks in Glasma, simulated with our particle solver, where the proper time evolution of (a) positions and (b) momenta are given by Eqs.~\eqref{eq:wongpos} and~\eqref{eq:wonginmilne} for $N_\mathrm{tp}=100$ test particles, all initialized with zero $p^{x,y,z}(\tau_\mathrm{form})$. The color represents the value of the proper time difference $\delta\tau \equiv \tau-\tau_\mathrm{form}$ at which the coordinates or momenta are evaluated.
}
\label{fig:glasma_noodles}
\end{figure*}

The positions and momenta of the partons are initialized using a toy model setup. Namely, the initial coordinates of the particles, chosen at formation time $\tau_\mathrm{form} \geq 0$, are randomly distributed in the transverse plane $x(\tau_\mathrm{form})$, $y\,(\tau_\mathrm{form}) \in [0,L]$ with $\eta(\tau_\mathrm{form})=0$. The particles initially only have transverse momenta $p_T=\sqrt{(p^x)^2+(p^y)^2}$ at formation time, with fixed $p_T(\tau_\mathrm{form})$ and $p^\eta(\tau_\mathrm{form})=0$ for heavy quarks, or an initial $p^x(\tau_\mathrm{form})$ and $p^{y,\eta}(\tau_\mathrm{form})=0$ for jets propagating along the $x$-axis. As will become evident in Sec.~\ref{sec:limcases}, we choose jets with initial momenta along $x$-direction in order to compare with previous studies having the same particle setup. Color charges are randomly sampled using Darboux variables for SU(2) or using the Haar measure for SU(3), and their associated classical Casimirs are fixed according to Eqs.~\eqref{eq:quad_casimir} and~\eqref{eq:cubic_casimir}. A complete description of how these classical color charges are constructed is given in Section \ref{sec:classiccharges} and Appendix \ref{appen:clascol}. 

Numerically, the Milne proper time evolution for positions and momenta from Eqs.~\eqref{eq:wongpos} along with~\eqref{eq:wonginmilne} and also~\eqref{eq:ptaueq} for the temporal constraint is solved with Euler's method. An example of the numerical solutions of these equations for particles propagating in Glasma fields is depicted in Fig.~\ref{fig:glasma_noodles}. The Glasma electric and magnetic fields from Eq.~\eqref{eq:glasmafields_main}, which appear in Wong's momenta equations given in Eq.~\eqref{eq:wongcurv}, have to be approximated on the lattice. This is because the electric fields reside on gauge links, the magnetic ones on plaquettes and we need to interpolate in order to get their value on a lattice site. Appropriate approximations that are accurate up to quadratic order in the lattice and time spacing are given by
\begin{widetext}
    \begin{equation}
        \begin{aligned}
            &E_{i}\left(\tau_n,{\boldsymbol{x}_n}\right)=\dfrac{1}{\tau_{n}} P^{i}\left(\tau_{n}, {\boldsymbol{x}_n}\right)\approx \dfrac{1}{4 \tau_{n}}\Bigg\{P_{\boldsymbol{x}_n}^{i}\left(\tau_{n}+\dfrac{\Delta \tau}{2}\right)+P_{\boldsymbol{x}_n}^{i}\left(\tau_{n}-\dfrac{\Delta \tau}{2}\right)\\
            &\phantom{E_{i}\left(\tau_n,{\boldsymbol{x}_n}\right)=\dfrac{1}{\tau_{n}} P^{i}\left(\tau_{n}, {\boldsymbol{x}_n}\right)\approx}+U_{{\boldsymbol{x}_n},-\hat{i}}\left(\tau_{n}\right)\left[P_{{\boldsymbol{x}_n}-\hat{i}}^{i}\left(\tau_{n}+\dfrac{\Delta \tau}{2}\right)+P_{{\boldsymbol{x}_n}-\hat{i}}^{i}\left(\tau_{n}-\dfrac{\Delta \tau}{2}\right)\right] U_{{\boldsymbol{x}_n},-\hat{i}}^{\dagger}\left(\tau_{n}\right)\Bigg\},\\
            &E_{\eta}\left(\tau_{n}, {\boldsymbol{x}_n}\right) =P^{\eta}\left(\tau_{n}, {\boldsymbol{x}_n}\right) \approx \dfrac{1}{2}\left[P_{{\boldsymbol{x}_n}}^{\eta}\left(\tau_{n}+\dfrac{\Delta \tau}{2}\right)+P_{{\boldsymbol{x}_n}}^{\eta}\left(\tau_{n}-\dfrac{\Delta \tau}{2}\right)\right],
        \end{aligned}
    \end{equation}
where $i=x,y$, and similarly
\begin{align}
    &B_{i}\left(\tau_{n}, {\boldsymbol{x}_n}\right)=-\dfrac{1}{\tau_{n}} D_{i} A_{\eta}\left(\tau_{n}, {\boldsymbol{x}_n}\right)\approx-\dfrac{1}{2 \tau_{n} a_{T}} \left[U_{{\boldsymbol{x}_n}, \hat{i}}\left(\tau_{n}\right) A_{{\boldsymbol{x}_n}+\hat{i}, \eta}\left(\tau_{n}\right) U_{{\boldsymbol{x}_n}, \hat{i}}^{\dagger}\left(\tau_{n}\right)-U_{{\boldsymbol{x}_n},-\hat{i}}\left(\tau_{n}\right) A_{x-\hat{i}, \eta}\left(\tau_{n}\right) U_{{\boldsymbol{x}_n},-\hat{i}}^{\dagger}\left(\tau_{n}\right)\right],\notag\\
    &B_{\eta}\left(\tau_{n}, {\boldsymbol{x}_n}\right)=-F_{x y}\left(\tau_{n}, {\boldsymbol{x}_n}\right)\approx- \dfrac{1}{4 g a_{T}^{2}}\left[U_{{\boldsymbol{x}_n}, \hat{x} \hat{y}}\left(\tau_{n}\right)+U_{{\boldsymbol{x}_n}, \hat{y}\,-\!\hat{x}}\left(\tau_{n}\right)+U_{{\boldsymbol{x}_n},-\!\hat{x}\,-\!\hat{y}}\left(\tau_{n}\right)+U_{{\boldsymbol{x}_n},-\!\hat{y} \hat{x}}\left(\tau_{n}\right)\right]_{\mathrm{ah}}.
\end{align}
\end{widetext}
When evaluating these expressions, the positions of the particles are approximated with the nearest grid point (NGP) on the transverse lattice of the Glasma. Thus, the color-electromagnetic fields that are exerted on the partons are computed at $\boldsymbol{x}_n\equiv \Big[\vec{x}_\perp(\tau_{n})\big]^\mathrm{NGP}$. 

\begin{figure}[t]
    \includegraphics[width=0.9\columnwidth]{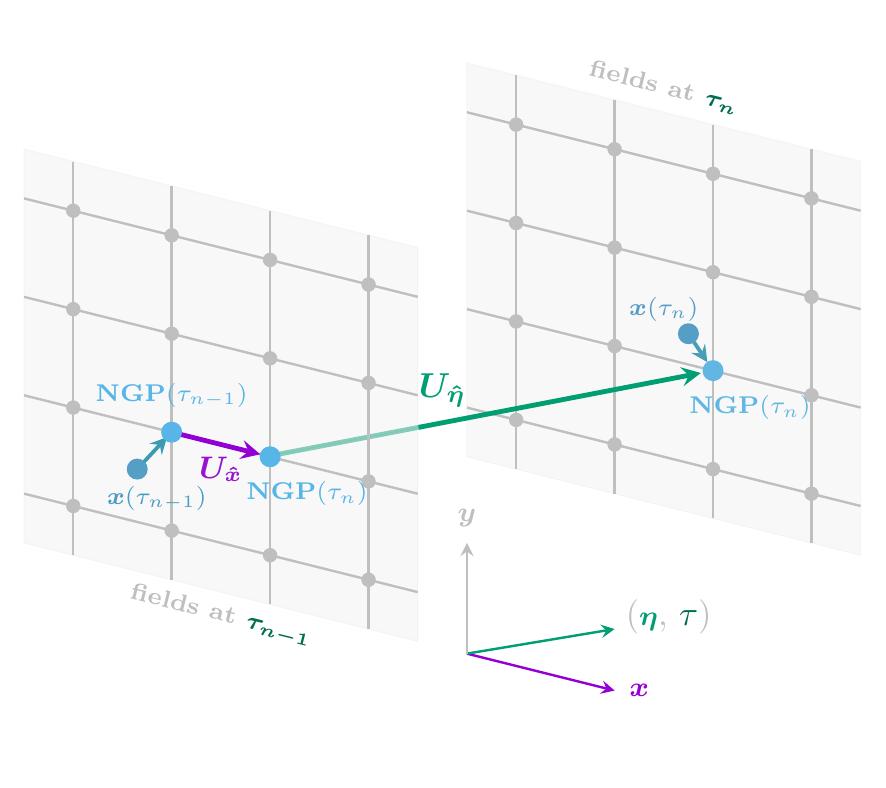}
    \caption{\label{fig:colorrotation} Diagram with the color rotation performed during a numerical time step from $\tau_{n-1}$ to $\tau_n$. The electric and magnetic Glasma fields reside on lattice points in the transverse plane $\boldsymbol{x}(\tau_n)\equiv\boldsymbol{x}_n$, while a particle may move at any location in the transverse plane. The particle position is approximated with the NGP on the lattice $\boldsymbol{x}(\tau_n)\mapsto \mathrm{NGP}(\tau_n)$ and when the transverse coordinates of the NGP change, one performs a color rotation with the corresponding transverse gauge link, in this case $U_{\hat{x}}$. Along the rapidity direction, a Wilson line $U_{\hat{\eta}}$ is computed via the matrix exponential and used in the color rotation, which here is simply given by $\mathcal{U}(\tau_{n-1},\tau_n)=U_{\hat{x}}(\tau_n) U_{\hat{\eta}}(\tau_n)$.
    }
\end{figure}

The numerical solution for the rotation of the color charge is more involved and relies on the same NGP approximation. It is inspired by colored particle-in-cell (CPIC) methods used in the context of particles in CYM plasmas \cite{Hu:1996sf,Moore:1997sn,Dumitru:2006pz,Schenke:2008hw}. In the CPIC method, the color charge of a particle is rotated with gauge links only when the NGP on the underlying simulation lattice of the background YM fields changes. It should be emphasized that the Glasma fields are discretized only in the transverse plane because of boost invariance, and the rapidity direction is left continuous. Thus, one needs to adapt the CPIC method to the Glasma lattice discretization with gauge links only in the transverse plane.
Numerically, one may approximate the Wilson line from Eq.~\eqref{eq:wilson_path}, namely $\mathcal{U}(\tau_i, \tau_f)$ at a given proper time $\tau_f$ as being comprised of subsequent products of ``short'' Wilson lines $\mathcal{U}(\tau_{n-1},\tau_n)$ as $\mathcal{U}(\tau_i, \tau_f)\approx \mathcal{U}(\tau_i,\tau_{i+1})\mathcal{U}(\tau_{i+1},\tau_{i+2})\dots\mathcal{U}(\tau_{f-1},\tau_{f})$. These short Wilson lines may be reduced to
\begin{equation}
\label{eq:colorrotnum}
    \begin{aligned}
        \mathcal{U}(\tau_{n-1},\tau_n) & \simeq \exp\Bigg(\mathrm{i}g \int\limits_{\boldsymbol{x}_{n-1}}^{\boldsymbol{x}_n}\mathrm{d}x^{\prime\,i} A_i\left(\boldsymbol{x}^\prime\right)\Bigg)\\
        & \times \exp\left(\mathrm{i}g \delta\eta_n A_\eta(\boldsymbol{x}_n)\right)\\
        & = U_{\boldsymbol{x}_{n-1},\hat{i}}(\tau_n) U_{\boldsymbol{x}_{n},\hat{\eta}}(\tau_n).
    \end{aligned}
\end{equation} 
Here, $U_{\boldsymbol{x}_{n},\hat{i}}(\tau_n)$ is a transverse gauge link along the direction $\hat{i}$ with $i=x,y$ evaluated at position $\boldsymbol{x}_n$, while $U_{\boldsymbol{x}_{n},\hat{\eta}}(\tau_n)$ represents a Wilson line along the $\hat{\eta}$ direction, which can be computed from $A_\eta$ via the matrix exponential. It should be noted that this approximation is only valid for small time steps $\delta\tau_n=\tau_n - \tau_{n-1}$. We have made use of the fact that the displacement in rapidity $\delta\eta_n$ is numerically small and it follows that $\left[\int\mathrm{d}x^iA_i,\delta\eta_n A_\eta\right]\simeq0$ such that higher order terms arising from the Baker–Campbell–Hausdorff formula are suppressed. This numerical color rotation is depicted in Fig.~\ref{fig:colorrotation}.

Alternatively, one may directly solve
\begin{equation}
    \frac{\d Q}{\d\tau}=\mathrm{i}g\left([Q,A_x]\frac{p^x}{p^\tau}+[Q,A_y]\frac{p^y}{p^\tau}+[Q, A_\eta]\frac{p^\eta}{p^\tau}\right),
\end{equation}
where the transverse gauge fields are numerically extracted from the gauge links using matrix logarithms
\begin{equation}
\label{eq:colorchargeamu}
    \begin{aligned}
        &\mathrm{i}gaA_x\left(x+\dfrac{a}{2},y\right)=\ln(U_{\hat{x}}(x,y)), \\ &\mathrm{i}gaA_y\left(x,y+\dfrac{a}{2}\right)=\ln(U_{\hat{y}}(x,y)).
    \end{aligned}
\end{equation}
We checked that these two distinct methods for solving the evolution of the color charge, either from Eqs.~\eqref{eq:chargepropertime} or~\eqref{eq:colorchargeamu}, are consistent with each other in the limit of small time steps and yield similar final results for momentum broadening. Nevertheless, the advantage of performing numerical color rotations with Wilson lines as in Eq.~\eqref{eq:chargepropertime} lies in ensuring that the color charge remains in the Lie algebra, i.e.~$Q\in \mathfrak{su}(N_c)$, and that the Casimir invariants are exactly conserved. The Casimirs, Eqs.~\eqref{eq:q2_main} and \eqref{eq:q3_main}, remain unchanged throughout the evolution: once the values of the Casimirs are fixed at formation time, color rotations with Wilson lines will not affect them.


\section{Classical color charges}
\label{sec:classiccharges}

In the previous sections we have outlined how to numerically solve the field and particle equations on a lattice. The question remains how to choose the classical color charges $Q$ in the ensemble of partons. Here, we largely follow the seminal works on classical non-Abelian transport theory \cite{Kelly:1994dh,Litim:1999id,Litim:1999ns,Litim:2001db}. There are three aspects to consider: first, what values to assign to the classical Casimir invariants of the color charges from Eqs.~\eqref{eq:q2_main} and \eqref{eq:q3_main}; second, how to distribute the charges in color space (the particular color charge a parton assumes after a random hard scattering is a priori unknown, hence additional considerations are required in order to construct their distribution); third, how fixing one affects the other. We address these by treating the color charge components $Q^a$ as stochastic variables with fixed values of $q_2=Q^aQ^a$ and $q_3=d_{abc}Q^aQ^bQ^c$. We emphasize that this is a choice and in our framework, where the Casimirs remain constant throughout the evolution, see Sec.~\ref{subsec:casimirwong}, the obvious choice is to fix the Casimirs. In analogy with the trace relations for operator-valued elements of the $\mathfrak{su}(N_c)$ color algebra
\begin{gather}
\label{eq:trqhat}
    \begin{aligned}
        \mathrm{Tr}\big[\widehat{Q}^a\big]&=0,\\
        \mathrm{Tr}\big[\widehat{Q}^a\widehat{Q}^b\big]&=T_R \delta^{ab},\\
        \mathrm{Tr}\big[\widehat{Q}^a \widehat{Q}^b \widehat{Q}^c\big]&=\frac{A_R}{4}(d_{abc}+i f_{abc}),
    \end{aligned}
\end{gather}
we choose to have color charges randomly distributed according to one-, two- and three-point functions 
\begin{subequations}
    \label{eq:qnpointfct_main}
    \begin{align}
        \langle Q^a \rangle&=0, \label{eq:qa}\\ 
        \langle Q^a Q^b \rangle &=T_R \delta^{ab}, \label{eq:qaqb}\\
        \langle Q^a Q^b Q^c \rangle &=\frac{A_R}{4}d^{abc},\label{eq:qaqbqc}
    \end{align}
\end{subequations}
where the representation-dependent coefficients $T_R$ and $A_R$ are given by
\begin{align}
    \label{eq:dynkinanomaly_main}
    T_R = \left.
      \begin{cases}
        \dfrac{1}{2}, & R=F \\
        N_c, & R=A  
      \end{cases}
    \right.,\qquad
    A_R = \left.
      \begin{cases}
        1, & R=F \\
        0, & R=A  
      \end{cases}\right..
\end{align}

The Casimir invariants from Eqs.~\eqref{eq:q2_main} and \eqref{eq:q3_main} constrain what values the color charges $Q^a$ can take. It may be shown that the ansatz for the two- and three-point functions from Eqs.~\eqref{eq:qaqb} and~\eqref{eq:qaqbqc} fixes the quadratic and cubic classical Casimirs, defined in Eqs.~\eqref{eq:q2_main} and~\eqref{eq:q3_main}, to
\begin{subequations}
    \label{eq:q23_main}
        \begin{align} 
    \label{eq:quad_casimir}
        q_2(R) &= \left.
          \begin{cases}
            \dfrac{N_c^2-1}{2}, & R=F \\
            N_c (N_c^2-1), & R=A  
          \end{cases}
        \right., \\ 
        \label{eq:cubic_casimir}
        q_3(R) &= \left.
          \begin{cases}
            \dfrac{(N_c^2-4)(N_c^2-1)}{4 N_c}, & R=F \\
            0, & R=A  
          \end{cases}\right..
    \end{align}
\end{subequations}

We point out that assigning the labels ``fundamental'' and ``adjoint'' to the classical Casimirs is inspired by the corresponding quantum representations and is inherited from the choice we made in Eqs.~\eqref{eq:qnpointfct_main}. A detailed derivation of the classical Casimirs from Eqs.~\eqref{eq:quad_casimir} and~\eqref{eq:cubic_casimir} is given in Appendix \ref{appen:relnptfctcasimirs}. Here, we provide a sketch of the derivation. If we take the two-point function from Eq.~\eqref{eq:qaqb}, choose the color component $a=b$ and perform a sum over it, we get the classical quadratic Casimir $q_2(R)=D_A\,T_R$. Similarly, we start from the three-point function in Eq.~\eqref{eq:qaqbqc}, multiply by $d^{abc}$ and sum over all color indices. Eventually, with the normalization we chose in Eqs.~\eqref{eq:qnpointfct_main}, the quadratic and cubic classical Casimirs may be recast in the following form
\begin{equation}
    \label{eq:q23c23}
    q_{2,3}(R)=D_R\,C_{2,3}(R)
\end{equation}
where $D_R$ denotes the dimension of the representation, namely
\begin{align}
    \label{eq:dr_main}
    D_R = \left.
      \begin{cases}
        N_c, & R=F \\
        N_c^2-1, & R=A  
      \end{cases}
    \right.,
\end{align}
and $C_{2,3}(R)$ are the group-theoretical quadratic and cubic Casimirs, given here for the fundamental and adjoint representations as
\begin{subequations}
    \label{eq:c23_main}
    \begin{align}
    &C_2(R) = \left.
      \begin{cases}
        \dfrac{N_c^2-1}{2N_c}, & R=F \\
        N_c, & R=A  
      \end{cases}\right.,\\[0.5em]
    &C_3(R) = \left.
      \begin{cases}
        \dfrac{(N_c^2-4)(N_c^2-1)}{4N_c^2}, & R=F \\
        0, & R=A  
      \end{cases}\right..
    \end{align}
\end{subequations}
Combining Eq.~\eqref{eq:q23c23} with the definitions in Eqs.~\eqref{eq:dr_main} and~\eqref{eq:c23_main} yield the values for the classical Casimirs as a function of number of colors and representation as written in Eqs.~\eqref{eq:quad_casimir} and~\eqref{eq:cubic_casimir}. More details about the classical and group-theoretical color algebras and their Casimir invariants, a discussion about why the choice in Eq.~\eqref{eq:q23c23} is made, along with other useful relations, are all collected in Appendix \ref{appen:clascol}.

With the chosen normalization from Eqs.~\eqref{eq:qnpointfct_main}, the resulting classical Casimirs differ from the group-theoretical ones, see Eq.~\eqref{eq:q23c23}. This immediately raises the question whether the factor of $D_R$ in Eq.~\eqref{eq:q23c23} can be absorbed in the normalization of the color charges from Eqs.~\eqref{eq:qnpointfct_main}, such that the classical Casimirs of $Q^a$ automatically match the group-theoretical ones $q_{2,3} \mapsto q_{2,3} / D_R = C_{2,3}$. We found that this is not always possible, i.e.~the classical Casimirs do not coincide with the quantum ones for any gauge group or representation. In particular, when we consider quarks in SU(3), we were not able to find a color charge vector $Q^a$ which satisfies Eqs.~\eqref{eq:q2_main} and \eqref{eq:q3_main} with $q_{2,3}(F)=C_{2,3}(F)$. Although we do not have a formal proof, we believe that there exist no solutions to the color charge constraints for these particular values of $q_2$ and $q_3$. In Sec.~\ref{sec:divisionbydr} we discuss in more detail how this difference between classical and quantum color charges affects some particular expectation values, for example the momentum broadening $\langle \delta p^2\rangle$ defined in Eq.~\eqref{eq:mombroad}, and how to address it.

The initial random classical color charges of the partons at formation time must satisfy the above relations in order to describe the physics of heavy quarks, and jets of quarks and gluons. In the following two subsections we show how random color charges satisfying the above $n$-point functions can be numerically realized for SU(2) and SU(3).

\subsection{SU(2) classical color charges}
\label{subsec:su2charges}
For generating SU(2) classical color charges, we rely on the Darboux variables parametrization \cite{Johnson:1988qm,Litim:2001db}. One may generically construct the classical limit of any semi-simple Lie algebra \cite{Bulgac_1990}. This is done by starting from the defining commutation relations 
\begin{equation}
    \big[\widehat{Q}_a,\widehat{Q}_b\big]=\mathrm{i} f_{abc}\widehat{Q}_c,
\end{equation}
where $f_{abc}$ denote the structure constants and $\{\widehat{Q}_a\}$ is the set of operator-valued generators. Taking the reverse of the quantum limit, the classical correspondent is given by the Poisson bracket
\begin{equation}
    \left\{Q_a,Q_b\right\}_{\mathrm{PB}}=f_{abc}Q_c.
\end{equation}

If one interprets the generators as classical variables depending on the symplectic structure of the underlying manifold through some phase-space coordinates $(\phi_i,\xi_i)$, the Poisson brackets may be expressed as
\begin{equation}
    \label{eq:liepb}
    \left\{Q_a,Q_b\right\}_{\mathrm{PB}}=\sum_k\left(\frac{\partial Q_a}{\partial\phi_k}\frac{\partial Q_b}{\partial\xi_k}-\frac{\partial Q_a}{\partial\xi_k}\frac{\partial Q_b}{\partial\phi_k}\right).
\end{equation}

The pair of conjugate variables obey the canonical Poisson bracket relations $\left\{\phi_i,\xi_j\right\}=\delta_{ij}$ and are called Darboux variables. For SU(2), whose generators $\{Q_a\}$ with $a\in\{1,2,3\}$ obey Eq.~\eqref{eq:liepb} with $f_{abc}=\epsilon_{abc}$, one identifies a single pair $\{\phi,\xi\}$ and the subsequent phase-space evolution is restricted to conserve the quadratic Casimir from Eq.~\eqref{eq:q2_main}.

Simply distributing the color charges uniformly on a three-dimensional sphere of fixed radius $J^2$ ensures that Eq.~\eqref{eq:liepb} is satisfied and that the Casimir is fixed by $q_2=J^2$. The SU(2) color charges are sampled according to the parametrization
\begin{equation}
    \begin{aligned}
        Q_1&=\cos\phi\sqrt{J^{2}-\xi^{2}},\\
        Q_2&=\sin\phi\sqrt{J^{2}-\xi^{2}},\\ 
        Q_3&=\xi,
    \end{aligned}
\end{equation}
where $\phi \in [0, 2\pi)$ and $\xi \in [-J, J]$ are uniformly distributed random numbers.

\subsection{SU(3) classical color charges}
\label{subsec:su3charges}
Similar to SU(2), one may construct a parametrization for classical SU(3) color charges in terms of the Darboux variables \cite{Johnson:1988qm}. Unfortunately, any parametrization of classical color charges only covers a portion of the underlying manifold of SU(3) \cite{Bulgac_1990}, leading to ill-defined one-, two- and three-point functions that will differ from the expected ones given in Eq.~\eqref{eq:qnpointfct_main}. For this reason, we rely on a different method to sample them, namely through the Haar measure of SU(3).
The main idea is that the integration over color charge configurations may be mapped to integration over the underlying manifold of the group. This is done by first constructing an initial color vector $Q_0=Q_0^a T^a$ such that the quadratic and cubic Casimirs $Q^a_0 Q^a_0$ and $d_{abc} Q^a_0 Q^b_0 Q^c_0$ satisfy Eqs.~\eqref{eq:quad_casimir} and \eqref{eq:cubic_casimir}. The exact choice of $Q_0$ is arbitrary, as long as the Casimir invariants $q_{2,3}(R)$ match the desired values. Once the initial color vector is fixed, random color charges are generated by performing color rotations as
\begin{align}
    Q(U)=U Q_0 U^\dagger,
\end{align}
with a random special unitary matrix $U\in\mathrm{SU(3)}$ distributed according to the Haar measure. From this color vector $Q$, color components are given by projecting onto the generators $T^a$
\begin{align}
    Q^a = \frac{1}{T_R} \tr{Q T^a } = Q_0^b U^{ab},
\end{align}
where we introduced the adjoint representation matrix 
\begin{align}
    \label{eq:uadj}
    U^{ab}\equiv \frac{1}{T_R} \tr{T^a U T^b U^\dagger}.
\end{align}
By construction, the quadratic and cubic Casimirs are invariant to these color rotations.

We have to verify whether replacing the integration over classical SU(3) color charges with that over the SU(3) group elements as $\int \d Q\rightarrow \int \d U$ yields equivalent results. For this purpose, it suffices to check that the $n$-point functions of the color charges computed with the Haar measure
\begin{subequations}
    \begin{align}
    \label{eq:nptfcthaar}
        \langle Q^a\rangle_U&\equiv \int \d U Q^a \notag \\
        &=Q_0^{a^\prime} \int\d U\, U^{a a^\prime},\\
        \langle Q^a Q^b\rangle_U&\equiv \int\d U Q^aQ^b \notag \\
        &=  Q_0^{a^\prime}Q_0^{b^\prime} \int\d U\, U^{a a^\prime} U^{b b^\prime},\\
        \langle Q^aQ^bQ^c\rangle_U&\equiv \int\d U Q^aQ^bQ^c \notag \\ 
        &=  Q_0^{a^\prime}Q_0^{b^\prime}Q_0^{c^\prime} \int\d U\, U^{a a^\prime} U^{b b^\prime} U^{c c^\prime},
    \end{align}
\end{subequations}
along with the classical Casimirs fixed by Eqs.~\eqref{eq:quad_casimir} and \eqref{eq:cubic_casimir}, exactly match the $n$-point functions of the classical colors charges from Eq.~\eqref{eq:qnpointfct_main}, namely
\begin{subequations}
\label{eq:matchqs}
    \begin{align}
        \langle Q^a\rangle&=\langle Q^a\rangle_U,\label{eq:match1pointfct}\\
        \langle Q^aQ^b\rangle&=\langle Q^aQ^b\rangle_U,\label{eq:match2pointfct}\\
        \langle Q^aQ^bQ^c\rangle&=\langle Q^aQ^bQ^c\rangle_U.\label{eq:match3pointfct}
    \end{align}  
\end{subequations}
This can be explicitly checked by carrying out the required integrals for SU(3). A detailed calculation can be found in Appendix \ref{appen:samplecolors}. In particular, we show that the $n$-point functions become independent of the initial color charge $Q_0$, except for the values of the two Casimirs $q_2$ and $q_3$. In this way, the generation of classical color charges for SU(3) can be replaced by sampling over the Haar measure.


\section{Limiting cases}
\label{sec:limcases}

In general, the dynamics of colored particles passing through Yang-Mills background fields are non-trivial and can only be solved numerically, e.g.~with the methods introduced in earlier sections. However, there are certain limiting cases where the dynamics become trivial and observables such as the momentum broadening defined in Eq.~\eqref{eq:mombroad} can be reduced to simple functionals of the background fields. These limiting cases are those of infinitely massive heavy quarks and highly energetic jets. In both cases, the particle trajectories are trivial in the sense that the particles are not deflected by the forces acting on them.
 
 These cases are of interest since there exist numerous studies which rely on the infinitely massive heavy quark approximation, with momentum broadening and diffusion coefficient $\kappa$ extracted from electric fields correlators computed on the lattice \cite{Banerjee:2011ra, Brambilla:2020siz, Boguslavski:2020tqz, Altenkort:2020fgs}, and the highly energetic jet scenario, with accumulated momentum and transport coefficient $\hat{q}$ related to light-like Wilson loops \cite{Liu:2006ug, Casalderrey-Solana:2007ahi, Panero:2013pla, Ipp:2020mjc, Ipp:2020nfu}. Moreover, they represent valuable numerical checks for our particle solver which, in these limiting cases, should give similar momentum broadenings as those extracted solely from Glasma fields.

 Using the formal solution for the evolution of the color charge from Eq.~\eqref{eq:chargepropertime}, one may recast Wong's equation in the Milne frame from Eq.~\eqref{eq:wongq} in the following form
\begin{align}
    \label{eq:dpmutransp_main}
    \frac{\d p_\mu}{\d\tau}=\dfrac{g}{T_R}Q_0^a\,\tr{T^a\widetilde{\mathcal{F}}_{\mu}},
\end{align}
where
\begin{align}
    \label{eq:transplorentzforce}
    \widetilde{\mathcal{F}}_{\mu}(\tau)\equiv \mathcal{U}^\dagger(\tau,\tau_0)\,\mathcal{F}_{\mu}(\tau)\,\mathcal{U}(\tau,\tau_0)
\end{align}
denotes the parallel transported color Lorentz force with
\begin{align}
    \label{eq:colorlorentz_main}
    \mathcal{F}_{\mu}\equiv F_{\mu\nu}\dfrac{p^\nu}{p^\tau},
\end{align}
and the initial color charge vector is expressed as $Q_0\equiv Q(\tau_\mathrm{form})=Q_0^a T^a$. Equipped with Eq.~\eqref{eq:dpmutransp_main}, the momentum broadening from Eq.~\eqref{eq:mombroad} may be written as
\begin{align}
    \label{eq:mombroadq2}
    \begin{split}
    &\langle\delta p_\mu^2(\tau)\rangle_R=\dfrac{g^2}{T_R^2} \int\mathrm{d}Q\,Q^a_0\,Q^b_0\\
    &\times\int\limits_{\tau_\mathrm{form}}^{\tau}\mathrm{d}\tau^\prime\int\limits_{\tau_\mathrm{form}}^{\tau}\mathrm{d}\tau^{\prime\prime}\,\Big\langle\tr{T^a \widetilde{\mathcal{F}}_{\mu}(\tau^\prime)}\tr{T^b \widetilde{\mathcal{F}}_{\mu}(\tau^{\prime\prime})}\Big\rangle_R,
    \end{split}
\end{align}
where no sum over $\mu$ is implied. Using the two-point function of the color charges chosen according to Eq.~\eqref{eq:qaqb}, along with the Fierz identity expressed as 
\begin{align}
    \tr{T^a \mathcal{X}}\tr{T^a \mathcal{Y}}=T_R\,\tr{\mathcal{X}\mathcal{Y}},
\end{align}
valid for traceless $N_c \times N_c$ complex matrices, we arrive at the formal solution for the momentum broadening
\begin{gather}
    \begin{aligned}
        \label{eq:mombroadforcecorr}
        &\big\langle\delta p_\mu^2(\tau)\big\rangle_R\\
        &=g^2 \int\limits_{\tau_\mathrm{form}}^{\tau}\mathrm{d}\tau^\prime\int\limits_{\tau_\mathrm{form}}^{\tau}\mathrm{d}\tau^{\prime\prime}\,\Big\langle\tr{\widetilde{\mathcal{F}}_{\mu}(\tau^\prime)\widetilde{\mathcal{F}}_{\mu}(\tau^{\prime\prime})}\Big\rangle_R.
    \end{aligned}
\end{gather}

\subsection{Infinitely massive heavy quarks}
 An infinitely massive heavy quark is static and remains at rest in the Milne frame. Due to temporal gauge, all temporal Wilson lines are unity $\mathcal{U}(\tau, \tau^\prime)=\mathbb{1}$. Therefore, no parallel transport is required, thus $\widetilde{\mathcal{F}}_{\mu}=\mathcal{F}_{\mu}$ according to Eq.~\eqref{eq:transplorentzforce}. Furthermore, in the infinite mass limit $m\rightarrow\infty$, the temporal component of the four-momentum simply behaves as $p^\tau\rightarrow\infty$. Thus, the Lorentz force contains only contributions from the electric fields
\begin{align}
    \mathcal{F}_{i}= F_{i\mu}\dfrac{p^\mu}{p^\tau}\xrightarrow{p^\tau\rightarrow\infty} F_{i\tau} = -E_i, \quad i \in \{x,y,\eta\}.
\end{align}
The momentum broadening for static particles thus reduces to an integral over electric field correlators
\begin{gather}
    \begin{aligned}
        \label{eq:mombroadhqs}
        &\big\langle\delta p_i^2(\tau)\big\rangle_{m\rightarrow\infty}\\
        &=g^2 \int\limits_{\tau_\mathrm{form}}^{\tau}\mathrm{d}\tau^\prime\int\limits_{\tau_\mathrm{form}}^{\tau}\mathrm{d}\tau^{\prime\prime}\,\Big\langle\tr{E_i(\tau^\prime)E_i(\tau^{\prime\prime})}\Big\rangle_R
    \end{aligned}
\end{gather}
where no sum over $i$ is implied and the fields are evaluated at some fixed transverse coordinate. This expression can be evaluated purely from color-electric fields, without the need to solve the dynamical particle equations of motion.

\subsection{Highly energetic light-like jets}
The case of a highly energetic jet moving through the Glasma has already been studied in \cite{Ipp:2020mjc,Ipp:2020nfu} and here only the final results are quoted, as the derivation is analogous to the case of static particles. The momentum broadening of a light-like parton traveling along the $x$-axis in Glasma fields is given by
\begin{align}
    \label{eq:mombroadjets}
    \big\langle \delta p_i^2(\tau)\big\rangle_{p^x\rightarrow\infty}=g^2 \int\limits_0^\tau \d \tau^{\prime} \int\limits_0^\tau \d \tau^{\prime \prime}\big\langle\tr{\widetilde{f}_i(\tau^{\prime}) \widetilde{f}_i(\tau^{\prime \prime})}\big\rangle_R,
\end{align}
since for jets we assume $\tau_\mathrm{form}=0$. The various components of the Lorentz force are evaluated using the Glasma color electric and magnetic fields as
\begin{align}
    f_x\equiv E_x, \quad f_y\equiv E_y-B_z, \quad f_z\equiv E_z+B_y.
\end{align}
These color field components have to be parallel transported according to
\begin{align}
    \widetilde{f}_i(\tau)\equiv \mathcal{U}_x^\dagger(\tau,\tau_0)f_i(\tau)\,\mathcal{U}_x(\tau,\tau_0),
\end{align}
using a Wilson line constructed along $x$ as
\begin{align}
    \mathcal{U}_x(\tau, \tau_0)=\mathscr{P} \exp \left(-\mathrm{i} g \int\limits_0^\tau \d \tau^{\prime} A_x(\tau^{\prime})\right).
\end{align}

\section{Mapping classical to quantum expectation values}
\label{sec:divisionbydr}

In our classical framework, we can express classical expectation values of arbitrary observables $\mathscr O[Q, A_\mu]$ via functional integrals
\begin{align}
    \langle \mathscr O[Q, A_\mu] \rangle^\mathrm{classic} = \int dQ \int \mathscr{D} A_\mu \, W[A_\mu] \, \mathscr O[Q, A_\mu].
\end{align}
The integration over initial color charges is replaced by the Haar measure over SU($N_c$), with particular choices for the Casimir invariants, as in ~Eqs. \eqref{eq:quad_casimir} and \eqref{eq:cubic_casimir}, and the functional integration over the background field is an average over the Glasma initial conditions encoded in a probability functional $W[A_\mu]$. In order to reproduce the correct physics using classical calculations, these classical expectation values should match quantum expectation values computed for example in pQCD in the limit where the classical approximation is appropriate. Due to the overoccupied nature of the gluon field in the early stages of the collision, this approximation is valid for the background field, but in a strict sense fails when we approximate quarks and gluons as classical color charges. The problem is that quarks and gluons are low dimensional representations of the color algebra, whereas classical color charges are obtained in the limit of high dimensional representations (for example, in the case of SU(3) quarks, we found no classical color charges whose Casimirs coincide with the group-theoretical ones, see Sec.~\ref{sec:classiccharges} and Appendix~\ref{appen:clascol}). However, as we shall see below, the classical framework can nevertheless reproduce quantum expectation values for certain observables of interest, such as the momentum broadening defined in Eq.~\eqref{eq:mombroad}, by constructing a meaningful quantity whose classical expectation value correctly gets mapped to its quantum version.

To establish this relationship, we focus on the case of a light-like parton moving along the $x^+$-axis in the eikonal approximation, i.e.~in the case where the trajectory is fixed. Within pQCD and holography, momentum broadening may be related to particular Wilson loops \cite{Liu:2006ug, Casalderrey-Solana:2007ahi, Majumder:2009cf, DEramo:2010wup}.
In particular, momentum broadening orthogonal to the trajectory may be evaluated from a rectangular Wilson loop with one side parallel to the trajectory (light-like extent $L$) and the other side chosen to be spatial and orthogonal to $x^+$ (transverse extent $L_\perp$). In the small transverse extent limit $L_\perp \rightarrow 0$ one finds
\begin{align}
    \frac{1}{D_{R}}\Big\langle \Re \big\{\tr{W_{i+}} \big\}\Big\rangle_R=\exp \left(-\frac{L_\perp^{2}}{2}\left\langle p_{i}^{2}\right\rangle_R^{\mathrm{quantum}}\right),
\end{align}
where $\langle p_i^2 \rangle_R$ is the momentum broadening of a parton in representation $R$ along the transverse direction $\hat i$. The expectation value is taken over an ensemble of background fields. It is relevant to notice the factor $1/D_R$ in front, which ensures that the identity holds for $L_\perp \rightarrow 0$, when the Wilson loop reduces to $W_{i+}(R)\rightarrow\mathbb{1}_{D_R}$. This factor will play an important role when matching with the classical computation.  Performing a Taylor expansion in $L_\perp$, where the Wilson loop is written as
\begin{align}
    W_{i+}=\mathbb{1}+L_\perp W_{i+}^{(1)}+\frac{L_\perp^{2}}{2} W_{i+}^{(2)}+\mathcal{O}\left(L_\perp^{3}\right),
\end{align}
and inspecting the second-order coefficient yields the momentum broadening \cite{Ipp:2020mjc}
\begin{align}
    \label{eq:mombroadwilson}
    \left\langle p_{i}^{2}\right\rangle^\mathrm{quantum}_R=-\frac{1}{D_R}\Big\langle \Re \big\{\tr{W_{i+}^{(2)}} \big\}\Big\rangle_R.
\end{align}
From this relation, one expects $\Re \big\{\tr{\,\dots}\big\}\propto D_R \,C_2(R)$, thus $\left\langle p_{i}^{2}\right\rangle^\mathrm{quantum}_R\propto C_2(R)$.

In the case of classical background fields such as the Glasma, the second order coefficient can be written in terms of the field strength tensor via
\begin{equation}
    \label{eq:momrbroadpqcd}
    \begin{aligned}
    &\left\langle \delta p_{i}^{2}(\tau)\right\rangle_R^\mathrm{quantum}\\
    &=\frac{2g^2}{D_R} \int\limits_{\tau_{\mathrm{form}}}^{\tau} \d \tau^{\prime} \int\limits_{\tau_{\mathrm{form}}}^{\tau} \d \tau^{\prime \prime}\left\langle\tr{\widetilde{F}_{i+}(\tau^\prime) \widetilde{F}_{i+}(\tau^{\prime\prime})}\right\rangle_R,
    \end{aligned}
\end{equation}
where $\widetilde{F}_{i+}(\tau) \equiv \widetilde{F}_{i+}\big(x(\tau)\big)$ denotes the parallel transported field strength tensor which is given by
\begin{align}
    \widetilde{F}_{i+}\big(x(\tau)\big)=W_{+}\left(0, x^{+}\right) F_{i+}\big(x(\tau)\big) W_{+}\left(x^{+}, 0\right),
\end{align}
and contains the light-like Wilson line
\begin{align}
    W_{+}(x^+_2, x^+_1)=\mathscr{P}^+ \exp \Bigg(-\mathrm{i} g \int\limits_{x_1^+}^{x_2^+} \d x^{+} A_{+}\left(x^{+}\right)\Bigg).
\end{align}

On the other hand, we can compute the same expectation value within the classical particle framework. For a light-like trajectory $x^+=\sqrt{2}t$, see also Eq.~\eqref{eq:mombroadjets}, the momentum broadening is given by 
\begin{equation}
    \label{eq:mombroadclassic}
    \begin{aligned}
        &\big\langle\delta p_i^2(\tau)\big\rangle_R^\mathrm{classic}\\
        &=2g^2 \int\limits_{\tau_\mathrm{form}}^{\tau}\mathrm{d}\tau^\prime\int\limits_{\tau_\mathrm{form}}^{\tau}\mathrm{d}\tau^{\prime\prime}\,\Big\langle\tr{\widetilde{F}_{i+}(\tau^\prime)\widetilde{F}_{i+}(\tau^{\prime\prime})}\Big\rangle_R.
    \end{aligned}
\end{equation}
Moreover, as may be inferred from Eq.~\eqref{eq:mombroadq2}, one expects $\left\langle p_{i}^{2}\right\rangle^\mathrm{classic}_R\propto q_2(R)$. A direct comparison of Eq.~\eqref{eq:momrbroadpqcd} with~\eqref{eq:mombroadclassic} suggests that the classical computation may be mapped to the quantum one by considering
\begin{align}  
    \label{eq:match}
    \big\langle\delta p_i^2\big\rangle_R^\mathrm{classic}\big/q_2(R)\mapsto\big\langle\delta p_i^2\big\rangle_R^\mathrm{quantum}\big/C_2(R).
\end{align}
or equivalently, the classical expectation value coincides with the quantum one after a division by the dimension of the representation, since $D_R=q_2(R)/C_2(R)$ according to the classical Casimirs from Eqs.~\eqref{eq:q23c23}. An analogous calculation can be performed for infinitely massive partons, which involves a time-like Wilson loop instead of a light-like loop. Repeating the same steps, we arrive at the same factor of $D_R$ to match the classical to the quantum expectation value. We note that this division is also performed in \cite{Majumder:2009cf, Carrington:2016mhd}.

It should not be surprising that calculations based on classical colored particles do not entirely match pQCD calculations since, already on a formal level, there is an important difference between the high-dimensional classical and the low-dimensional quantum representations (see Appendix \ref{appen:clascol}), namely they are not labeled by the same Casimir invariants. Our choice given in Eq.~\eqref{eq:q23c23} shows that the classical Casimirs $q_{2,3}(R)$ are the dimension of the representation $D_R$ times the group-theoretical ones $C_{2,3}(R)$, which comes from how we choose to distribute the classical color charges, see Eq.~\eqref{eq:qnpointfct_main}. Thus, the origin of the difference between the \emph{classical} and the \emph{quantum} expectation values may be traced back to the statistical properties of the ensemble of classical color charges. Specifically, $\langle \delta p^2 \rangle$ is directly related to the classical two-point function $\langle Q^a Q^b \rangle$, see Eq.~\eqref{eq:mombroadforcecorr}, and thus, the quadratic classical Casimir $\langle \delta p^2 \rangle^\mathrm{classical}\propto q_2$. On the other hand, the quantum correspondent satisfies $\langle \delta p^2 \rangle^\mathrm{quantum}\propto C_2$. Therefore, when mapping classical to quantum expectation values, the meaningful quantity to compare is actually $\langle \delta p^2 \rangle/\mathscr{C}_2$, where $\mathscr{C}_2$ denotes either the classical or quantum quadratic Casimir, as stated in Eq.~\eqref{eq:match}.

Moreover, a similar argument which leads to Eq.~\eqref{eq:match} also works for $\langle\delta p^3\rangle$, namely $\langle \delta p^3\rangle/\mathscr{C}_3$ is the correct quantity to map from classical to quantum, where $\mathscr{C}_3$ denotes the classic or group-theoretical cubic Casimir. Nevertheless, it fails for $\langle\delta p^4\rangle$ or higher-order moments. As shown in \cite{Laine:2001my}, where the computation of the averages over the classical color charges is performed for SU(2), such a matching fails for the four-point function of the gauge field when compared to the 1-loop quantum effective action. It is only in the limit of high dimensional representations, where such a matching is exact. 
More concretely, the previous arguments generalize to $\langle \delta p^n\rangle^\mathrm{classic}\propto \langle Q^{a_1}\dots Q^{a_n}\rangle$. In analogy with Eq.~\eqref{eq:qnpointfct_main}, we choose 
\begin{equation}
    \int\d Q \, Q^{a_1}\dots Q^{a_n}=\tr{T^{(a_1}\dots T^{a_n)}}
\end{equation}
where $T^{(a} T^{b)}$ denotes the symmetric part of $T^aT^b$. As previously shown, such relations are satisfied for $n=1,2,3$ with $Q^a$ obeying the classical Casimir constraints in Eqs.~\eqref{eq:q23c23}, but are violated for $n \geq 4$. For consistency with pQCD calculations, the classical framework is thus limited to the quadratic and cubic moments of the momenta in a strict sense.

There is another important property of highly energetic jets that suggests that the matching condition in Eq.~\eqref{eq:match} is appropriate, namely Casimir scaling. 
The ratio of the accumulated momentum of the adjoint and fundamental representation must yield the ratio 
\begin{align}
    \label{eq:casimirscaling}
    \left\langle \delta p_{\mu}^{2}\right\rangle_{A}^\mathrm{quantum}\big/\left\langle \delta p_{\mu}^{2}\right\rangle_{F}^\mathrm{quantum}=C_2(A)\big/C_2(F),
\end{align}
as was already noted for Eq.~\eqref{eq:mombroadwilson}. This scaling with the ratios of the Casimir invariants is observed in many systems. For example, the Casimir scaling of the transverse momentum broadening coefficient $\hat{q}$ is a result inherited from pQCD computations of partons in weakly-coupled QGP \cite{Arnold:2008vd, Caron-Huot:2008zna} or in weakly-coupled $\mathcal{N}=4$ SYM \cite{Ghiglieri:2018ltw} and holds in the eikonal limit as in Eq.~\eqref{eq:momrbroadpqcd}.
Moreover, it is also a direct consequence of Wong's equations and the properties of the color charges provided that we use the matching condition in Eq.~\eqref{eq:match}.
To see this, we first rewrite Eq.~\eqref{eq:mombroadclassic} into
\begin{equation}
\label{eq:mombroadrepr}
    \begin{aligned}
        &\big\langle\delta p_\mu^2(\tau)\big\rangle^{\mathrm{classic}}_R\\
        &=T_R\underbrace{g^2 \int\limits_{\tau_{\mathrm{form}}}^{\tau}\mathrm{d}\tau^\prime\int\limits_{\tau_{\mathrm{form}}}^{\tau}\mathrm{d}\tau^{\prime\prime}\,\Big\langle\widetilde{\mathcal{F}}_{\mu}^{a}(\tau^\prime)\widetilde{\mathcal{F}}_{\mu}^{a}(\tau^{\prime\prime})\Big\rangle}_{\mathclap{\mathrm{independent\,of\,}R}}.
    \end{aligned}
\end{equation}
This expression states that the classical accumulated momentum for a colored parton in representation $R$ is simply proportional to the representation-dependent factor $T_R$. Consequently, since $T_{A}/T_{F}=q_2(A)/q_2(F)$, the classical accumulated momenta behave as
\begin{align}
    \label{eq:diffcasscaling}
    \left\langle \delta p_{\mu}^{2}\right\rangle_{A}^\mathrm{classic}\big/\left\langle \delta p_{\mu}^{2}\right\rangle_{F}^\mathrm{classic}=q_2(A)\big/q_2(F),
\end{align}
which resembles the Casimir scaling of Eq.~\eqref{eq:casimirscaling} but in terms of the classical Casimir from Eqs.~\eqref{eq:q2_main}. The division by $D_R$ of the classical momentum broadening, see Eq.~\eqref{eq:match}, restores it to a Casimir scaling with group-theoretical Casimirs
\begin{gather}
    \begin{aligned}
        \big(\left\langle \delta p_{\mu}^{2}\right\rangle_{A}^\mathrm{classic}\big/D_A\big)&\Big/\big(\left\langle \delta p_{\mu}^{2}\right\rangle_{F}^\mathrm{classic}\big/D_F\big)\\
        =\left\langle \delta p_{\mu}^{2}\right\rangle_{A}^\mathrm{quantum}&\big/\left\langle \delta p_{\mu}^{2}\right\rangle_{F}^\mathrm{quantum}
    \end{aligned}
\end{gather}
for a given SU($N_c$) group. Such a relation can equivalently be seen from the mapping proposed in Eq.~\eqref{eq:match}. We checked, with our particle solver, that the group-theoretical Casimir scaling for momentum broadenings divided by $D_R$ is satisfied throughout the evolution, see Fig.~\ref{fig:casimirscaling} and the discussion in Appendix \ref{appen:numchecks}.


\section{Results}
\label{sec:results}

In this section we apply the previously developed numerical methods to study momentum broadening of heavy quarks and jets in the early Glasma stage of heavy-ion collisions. To gain trust in our methods, we first compare our particle simulations to the limiting cases of infinitely heavy quarks and infinitely energetic jets. In Subsection \ref{subsec:hqsresults} we proceed with realistic simulations of dynamical heavy quarks, such as charm and beauty. Similarly, in Subsection \ref{subsec:jetsresults} realistic jet momentum broadenings are extracted and the jet transport coefficient is computed. 

\subsection{Choice of parameters}
\label{subsec:parameters}
For the Glasma, the saturation momentum is chosen as $Q_s=2\,\mathrm{GeV}$, while the MV model parameter is fixed through $g^2\mu\approx 0.8\,Q_s$ for $N_s=50$ color sheets and the IR regulator as $m=0.1 \,g^2\mu$, according to \cite{Lappi:2007ku}, and the UV regulator as $\Lambda=10\,\mathrm{GeV}$. The coupling constant is evaluated from the running coupling constant as $g^2=4\pi \alpha_s(Q_s)$ computed at a given saturation momentum
\begin{align}
    \alpha_s(Q_s)=\dfrac{1}{\dfrac{33-3 N_{f}}{12 \pi} \ln \dfrac{Q_s^2}{\Lambda_{\mathrm{QCD}}^2}} \approx 0.341,
\end{align}
with $N_f=3$ and $\Lambda_\mathrm{QCD}=200\,\mathrm{MeV}$, which yields $g \approx 2.07$. The rest of the numerical parameters of the Glasma are set as follows: the length of the simulation domain in the transverse plane is $L=10\,\mathrm{fm}$, the number of lattice points is $N=512$ for heavy quarks or $N=1024$ for jets. The time step $\Delta \tau$, which is used in the leapfrog scheme for the Glasma fields, is given in terms of the transverse lattice spacing $a_\perp = L / N$: for heavy quarks we use $\Delta \tau = a_\perp / 8$ and for jets we use $\Delta \tau = a_\perp / 16$. The numerical code for this work is an extension of an earlier Glasma code used in \cite{Ipp:2020mjc, Ipp:2020nfu} and is hosted publicly \cite{curraun}.

As initial conditions for classical particles, we rely  on a toy model initialization of positions and momenta. Namely, all partons are randomly distributed in the transverse plane, at mid-rapidity, and have a fixed initial transverse momentum. For heavy quarks, their formation time is given by $\tau_\mathrm{form}\approx 1/(2m_\mathrm{HQ})$, with $m_\mathrm{charm}=1.27\,\mathrm{GeV}$ and $m_\mathrm{beauty}=4.18\,\mathrm{GeV}$ \cite{ParticleDataGroup:2022pth}. All jets are formed instantaneously, at the same time as the Glasma fields. A single Glasma event contains $N_\mathrm{tp}=10^5$ test particles and most of the results are obtained for $N_\mathrm{events}=30$ Glasma events, although convergence was reached for fewer events. The transverse simulation region has periodic boundary conditions for the particles, whereas the rapidity direction is left continuous. We emphasize that the nuclei we simulate are not finite in size (their realistic geometry is not taken into account) and occupy a square lattice in the transverse plane, with periodic boundary conditions. Moreover, expecting that the Glasma picture holds up to $\tau\lessapprox 0.3\,\mathrm{fm/c}$, the details of the geometry and the transverse expansion are expected to be less relevant in the extraction of less sensitive quantities, for example the momentum broadening.

\subsection{Comparison with limiting cases}
\label{subsec:complimcases}

\begin{figure*}[!hbt]
\centering
\subfigure[Infinitely massive heavy quarks]{\includegraphics[width=0.9\columnwidth]{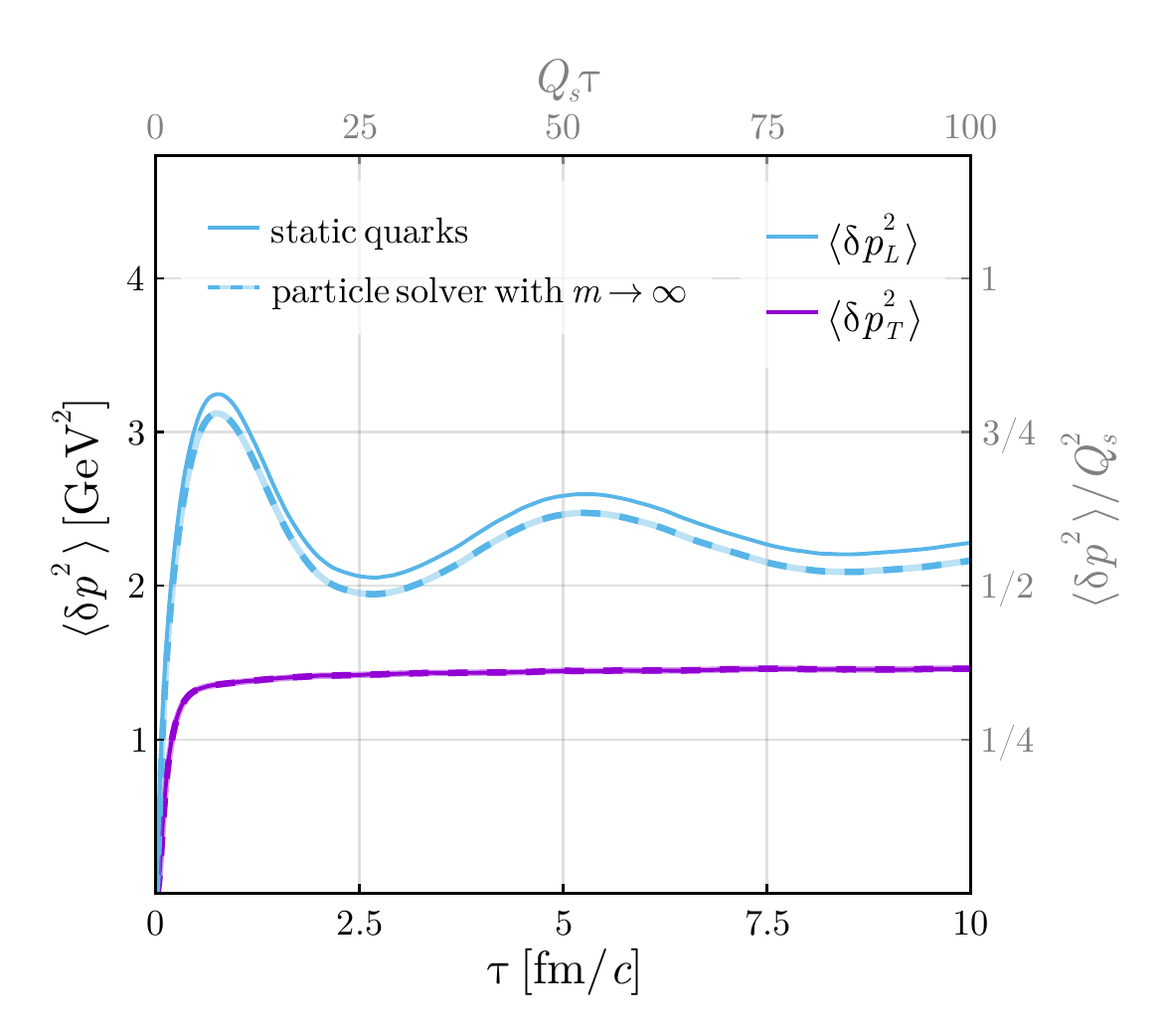}}
\hspace{0.1\columnwidth}
\subfigure[Highly energetic light-like jets]{\includegraphics[width=0.9\columnwidth]{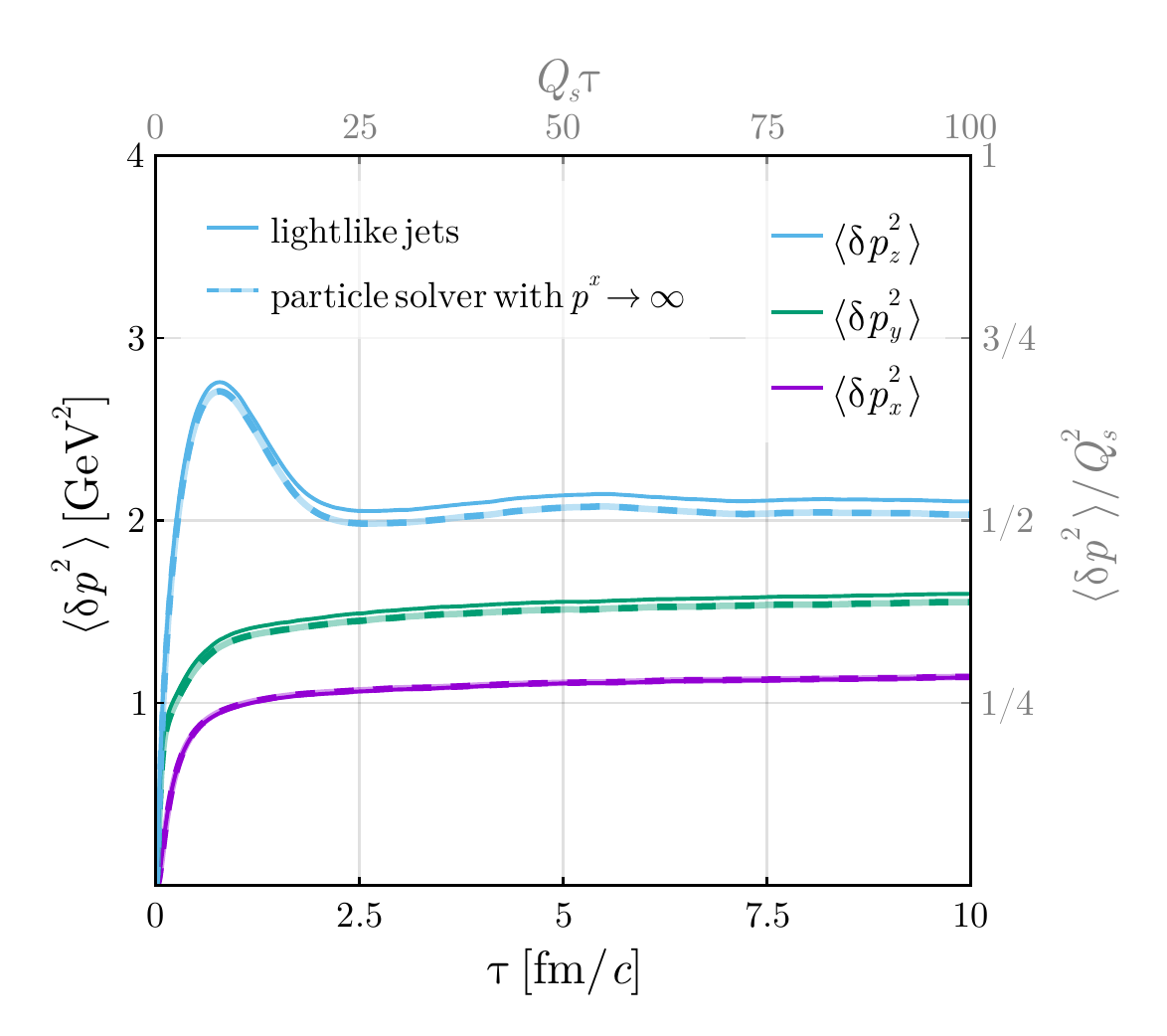}}
\caption{Comparison of the proper time evolution for longitudinal and transverse momentum broadenings, computed from expressions for limiting cases  \textit{(full lines)} or using the particle solver \textit{(dashed lines)}: (a) infinitely massive heavy quarks with accumulated momenta extracted from Eq.~\eqref{eq:mombroadhqs} vs.~particle solver with $m\rightarrow\infty$; (b) highly energetic light-like jets with momentum broadenings computed from Eq.~\eqref{eq:mombroadjets} vs.~particle solver with $p^x\rightarrow\infty$. For an easy comparison with previous works \cite{Ipp:2020mjc}, the corresponding dimensionless quantities are labeled \textit{(grey color)} on the remaining \textit{(upper and right)} axes. 
}
\label{fig:wong_kappa_qhat}
\end{figure*}

In the limit of infinite particle mass $m\rightarrow \infty$ (infinitely heavy quarks) or infinite spatial momentum  $p^x\rightarrow\infty$ (infinitely energetic jets), the dynamics of the particles become trivial and the accumulated momenta reduce to Eqs.~\eqref{eq:mombroadhqs} and \eqref{eq:mombroadjets} respectively. In order to validate our simulations, we compare results from the limiting cases (which were already used in \cite{Ipp:2020mjc,Ipp:2020nfu}) to simulations with classical particles in these particular limits. Our results are shown in Fig.~\ref{fig:wong_kappa_qhat}. For heavy quarks, we show the transverse $\delta p_T^2=\delta p_x^2+\delta p_y^2$ and the longitudinal $\delta p_L^2$ momentum components. We note that ``longitudinal'' refers to the component along the beam axis, whereas ``transverse'' denotes the orthogonal direction. For jets, we show all three independent components $\delta p_i^2$ with $i \in \{x, y, z\}$. In both cases, the longitudinal momentum broadening increases much faster than the transverse one at early times and reaches a maximum around $Q_s \tau \approx 10 $. Peculiarly, the longitudinal component for heavy quarks undergoes multiple damped oscillations before settling to a constant value at late times.  In contrast, the longitudinal component for light-like jets has only a single pronounced peak and then quickly saturates.   For both heavy quarks and jets, the transverse momentum broadening components increase rapidly at early times, due to strong coherent fields, but become essentially constant within $Q_s \tau \lesssim 10$. Similar phenomena have been observed in \cite{Boguslavski:2020tqz}, where heavy quark diffusion was studied in overoccupied gluonic systems without expansion. In these systems, the accumulated momenta of heavy quarks exhibit damped oscillations with the plasmon frequency. It is likely that the longitudinal component $\langle\delta p_z^2\rangle$ in the Glasma oscillates for a similar reason (plasmon excitations), although it is not clear why only the longitudinal component is affected.

As is evident from our data, both approaches yield the same results to a large degree. The slight numerical difference is due to the fact that for the limiting case result in terms of field correlators, we discretize over time the integrals in Eqs.~\eqref{eq:mombroadhqs} and \eqref{eq:mombroadjets} in steps of the transverse lattice spacing $a_\perp$. For our particle solver we typically use much smaller time steps $\Delta \tau \ll a_\perp$ leading to a slightly more accurate result. More concisely, we numerically checked that reducing the lattice spacing used in the particle solver (in order to make it ``less accurate'') lead to a better agreement with the limiting case result.

\subsection{Heavy quark momentum broadening}
\label{subsec:hqsresults}

\begin{figure*}[!hbt]
\centering
\subfigure[Momentum broadening of beauty quarks]{\includegraphics[width=0.8\columnwidth]{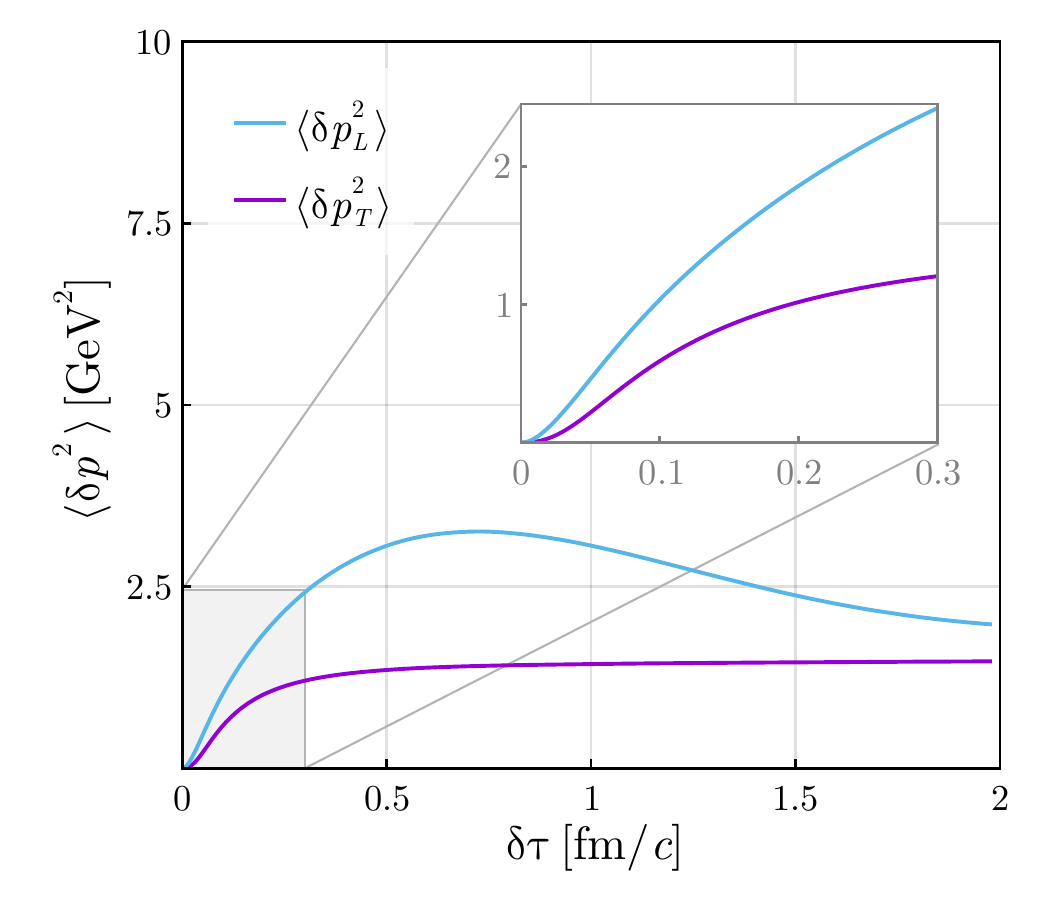} \label{fig:beautyearly_a}}
\hspace{0.1\columnwidth}
\subfigure[Derivative of momentum broadening]{\includegraphics[width=0.8\columnwidth]{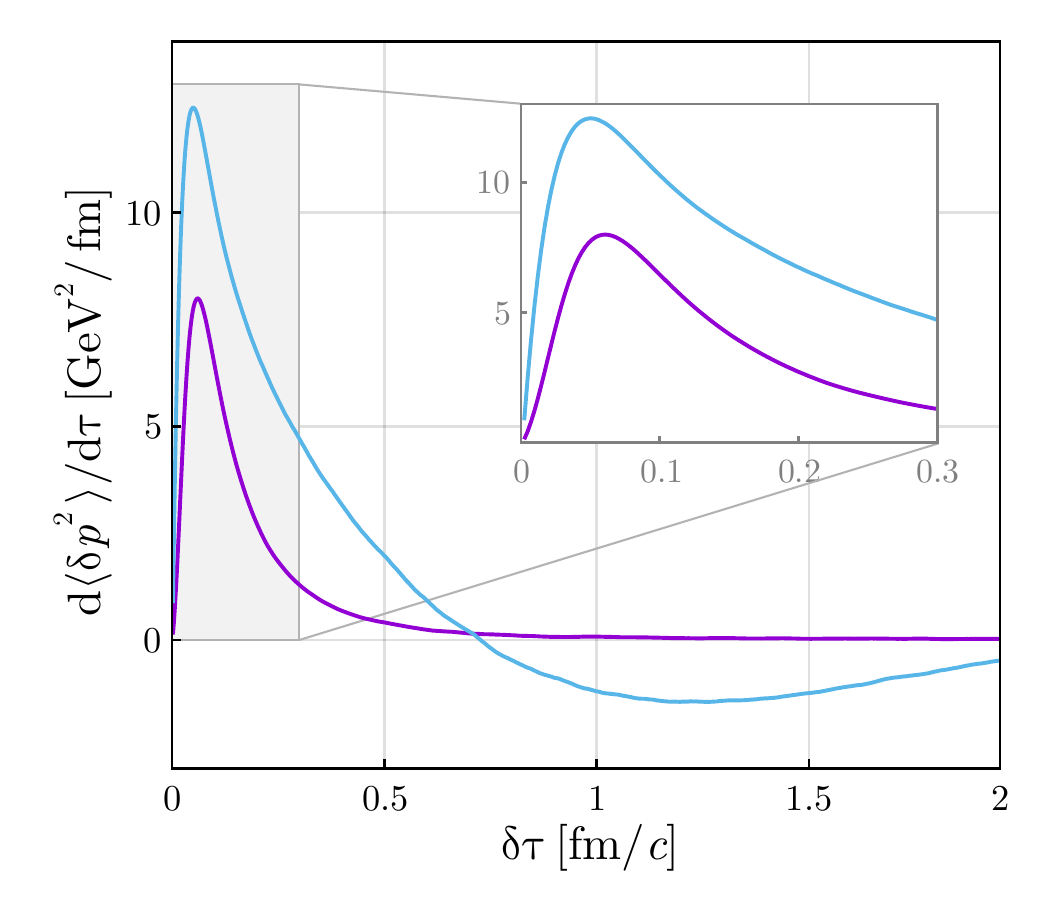} \label{fig:beautyearly_b}}
\caption{\label{fig:beautyearly} (a) Longitudinal and transverse momentum broadening components of beauty quarks formed at $\tau_\mathrm{form}\approx 0.02\,\mathrm{fm/}c$, initialized with $p_T(\tau_\mathrm{form})=0\,\mathrm{GeV}$, as a function of the time difference $\delta\tau\equiv\tau-\tau_\mathrm{form}$. (b) Derivatives of the accumulated momenta which give the transport coefficients according to Eq.~\eqref{eq:kappalt}.}
\end{figure*}

Having established that our particle simulations correctly reproduce limiting cases, we can now focus on more realistic simulations of heavy quarks with finite masses and finite formation times. Moreover, we can use our simulations to extract the heavy quark transport coefficient, which we define as
\begin{align}
    \label{eq:kappalt}
    \kappa^{\mathrm{inst}}_{L,T}(\tau)\equiv \frac{\d }{\d\tau}\langle\delta p^2_{L, T}(\tau)\rangle.
\end{align}
This is the instantaneous heavy quark coefficient and may be interpreted as a diffusion coefficient in the limit of large proper times, namely $\kappa^\mathrm{diffusion}=\lim_{\tau\rightarrow\infty }\kappa^{\mathrm{inst}}(\tau)$. Our results for beauty quarks with vanishing initial transverse momentum are shown in Fig.~\ref{fig:beautyearly}, where we plot the accumulated momenta and their time derivatives. As in the case of infinitely heavy quarks, the longitudinal momentum broadening component $\langle\delta p_L^2\rangle$ increases more rapidly than the transverse component $\langle\delta p_T^2\rangle$ at early times. Even though not shown here, we checked that the longitudinal and transverse momentum broadenings have the same behaviors at larger proper times $\tau\gg 2\,\mathrm{fm/}c$, as already noticed in Fig.~\ref{fig:wong_kappa_qhat} for static quarks. The first peak of the oscillations in $\langle\delta p_L^2\rangle$ happens at around $\delta \tau = 0.8 \, \mathrm{fm/}c$, giving rise to a temporary negative heavy quark diffusion coefficient $\kappa_L$. In contrast, the transverse component approaches a constant value after $\delta \tau \approx 0.5 \mathrm{fm} / c$. A qualitatively similar picture emerges for charm quarks. 

\begin{figure*}[!hbt]
\includegraphics[width=0.8\textwidth]{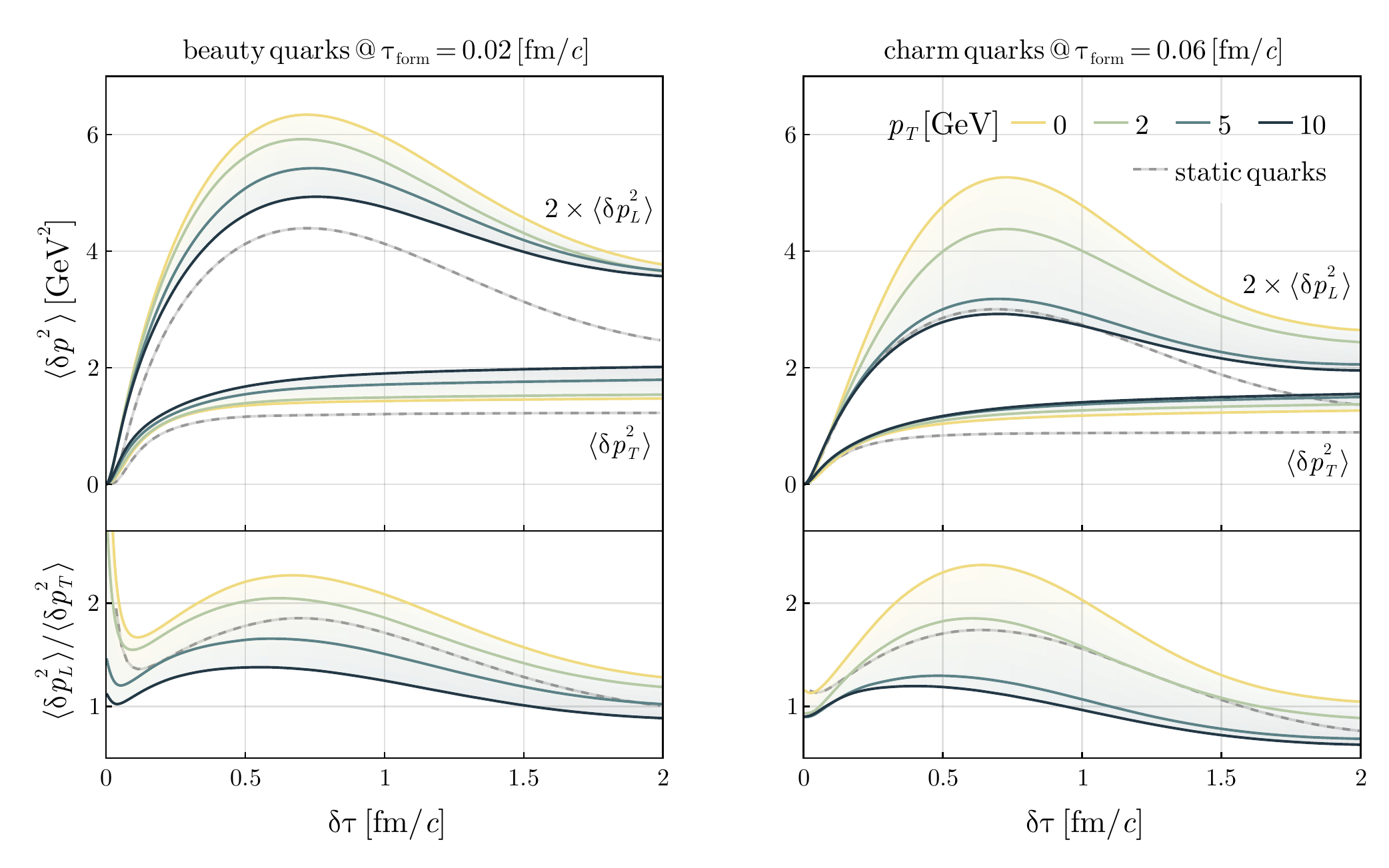}
\caption{\label{fig:wong_kappa} \textit{(Top)} Longitudinal and transverse momentum broadening components, along with  their ratio \textit{(bottom)}. The simulations are performed for \textit{(left)} beauty and \textit{(right)} charm quarks for various values of initial transverse momentum \textit{(colored full lines)}. We compare to the static case \textit{(grey dashed line)}, when the quarks are considered infinitely massive and the accumulated momentum is extracted solely from color-electric correlator, see Eq.~\eqref{eq:mombroadhqs}.
}
\end{figure*}

In general, the accumulation of momentum of heavy quarks depends not only on their mass (and thus formation time), but also their initial transverse momentum $p_T$. Figure~\ref{fig:wong_kappa} shows the numerical results for beauty and charm quarks for various values of the initial $p_T\in\{0,2,5,10\}\,\mathrm{GeV}$ \footnote{It should be noted that at the highest initial transverse momenta, these heavy quarks essentially behave like jets.}.  For comparison, we include the static quark limit as a dashed curve.  Since the Glasma affects the heavy quarks in an anisotropic manner, we also plot the heavy quark anisotropy coefficient, which we define as
\begin{align}
    \mathrm{heavy\,quark\,anisotropy}\equiv\dfrac{\langle\delta p_L^2\rangle}{\langle\delta p_T^2\rangle}.
\end{align}
Beauty quarks, due to their early formation time, experience the initial strong and coherent Glasma fields more than charm quarks. For this reason, their momentum broadening is generally larger than that of charm quarks. On average, beauty quarks acquire $30$-$50\%$ more momentum than charm quarks. A similar observation was also emphasized in \cite{Khowal:2021zoo}. 

Focusing on the heavy quark anisotropy, we find that as the initial $p_T$ increases, $\langle\delta p_L^2\rangle$ decreases and $\langle\delta p_T^2\rangle$ increases. Consequently, the corresponding anisotropy $\langle\delta p_L^2\rangle/\langle\delta p_T^2\rangle$ becomes smaller. Compared to the static quark accumulated momentum (dashed lines), beauty quarks with zero initial $p_T$ have an increase in $\langle\delta p_L^2\rangle$ of $50\%$ and charm quarks $50$-$80\%$ throughout the proper time evolution. For the maximum initial $p_T$ taken in our simulations, $\langle\delta p_L^2\rangle$ for dynamic quarks differs from that for static quarks by $50\%$ for beauty and $30$-$70\%$ for charm quarks, whereas $\langle\delta p_T^2\rangle$ increases only by $20$-$30\%$ compared to static quarks. These will have a complementary effect on the anisotropy. Namely, $\langle\delta p_L^2\rangle/\langle\delta p_T^2\rangle$ for beauty or charm quarks is $20$-$40\%$ larger or smaller, depending on the initial $p_T$, than that of infinitely massive heavy quarks formed at the same formation time. The anisotropy is higher for small initial $p_T$ heavy quarks and lower for quite large initial $p_T$. The anisotropy is more pronounced for the zero initial $p_T$ heavy quarks. Therefore, there are slight differences between static quarks that ``see'' only the Glasma electric fields, and quarks initialized with vanishing momentum but allowed to move in the Glasma. 

\begin{figure*}[!hbt]
\includegraphics[width=\textwidth]{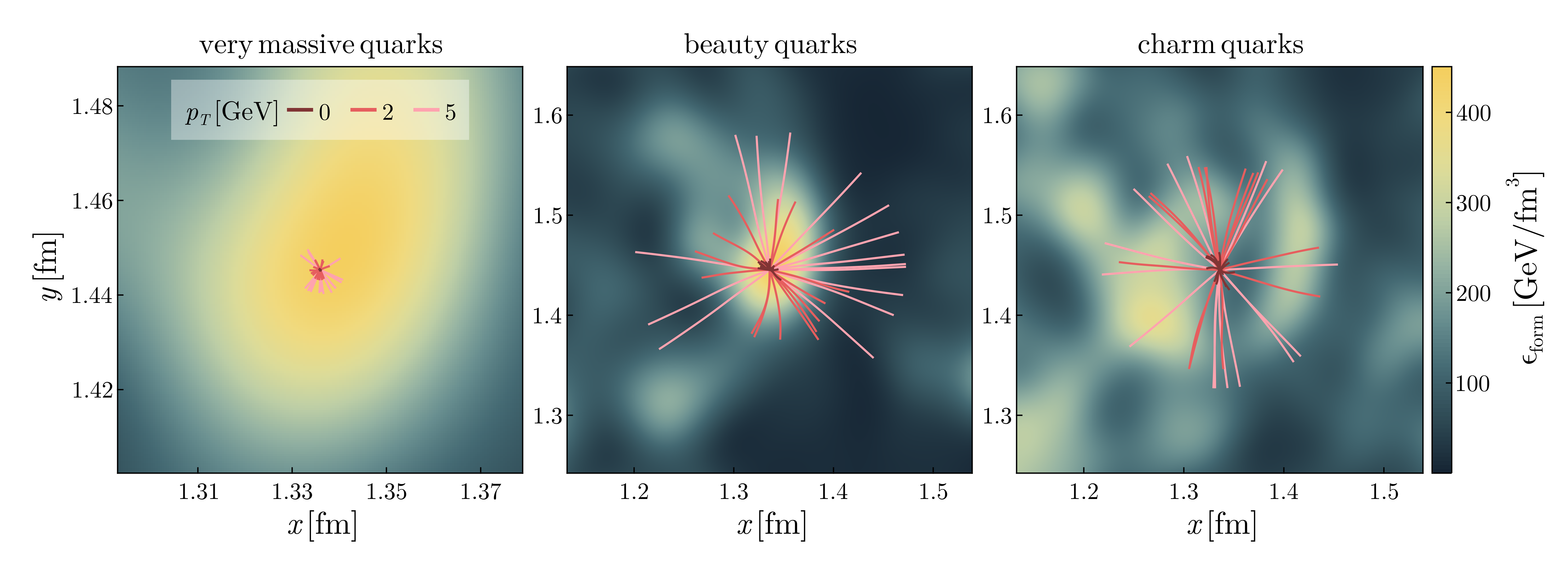}
\caption{\label{fig:hqs_flux_tubes} \textit{(Colored lines)} trajectories of heavy quarks propagating in a single Glasma flux tube evolved up to $\tau=0.2\,\mathrm{fm/}c$. All partons are produced at the center of a flux tube, where the energy density was locally maximal at the creation time of the Glasma. We consider three cases: \textit{(left)} very massive quarks with $m=200\,\mathrm{GeV}$ (approaching the static quark limit) with $\tau_\mathrm{form}=0\,\mathrm{fm/}c$, \textit{(middle)} beauty quarks with $\tau_\mathrm{form}=0.02\,\mathrm{fm/}c$ and \textit{(right)} charm quarks with $\tau_\mathrm{form}=0.06\,\mathrm{fm/}c$. The initial transverse momentum $p_T$ is varied between $0$ and $5 \, \mathrm{GeV}$. The background shows the energy density at formation time of the respective particle type, namely $\epsilon_\mathrm{form}\equiv\epsilon(\tau_{\mathrm{form}})$.}
\end{figure*}

Understanding the dynamics of heavy quarks in the Glasma in terms of the electric and magnetic color fields is generally not trivial, but some of their properties may be inferred from particular features of the background Glasma fields. Initially at $\tau = 0 \, \mathrm{fm}/c$, the Glasma consists of correlated domains of longitudinal color-electric and -magnetic flux tubes with a typical size of $\approx 1/Q_s^2$. This shortly lived initial phase is probed by heavy quarks with very high mass $m_\mathrm{HQ}$ due to their early formation time $\tau_\mathrm{form} = 1 / (2 m_\mathrm{HQ})$. These heavy quarks are accelerated due to the strong longitudinal color-electric fields of the Glasma, leading to the rapid increase of the  longitudinal momentum broadening component as seen in Figs.~\ref{fig:beautyearly} and \ref{fig:wong_kappa}. If heavy quarks have a non-negligible initial transverse momentum, there is additional transverse acceleration due to longitudinal color-magnetic flux tubes. This effect, albeit small, is seen in Fig.~\ref{fig:wong_kappa} for both beauty \textit{(left panel)} and charm \textit{(right panel)} quarks, where the transverse momentum broadening component increases with the initial transverse momentum. Remarkably, the opposite occurs for the longitudinal component: larger initial transverse momentum leads to reduced longitudinal broadening, but we have found no simple explanation in terms of the field structure of the initial Glasma for this effect.

Immediately after their initial formation, the flux tubes start to expand in the transverse plane, which generates transverse color-electric and -magnetic field components. These transverse electric fields lead to a slightly delayed increase of the transverse momentum broadening component. At the same time, the highly correlated regions within the Glasma are lost, and the longitudinal acceleration becomes less efficient. Eventually, the Glasma transitions to the free-streaming regime at around $\tau_\mathrm{free}\approx 1/Q_s \approx 0.1 \, \mathrm{fm}/c$, after which the fields become more dilute and the mean energy density falls off as $1/\tau$. As seen in Fig.~\ref{fig:beautyearly_b}, the heavy quark diffusion coefficient $\kappa$ has already peaked by then and falls off quickly. The formation time of heavy quarks has a large influence on the accumulated momenta in the Glasma stage. As can be seen from Fig.~\ref{fig:wong_kappa} \textit{(right panel)}, charm quarks accumulate less momentum because they ``skip'', at least in part, the initially highly correlated phase of the Glasma at $\tau \ll Q_s^{-1}$.  

Even though in our current setup we initialize heavy quarks homogeneously in the transverse plane, it is more likely that partons are formed inside the Glasma flux tubes, where the energy density is larger, and thus the particle production is more favorable. Thus, for illustrative purposes, we look at trajectories of almost static or dynamic beauty and charm quarks, initialized in the ``center'' of such a Glasma correlation domain, where the energy density reaches its maximum value.

The results are shown in Fig.~\ref{fig:hqs_flux_tubes}, where the different colors of the trajectory lines correspond to various values of initial $p_T\in\{0,2,5\}\,\mathrm{GeV}$ and the background shows the energy density at the formation time of the corresponding heavy quark. 
Almost static quarks with very high mass barely move during the evolution and thus essentially remain where they were originally produced at formation time. On the other hand, quarks with realistic masses are able to move further and probe larger spatial regions of the Glasma. Moreover, the quark mass determines when the particles are being introduced into the system and what regime of the evolution the partons are able to ``see''. For example, as shown in Fig.~\ref{fig:hqs_flux_tubes}, charm quarks are produced close to the transition to the free-streaming regime, where the color flux tubes already started to expand. Slow heavy quarks spend more time in the correlation domains before they expand, whereas fast quarks escape them more quickly, and thus lose the correlation faster. Even though the picture of heavy quarks probing the Glasma correlation domains as illustrated in Fig.~\ref{fig:hqs_flux_tubes} describes an over-simplified scenario, it still offers a valuable qualitative understanding. Moreover, within the approximations we use for particle initialization, it offers hints that beauty quarks might be more viable probes of the Glasma than charm quarks.

\subsection{Jet momentum broadening}
\label{subsec:jetsresults}
In recent years, jets in the Glasma have been investigated using classical simulations \cite{Ipp:2020mjc, Ipp:2020nfu} and the small $\tau$ expansion \cite{Carrington:2021dvw,Carrington:2022bnv}. In all of these works, the initial energy of the jet has been assumed to be very large, such that the trajectory can be approximated as essentially light-like. Since we account for particle dynamics via Wong's equations, we can use our particle solver to go beyond the light-like jet case and consider the effect of finite initial momentum along the propagation axis and different jet masses. For simplicity, we choose the jets to be  initialized with finite $p^x$ values.  

\begin{figure*}[t]
\centering
\subfigure[Momentum broadening of jets]{\includegraphics[width=0.8\columnwidth]{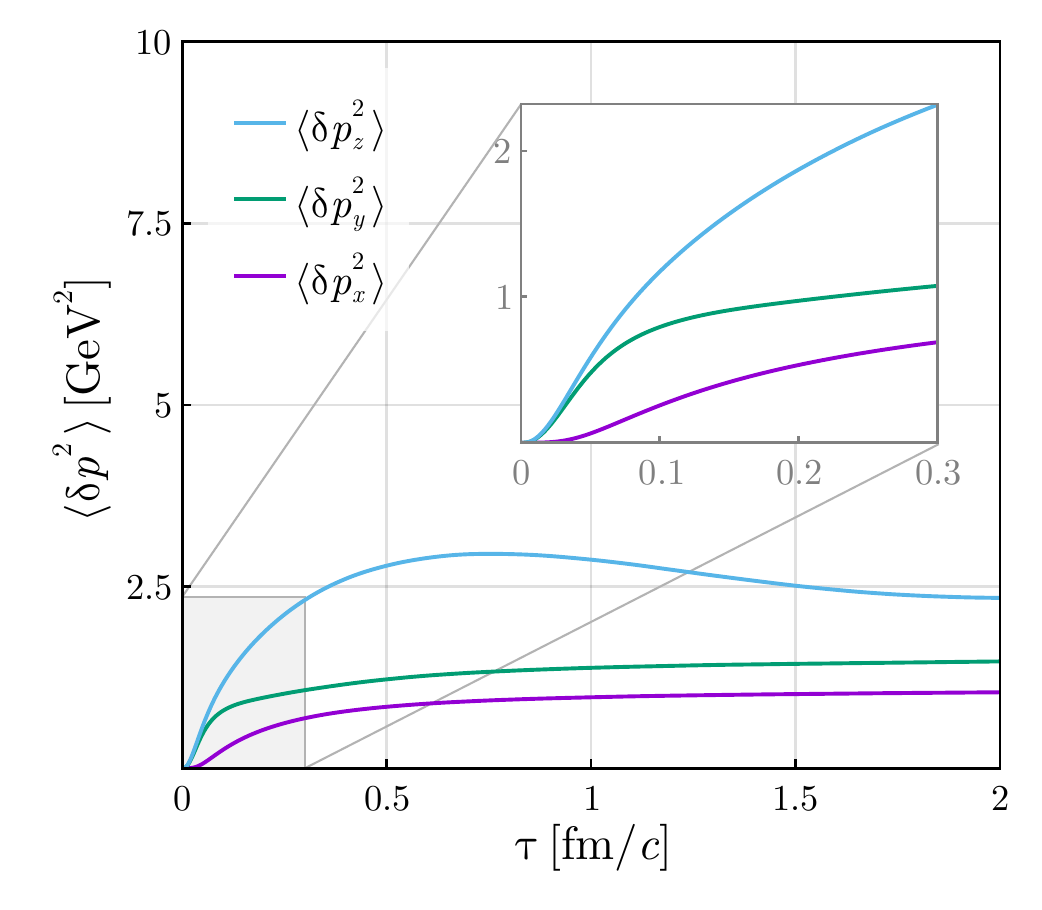} \label{fig:jetearly_a}}
\hspace{0.1\columnwidth}
\subfigure[Derivative of momentum broadening]{\includegraphics[width=0.8\columnwidth]{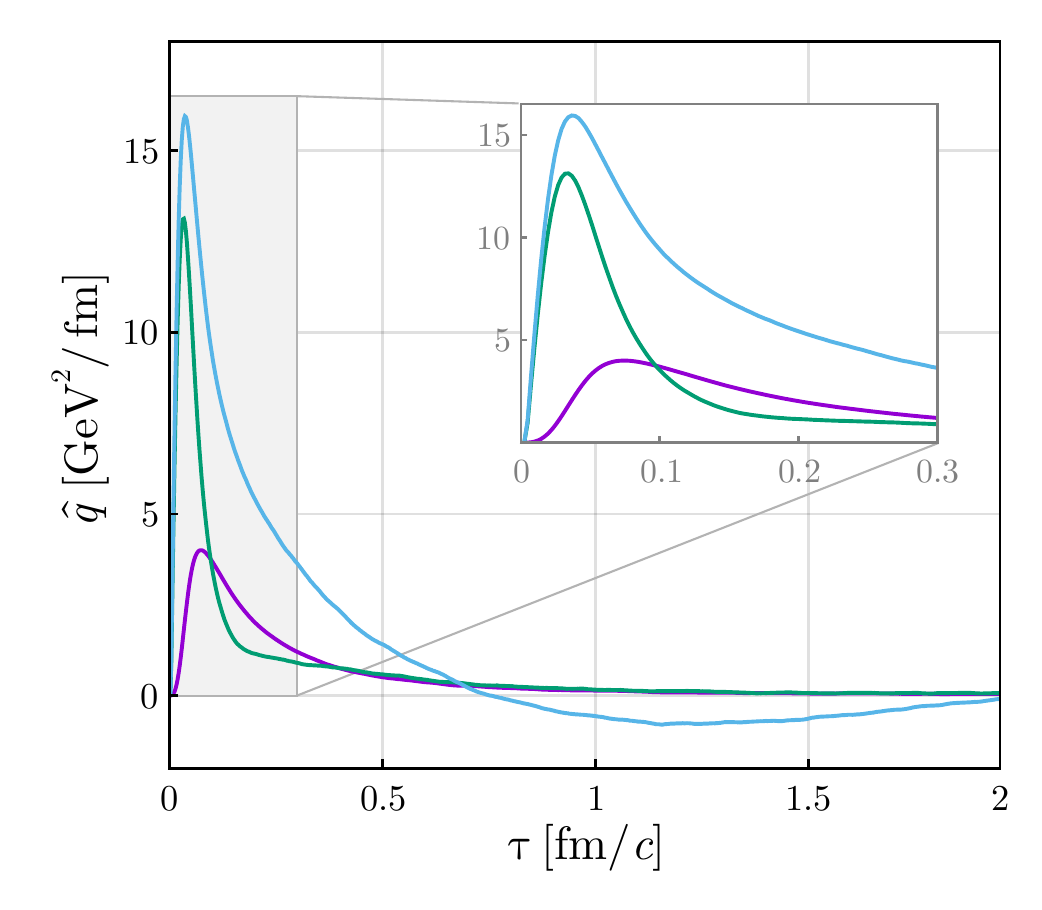} \label{fig:jetearly_b}}
\caption{\label{fig:jetearly} (a) Momentum broadening components of jets with $m=1\,\mathrm{GeV}$ and initial $p^x=10\,\mathrm{GeV}$, along the $x,y,z$-axes, as a function of proper time and (b) the derivative of the accumulated momenta that produces components of the jet transport coefficients according to Eq.~\eqref{eq:qhati}. \textit{(Insets)} Zoom-in on the very early stage.
}
\end{figure*}

Similarly to the heavy quark transport coefficient $\kappa$, we distinguish between various components of the jet transport coefficient $\hat{q}$. We define the instantaneous jet broadening coefficient
\begin{align}
    \label{eq:qhati}
    \hat{q}_i(\tau)\equiv \dfrac{\d }{\d \tau}\langle\delta p^2_{i}(\tau)\rangle
\end{align}
with $i\in\{x, y,z\}$. This is different from the collisional energy loss $\mathrm{d}E/\mathrm{d}x$. Since the jet propagates along the $x$-axis, we introduce the transverse $\hat{q}_T\equiv \hat{q}_y$ and longitudinal $\hat{q}_L\equiv \hat{q}_z$ jet transport coefficients.

\begin{figure}[t]
\includegraphics[width=0.8\columnwidth]{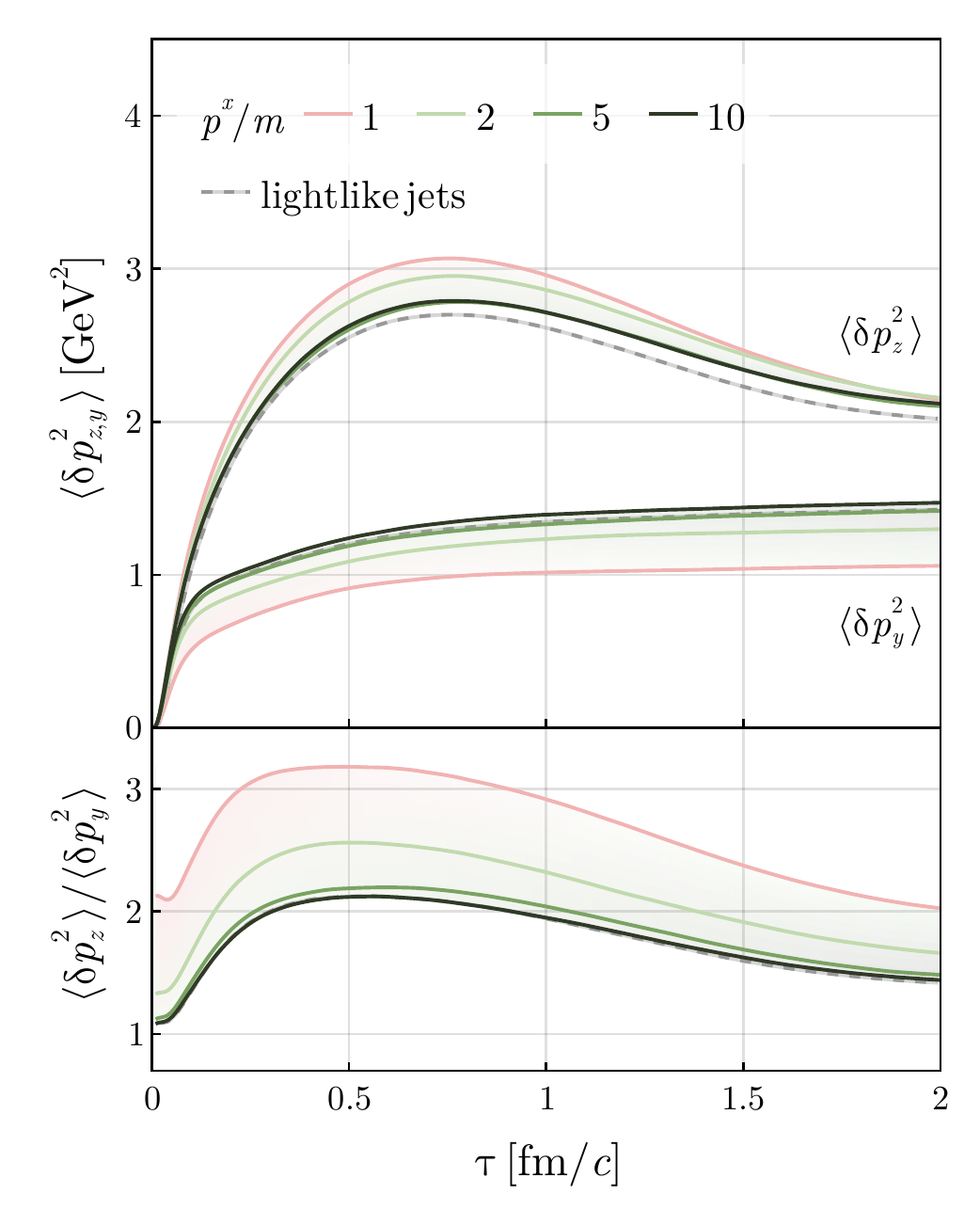}
\caption{\label{fig:wong_qhat} Momentum broadening along \textit{(top)} $z$-axis and $y$-axis, together with \textit{(bottom)} their ratio, which is a measure of the momentum broadening anisotropy. The simulations are performed for various values of $p^x/m\in\{1,2,5,10\}$ \textit{(colored full lines)}, compared to light-like jets moving along the $x$-axis \textit{(grey dashed line)}. For large $p^x/m$, the jet becomes light-like and our particle simulations approach the limiting case.}
\end{figure}

In addition, we are interested in deviations from the light-like jet scenario by considering a finite jet mass $m$ and an initial $p^x$. We find that jet momentum broadening essentially only depends on the ratio $p^x/m$. In the limit of $p^x / m \rightarrow \infty$, we can compare to the limiting case given by Eq.~\eqref{eq:mombroadjets}. Similarly to the heavy quark anisotropy, we also introduce a measure of how the Glasma anisotropy affects the jets by defining the ratio
\begin{align}
    \label{eq:jetanisotropy}
    \mathrm{jet\,anisotropy}\equiv\dfrac{\langle\delta p_z^2\rangle}{\langle\delta p_y^2\rangle},
\end{align}

Our numerical results for jets are shown in Figs.~\ref{fig:jetearly} and \ref{fig:wong_qhat}. Figure~\ref{fig:jetearly_a} shows the accumulated momentum broadening for a quark jet with  $m=1\,\mathrm{GeV}$ and initial $p^x=10\,\mathrm{GeV}$ as a function of Milne proper time $\tau$. The longitudinal component $\langle\delta p_z^2\rangle$ (along the beam axis) shows similar behavior as in the case of heavy quarks. After reaching a maximum at roughly $\tau \approx 0.8 \, \mathrm{fm}/c$, the longitudinal component $\langle\delta p_z^2\rangle$ starts to decrease at late times  $\tau \gtrsim 1 \, \mathrm{fm}/c$. The same early-time behavior is observed for heavy quarks, see Fig.~\ref{fig:beautyearly_a}. Nevertheless, at later proper times, the jets do not appear to undergo multiple oscillations. This was noticed in the limiting cases shown in Fig.~\ref{fig:wong_kappa_qhat} and we have checked that it is still present in realistic heavy quark and jets simulations. The other components, $\langle\delta p_y^2\rangle$ and $\langle\delta p_x^2\rangle$, show a steady monotonic increase at late times. We also plot the jet broadening coefficient $\hat{q}_i$ as the time derivative of the momentum broadening in Fig.~\ref{fig:jetearly_b}. Similar to the case of heavy quarks, there is a strong peak at very early stages with $\tau < 0.1 \, \mathrm{fm} / c$ and a quick decay afterwards. Due to the decrease of $\langle\delta p_z^2\rangle$ at later times ($\tau \gtrsim 0.6 \, \mathrm{fm} / c$), the longitudinal component $\hat{q}_z$ becomes negative. 

Results for jets with various values of $p^x/m$ are shown in Fig.~\ref{fig:wong_qhat}, where we also plot the momentum broadening anisotropy. The values of the initial jet momentum are chosen such that $p^x>5\,\mathrm{GeV}$. We also include the results for lightlike jets. As expected, one recovers the highly energetic jet limit by choosing a sufficiently large value for $p^x/m$ in the particle solver. Compared to heavy quarks, there is little difference in the results when accounting for finite masses and momenta. Increasing $p^x/m$ leads to a slight decrease in the longitudinal component $\langle\delta p_z^2\rangle$ (at most $15\%$). The transverse component is affected in the opposite way: $\langle\delta p_y^2\rangle$ increases with $p^x/m$ (at most $25\%$). Remarkably, while the momenta are not strongly affected, the anisotropy is enhanced (up to $40$-$60\%$) for less relativistic jets with $p^x/m \approx 1$ as can be seen from the \textit{lower panel} in Fig.~\ref{fig:wong_qhat}. 

\begin{figure}[t]
\includegraphics[width=0.95\columnwidth]{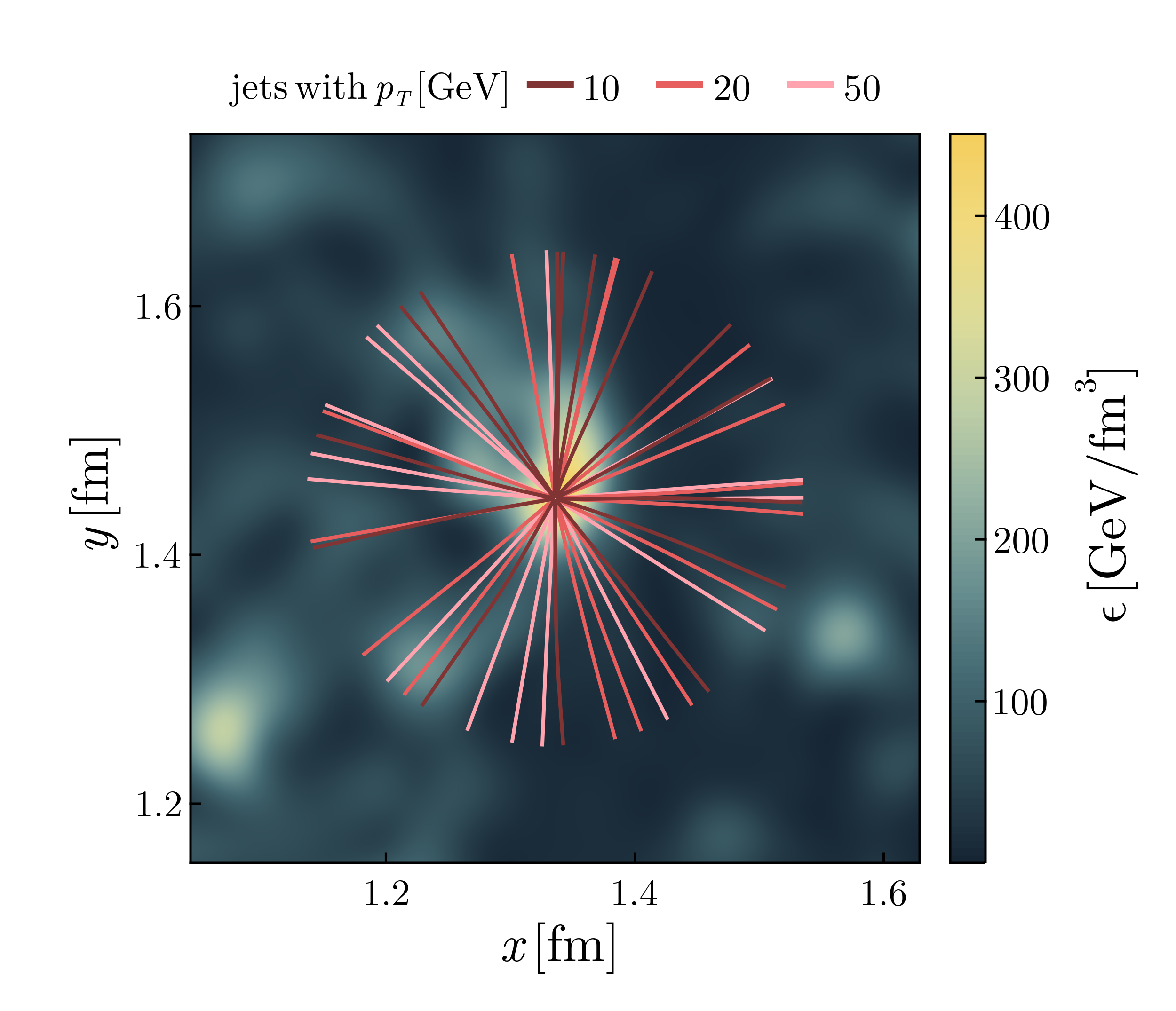}
\caption{\label{fig:jets_flux_tubes} \textit{(Colored lines)} Trajectories of jets propagating out of a single Glasma flux tube evolved up to $\tau_\mathrm{sim}=0.2\,\mathrm{fm/}c$. The colors of the lines indicate the initial momentum $p_T\in\{10, 20, 50\}\,\mathrm{GeV}$. All jets are initialized with $m=1\,\mathrm{GeV}$. The jet trajectories are essentially straight and are barely affected by the color fields of the Glasma.
}
\end{figure}

Figure~\ref{fig:jets_flux_tubes} depicts jet trajectories overlaid on top of the initial energy density of the Glasma for various initial momenta $p_T\in\{10, 20, 50\}\,\mathrm{GeV}$. Here, instead of fixing $p^x$ as the initial direction, we choose the direction of the initial transverse momentum randomly. Unlike slow heavy quarks (see Fig.~\ref{fig:hqs_flux_tubes}), jets propagate on straight lines, due to their high initial momentum. Similar to Fig.~\ref{fig:wong_qhat}, the initial value for $p_T$ only weakly affects the jet trajectories. 

\section{Summary and outlook}
\label{summary}

We have investigated the impact of the early stages of heavy-ion collisions, namely the Glasma, on hard probes such as heavy quarks and jets. To accomplish this, we approximate these hard probes as classical colored particles and simulate their dynamics using Wong's equations on top of the non-Abelian background field of the boost-invariant Glasma.
This work can be understood as an extension of earlier studies on highly energetic jets \cite{Ipp:2020mjc, Ipp:2020nfu}
and heavy quarks \cite{Das:2015aga, Das:2017dsh, Ruggieri:2018rzi, Sun:2019fud, Liu:2019lac, Liu:2020cpj, Khowal:2021zoo} in the pre-equilibrium medium, which were limited in different ways. Simulations of jets in the Glasma were based on the ultrarelativistic limit, i.e.~the jets were assumed to be lightlike. Thus, these simulations only apply to jets at extremely high energies. On the other hand, studies of heavy quarks in the Glasma relied on using SU(2) \cite{Ruggieri:2018rzi, Liu:2019lac, Sun:2019fud, Liu:2020cpj} instead of SU(3) as the gauge group, which can only provide a qualitative picture. Thus, to improve upon these earlier studies, we have developed a fully non-perturbative simulation of classical particles with SU(3) color charges based on Wong's equations. The background field in which the charges are moving in is provided by classical real-time simulations of the Glasma. As such, we have realized a unified numerical setup where the effects of the Glasma on both heavy quarks and jets can be studied quantitatively.

To measure the impact of the Glasma on hard probes, we focused on the momentum broadening components $\langle \delta p_i^2 (\tau) \rangle$, which describe how much momentum is accumulated by heavy quarks and jets as they pass through the medium. This observable is particularly interesting, because it can be related to transport coefficients such as the heavy quark diffusion coefficient $\kappa$ and the jet momentum broadening coefficient $\hat{q}$. Additionally, we studied anisotropy ratios of different components of $\langle \delta p_i^2 (\tau) \rangle$.

As a consistency check for our simulations, we have performed non-trivial numerical checks of our code by comparing to certain limiting cases, where the dynamics of the hard probes become trivial. These cases are heavy quarks with infinite mass (static quarks) and jets at very high energies (lightlike jets), where the particle trajectories are fixed and the eikonal approximation applies. Consequently, it is possible to compute momentum broadening directly from Wilson loops of the background field, which provides a benchmark result that our particle simulations must be able to reproduce. By taking these limits in the particle solver and performing extensive numerical checks, we have verified that our numerical solutions to Wong's equations are indeed consistent with the calculation from Wilson loops.

Going towards more realistic settings, we then considered the effects of finite mass and initial momentum of the hard probes. In particular, we performed simulations for beauty and charm quarks. In both cases, we notice deviations from the static quark limit.
We found that there is strong initial acceleration at early times which results in a strongly time-dependent diffusion coefficient $\kappa$, with a characteristic peak at early times $\tau \lesssim 0.1 \, \mathrm{fm} / c$ and a subsequent quick decay. This behavior differs from the standard Langevin or Boltzmann approaches, in which the momentum broadening grows slowly, is generally smaller and does not exhibit a peak \cite{Liu:2020cpj}. Following \cite{Sun:2019fud}, it is of future interest to investigate the impact of such large broadening induced by the Glasma on observables such as elliptic flow or nuclear modification factors in both proton-nucleus or nucleus-nucleus collisions. Moreover, our calculations showed that beauty quarks, even though they are heavier, accumulate more momentum compared to charm quarks. This is due to their larger mass, which allows them to be formed slightly earlier in the evolution of the Glasma, where the color fields are particularly strong. 
Regardless of quark species, there is a sizable momentum broadening anisotropy with $\langle \delta p_L^2 \rangle > \langle \delta p_T^2 \rangle$, i.e.~more accumulation along the beam axis compared to the transverse plane at early times. Curiously, this effect is reversed at late times for charm quarks, where $\langle \delta p_L^2 \rangle < \langle \delta p_T^2 \rangle$. Most remarkably, we observed that the longitudinal component $\langle \delta p_L^2 \rangle$ oscillates as a function of time. It is possible that this effect could be traced back to the existence of plasmon modes in the Glasma. The plasmon modes are a collective feature of the Glasma color fields themselves. They could further be transmitted to the particles propagating in these fields, thus causing oscillations in their accumulated momenta. Moreover, plasmon frequency oscillations were already observed in a study involving a Yang-Mills plasma with large occupation numbers \cite{Boguslavski:2020tqz}. The emergence of such oscillations only in the longitudinal direction and not in the transverse plane is intriguing and requires further investigation.
Thus, a possible extension of the current work would be to determine the Glasma plasmon frequency using methods similar to \cite{Krasnitz:2000gz, Lappi:2016ato, Lappi:2017ckt}. 

We have performed analogous calculations for jets with finite mass and finite initial momenta. Similar to heavy quarks and also confirming previous studies \cite{Ipp:2020nfu}, we found that the jet momentum broadening coefficient $\hat{q}$ is highly peaked at early times $\tau \lesssim 0.1 \, \mathrm{fm} / c$. There is a rapid increase of both longitudinal and transverse components at early times, and a sizable momentum broadening anisotropy at later times with $\langle \delta p_L^2 \rangle > \langle \delta p_T^2 \rangle$. For less relativistic jets with $|p| \sim m$,
this anisotropy is more pronounced compared to the ultrarelativistic limit. In contrast to heavy quarks, the effects of finite masses and initial momentum are quantitatively less important. Remarkably, there is a notable absence of oscillatory behavior in the longitudinal (beam axis) component. Instead, $\langle \delta p_L^2 \rangle $ exhibits a single peak around $\tau \approx 0.8 \, \mathrm{fm} / c$. It would be interesting if this behavior could also be understood in terms of the excitation spectrum of the Glasma.

Besides determining the origin of the oscillations of $\langle \delta p_L^2 \rangle$, there are multiple other ways to extend our current work. Concerning the Glasma itself, a possible extension is to consider more complicated initial conditions beyond the McLerran-Venugopalan model used here.  In particular, it would interesting to see the effects of more realistic transverse structure (such as in the IP-Glasma model \cite{Schenke:2012wb, Schenke:2012hg}) or
 hot spots \cite{Mantysaari:2016jaz, Mantysaari_2022, Demirci:2022wuy}. Another extension, related to the longitudinal structure of the colliding nuclei, could be to go beyond the boost-invariant approximation and consider the full 3+1 dimensional structure of the Glasma, either due to finite extent along the beam axis \cite{Gelfand:2016yho, Ipp:2017lho, Ipp:2020igo, Schlichting:2020wrv}  or due to the JIMWLK evolution \cite{Schenke:2016ksl, McDonald:2018wql, McDonald:2020oyf}. Although generalizing our numerical setup to 3+1 dimensions is in principle trivial, a large amount of computational resources would be required to carry out such simulations. In practice, this generalization might still be possible through the weak field approximation \cite{Ipp:2021lwz}, which exhibits significantly reduced computational costs compared to lattice simulations at the expense of neglecting non-perturbative effects.
 
Regarding the dynamics of the hard probes, an immediate improvement would be the inclusion of the color current generated by the color charges as they propagate through the Glasma. This would induce a back reaction of the hard particles onto the Glasma. It has already been demonstrated in \cite{Liu:2020cpj} that including the color current of heavy quarks does not significantly modify momentum broadening, spectra, or nuclear modification factor at early times. However, one would expect the back reaction to be more significant for jets, in particular regarding (classical) gluon radiation and energy loss. Unfortunately, fast moving charged particles in lattice simulations are plagued by the numerical Cherenkov instability which is not tractable in the current setup without significant changes to the numerical scheme \cite{Ipp:2018hai}. 

Another interesting aspect, unrelated to classical particle simulations, would be a more detailed study of large temporal and lightlike Wilson loops in the Glasma. Beyond just the lowest moments $\langle \delta p^2 \rangle$, the Wilson loops encode information about the probability $P(p_\perp)$ that a hard parton picks up transverse momentum $p_\perp$ during its evolution \cite{Casalderrey-Solana:2007ahi, DEramo:2010wup}. The Wilson loop formulation therefore allows for the extraction of the collision kernel for momentum broadening. Such a quantity was computed in the context of anisotropic plasmas within a kinetic theory approach \cite{Hauksson:2021okc} or using perturbative computations \cite{Caron-Huot:2008zna} or non-perturbative lattice techniques \cite{Moore:2021jwe}. Computing the collision kernel in the Glasma, which is an anisotropic and out-of-equilibrium medium, is an exciting prospect.

Lastly, there are additional observables which describe the effect of the Glasma on heavy quarks and jets, namely two-particle correlations that may be significantly affected by the large momentum broadening. In principle, these are possible observables within the available setup, which could be extended by off-central collisions, more sophisticated nuclear models, and more realistic ways of initializing particles in our simulation.
We plan to include such features in our code and study the angular correlations of quark-antiquark pairs and how they are affected by the early stages of heavy-ion collisions.

\begin{acknowledgments}
D.~A.~acknowledges funding from the Academy of Finland, Center of Excellence in Quark Matter project 346324. V.~G.~acknowledges funding from UniCT under ``Linea di intervento 2'' (HQCDyn Grant). D.~M.~acknowledges funding from the Austrian Science Fund (FWF) projects P~34455 and P~34764. All simulations were performed using the GPU nodes of the Center of Theoretical Physics, University of Bucharest.  D.~A.~, D.~M.~ and M.~R.~ acknowledge S.~Mr\'{o}wczy\'{n}ski and C.~Manuel for discussions regarding classical color charges, and K.~Boguslavksi, H.~M\"antysaari, and T.~Lappi for many insightful discussions regarding the early stages of heavy-ion collisions. M.~R.~acknowledges John Petrucci for inspiration. We are grateful to T.~Lappi for reviewing the manuscript.
\end{acknowledgments}

\appendix


\section{Some details regarding Wong's equations} 
\label{appen:wong_details}

In this part of the appendix we collect some derivations and technical details regarding Wong's equations.

\subsection{Wong's equations in Milne coordinates} \label{appen:wong_milne}

Here, we provide an explicit derivation of Wong's equations in the Milne frame. We start from the covariant form given by Eq.~\eqref{eq:wongcurv}. 

The coordinate vector of the Milne frame is $\widetilde{x}^\mu=(\tau, x, y, \eta)$ with Milne proper time $\tau$ and longitudinal space-time rapidity $\eta$. The coordinate change from the laboratory to the Milne frame is described by
\begin{align}
    \label{eq:taueta_appen}
    \tau=\sqrt{t^2 - z^2},\quad \eta=\frac{1}{2}\ln{\left(\frac{t+z}{t-z}\right)}.
\end{align}
The inverse transformations are $t=\tau\cosh\eta$ and $z=\tau\sinh\eta$. The components of the metric are $\widetilde{g}_{\mu\nu}=\mathrm{diag}(1, -1, -1, -\tau^2)$. Consequently, the only non-vanishing Christoffel symbols are 
\begin{equation}
    \Gamma^\tau_{\eta\eta}=\tau,\quad \Gamma^\eta_{\tau\eta}=\Gamma^\eta_{\eta\tau}=\dfrac{1}{\tau}.
\end{equation}
The Christoffel symbols of the second kind are related to the first-kind Christoffel symbols through
\begin{equation*}
    [ab,c]=g_{cd}\Gamma^d_{a b},
\end{equation*}
which in Milne coordinates read
\begin{equation}
    \label{eq:christoffels}
    [\eta\eta,\tau]=\tau,\quad[\eta\tau,\eta]=[\tau\eta,\eta]=-\tau.
\end{equation}
The Christoffel symbols are used to relate the covariant derivative along the worldline of a particle, denoted by $\mathrm{D}/\mathrm{d}\boldsymbol{\tau}$, to the usual derivative $\mathrm{d}/\mathrm{d}\boldsymbol{\tau}$. For the four-velocity $u^\mu$ of a particle, this relationship is given by
\begin{equation}
    \label{eq:covdercurved}
    \frac{\mathrm{D}u_\mu}{\mathrm{d}\boldsymbol{\tau}}=g_{\mu\nu}\dfrac{\mathrm{d} u^{\nu}}{\mathrm{d} \boldsymbol{\tau}}+[\nu\lambda,\mu]u^{\nu} u^{\lambda}.
\end{equation}
Note that $\boldsymbol{\tau}$ denotes the proper time in the rest frame of the particle, which should not be confused with Milne proper time $\tau$. The transformations of the four-velocity components are  $u^\tau=\cosh\eta\,u^t-\sinh\eta\,u^z$ along with $u^\eta=-(\sinh\eta\,u^t+\cosh\eta\,u^z)/\tau$. The inverse transformations are $u^t=\cosh\eta\,u^\tau+\sinh\eta\,\tau u^\eta$ and $u^z=\sinh\eta\,u^\tau+\cosh\eta\,\tau u^\eta$. 

The next step is to express derivatives with respect to $\boldsymbol{\tau}$ in terms of $\tau$-derivatives. In particular, we use $m\,\mathrm{d}/\mathrm{d}\boldsymbol{\tau}=p^\tau\,\mathrm{d}/\mathrm{d}\tau$. This allows us to write the 
$\tau$-evolution of the particle coordinates from Eq.~\eqref{eq:wongcurv} as
\begin{equation}
    \label{eq:xmumilne}
   \frac{\mathrm{d}x}{\mathrm{d}\tau}=\frac{p^x}{p^\tau},\quad \frac{\mathrm{d}y}{\mathrm{d}\tau}=\frac{p^y}{p^\tau},\quad \frac{\mathrm{d}\eta}{\mathrm{d}\tau}=\frac{p^\eta}{p^\tau},
\end{equation}
where the temporal component $p^\tau$ is given by 
\begin{align}
    \label{eq:ptau}
    p^\tau=\sqrt{p_T^2+\tau^2 (p^\eta)^2+m^2},
\end{align} 
with $p_T^2=(p^x)^2+(p^y)^2$. The Milne proper time evolution of the momenta is given by
\begin{equation}
    \label{eq:wongmompropertime}
    \frac{\mathrm{D}p_\nu}{\mathrm{d}\tau}=\frac{g}{T_R}\tr{QF_{\nu\mu}}\frac{p^\mu}{p^\tau},
\end{equation}
where the covariant derivatives can be written as
\begin{align}
\dfrac{\mathrm{D} p_{\tau}}{\mathrm{d} \tau}&=\dfrac{\mathrm{d}p^\tau}{\mathrm{d}\tau}+\tau \frac{ (p^\eta)^2}{p^\tau}, \\
   \dfrac{\mathrm{D} p_{i}}{\mathrm{d} \tau}&=-\dfrac{\mathrm{d} p^{i}}{\mathrm{d} \tau}, \\
   \dfrac{\mathrm{D} p_{\eta}}{\mathrm{d} \tau}&=-\tau^2\dfrac{\mathrm{d}p^\eta}{\mathrm{d}\tau}-2\tau p^\eta. 
\end{align}
The last set of equations follows from Eq.~\eqref{eq:covdercurved} and $p_\mu = m u_\mu$.

Finally, we write the components of the field strength tensor in terms of color-electric and -magnetic fields. Using the relations
\begin{equation}
    \label{eq:glasmafields}
    \begin{aligned}
        &E_i\equiv F_{\tau i},  &&B_i\equiv \epsilon_{ij}\dfrac{1}{\tau}F_{\eta j},\\
        &E_\eta\equiv \dfrac{1}{\tau}F_{\tau\eta}, &&B_\eta\equiv -F_{xy},
    \end{aligned}
\end{equation}
the spatial components of the momentum equations become
\begin{align}
    \begin{split}
    &\tau\dfrac{\mathrm{d}p^\eta}{\mathrm{d}\tau}+2 p^\eta\\
    &=\frac{g}{T_R}\left(\tr{Q E_\eta}-\tr{QB_x}\frac{p^y}{p^\tau}+\tr{QB_y} \frac{p^x}{p^\tau}\right),\\
    &\frac{dp^x}{d\tau}=\frac{g}{T_R}\left(\tr{QE_x}+\tr{QB_\eta}\frac{p^y}{p^\tau} -\tr{QB_y} \frac{\tau p^\eta}{p^\tau}\right),\\
    &\frac{dp^y}{d\tau}=\frac{g}{T_R}\left(\tr{QE_y}-\tr{QB_\eta} \frac{p^x}{p^\tau}+\tr{QB_x}\frac{\tau p^\eta}{p^\tau}\right).
    \end{split}
\end{align}
The temporal component is given by
\begin{align}
    \begin{split}
    &\dfrac{\mathrm{d}p^\tau}{\mathrm{d}\tau}+\frac{\tau p^\eta}{p^\tau}p^\eta\\
    &=\frac{g}{T_R}\left(\tr{Q E_\eta}\frac{\tau p^\eta}{p^\tau}+\tr{QE_x} \frac{p^x}{p^\tau}+\tr{QE_y} \frac{p^y}{p^\tau}\right).
    \end{split}
\end{align}

\subsection{Conservation of Casimirs by Wong's equations}
\label{subsec:casimirwong}
Wong's equations from Eq.~\eqref{eq:wongcurv} preserve the classical Casimir values of the color charge $Q^a$. The quadratic and cubic Casimirs are given by
\begin{align}
    Q^a Q^a &= q_2, \\
    d_{abc} Q^a Q^b Q^c &= q_3.
\end{align}
Taking the time derivative of the quadratic Casimir and using the anti-symmetry of the structure constants immediately yields
\begin{equation}
    \begin{aligned}
        \dfrac{\mathrm{d}q_2}{\mathrm{d}\boldsymbol{\tau}}&=2Q^a\dfrac{\mathrm{d}Q^a}{\mathrm{d}\boldsymbol{\tau}}=-2g\dfrac{p^\mu}{m}A_\mu^b f_{abc}Q^aQ^c=0.
    \end{aligned}
\end{equation}
Using the identity (see e.g.~\cite{Haber:2019sgz}) valid for the fundamental generators
\begin{align}
    \tr{T^a T^b T^c} = \frac{1}{4}\left( d_{abc} + \mathrm{i} f_{abc}\right),
\end{align}
the cubic Casimir can be written as
\begin{align}
    q_3 = 4\, \tr{Q^3},
\end{align}
where $Q = Q^a T^a$. The conservation of $q_3$ then follows from
\begin{align}
    \dfrac{\mathrm{d}}{\mathrm{d} \boldsymbol{\tau}} \tr{Q^3} &= 3\, \tr{Q^2 \dfrac{\mathrm{d} Q}{\mathrm{d} \boldsymbol{\tau}} } \notag \\
    &= -3\,\mathrm{i} g \,\tr{Q^2 \left[ A_\mu, Q \right] } \frac{p^\mu}{m} \notag \\
    &= 0,
\end{align}
where we have used Eq.~\eqref{eq:wongq} in the second line. The last line vanishes for any $Q$ and $A_\mu$ due to the cyclic property of the trace.

\section{Properties of classical color charges}
\label{appen:clascol}

Here we provide some additional mathematical details regarding classical color charges such as the choice of Casimir invariants and the integration measure. 

\subsection{Casimir invariants for generators $T^a$ and classical color charges $Q^a$}
Let $\{T^a\}$ with $a \in \{1,2, \dots D_A \}$ be the generators of SU($N_c$) with $D_A = N_c^2 - 1$. Common choices for generators in the fundamental (quark) representation are $T^a=\sigma^a/2$ for SU(2) and $T^a=\lambda^a/2$ for SU(3), with Pauli matrices $\sigma^a$ and Gell-Mann matrices $\lambda^a$. 
Regardless of a particular representation, the generators satisfy $[T^a,T^b]= i f_{abc}T^c$, where $f_{abc}$ are the totally anti-symmetric structure constants of the group. For SU(2) these constants are given by 
$f_{abc}=\epsilon_{abc}$. For SU(3) the non-vanishing structure constants are listed in Table \ref{tab:fabc}.
The other representation that we are interested in is the adjoint (gluon) representation, whose generators are given by
\begin{align}
    (T^a)_{bc} = - i f^{abc}.
\end{align}
The dimensions of the fundamental ($F$) and adjoint ($A$) representations are 
\begin{align}
    \label{eq:dr}
    D_R = \left.
      \begin{cases}
        N_c, & R=F \\
        N_c^2-1, & R=A  
      \end{cases}
    \right..
\end{align}

The generators are orthonormal in the sense that
\begin{align}
    \tr{T^a T^b}_R = T_R \delta^{ab},
\end{align}
where the Dynkin index $T_R$ depends on the chosen representation $R$. For the fundamental and adjoint representations it is given by
\begin{align}
T_R = \left.
      \begin{cases}
        \dfrac{1}{2}, & R=F \\
        N_c, & R=A  
      \end{cases}
    \right..    
\end{align}

\begin{table}[t]
\centering
\begin{ruledtabular}
    \begin{tabular}{llllllllll}
        $f_{abc}$ & $f_{123}$ & $f_{147}$ & $f_{156}$ & $f_{246}$ & $f_{257}$ & $f_{345}$ & $f_{367}$ & $f_{458}$ & $f_{678}$ \\
        \midrule
        & 1 & $\dfrac{1}{2}$ & $-\dfrac{1}{2}$ & $\dfrac{1}{2}$ & $\dfrac{1}{2}$ & $\dfrac{1}{2}$ & $-\dfrac{1}{2}$ & $\dfrac{\sqrt{3}}{2}$ & $\dfrac{\sqrt{3}}{2}$ \\ 
    \end{tabular}
\end{ruledtabular}
\caption{Anti-symmetric structure constants for SU(3).}
\label{tab:fabc}
\end{table}

The representations of the algebra $\mathfrak{su}(2)$ are uniquely labeled by the value of the quadratic Casimir 
\begin{align}
    \label{eq:c2}
    \sum_a T^a T^a=\mathbb{1}_{D_R}C_2(R),
\end{align}
with $D_R$ the dimension of the representation. For the $\mathfrak{su}(3)$ algebra, representations are additionally labeled by the cubic Casimir
\begin{align}
    \label{eq:c3}
    \sum_{abc}d_{abc}T^aT^bT^c=\mathbb{1}_{D_R}C_3(R),
\end{align}
where $d_{abc}$ denote the symmetric structure constants (see Tab.~\ref{tab:dabc}). For SU(2), $d_{abc}$ can be taken as zero.
\begin{table*}[t]
    \centering
    \begin{ruledtabular}
        \begin{tabular}{lllllllllllllllll}
        $d_{abc}$ & $d_{118}$ & $d_{146}$ & $d_{157}$ & $d_{228}$ & $d_{247}$ & $d_{256}$ & $d_{338}$ & $d_{344}$ & $d_{355}$ & $d_{366}$ & $d_{377}$ & $d_{448}$ & $d_{558}$ & $d_{668}$ & $d_{778}$ & $d_{888}$\\ 
        \midrule
        & $\dfrac{1}{\sqrt{3}}$ & $\dfrac{1}{2}$ & $\dfrac{1}{2}$ & $\dfrac{1}{\sqrt{3}}$ & $-\dfrac{1}{2}$ & $\dfrac{1}{2}$ & $\dfrac{1}{\sqrt{3}}$ & $\dfrac{1}{2}$ & $\dfrac{1}{2}$ & $-\dfrac{1}{2}$ & $-\dfrac{1}{2}$ & $-\dfrac{1}{2\sqrt{3}}$ & $-\dfrac{1}{2\sqrt{3}}$ & $-\dfrac{1}{2\sqrt{3}}$ & $-\dfrac{1}{2\sqrt{3}}$ & $-\dfrac{1}{\sqrt{3}}$ \\ 
        \end{tabular}
    \end{ruledtabular}
    \caption{Symmetric structure constants for SU(3).}
    \label{tab:dabc}
\end{table*} 
The values of the quadratic and cubic Casimirs are (see \cite{Haber:2019sgz})
\begin{subequations}
    \label{eq:c23}
    \begin{align}
    &C_2(R) = \left.
      \begin{cases}
        \dfrac{N_c^2-1}{2N_c}, & R=F \\
        N_c, & R=A  
      \end{cases}\right.,\\[0.5em]
    &C_3(R) = \left.
      \begin{cases}
        \dfrac{(N_c^2-4)(N_c^2-1)}{4N_c^2}, & R=F \\
        0, & R=A  
      \end{cases}\right..
    \end{align}
\end{subequations}

Classical color charges $Q^a$ can be understood as the limit of high dimensional representations of $\mathfrak{su}(N_c)$. By analogy, it is then possible to also define Casimir invariants similar to the finite dimensional case in Eqs.~\eqref{eq:c2} and~\eqref{eq:c3}. Thus, the quadratic and cubic Casimirs of $Q^a$ are given by%
\begin{align}
    \label{eq:q2}
    \sum_a Q^aQ^a \equiv q_2(R),
\end{align}
and similarly
\begin{align}
    \label{eq:q3}
    \sum_{abc} d_{abc}Q^aQ^bQ^c\equiv q_3(R).
\end{align}
The color charges $Q^a$ are real-valued numbers and there are $D_A = N^2_c-1$ components for a given SU($N_c$) group. The Casimir invariants in Eqs.~\eqref{eq:q2} and \eqref{eq:q3} may be viewed as constraints on the set of possible color charges. Evidently, the manifold of admissible color charge vectors depends on the number of colors. For example, in SU(2) only the quadratic Casimir invariant applies because $d_{abc} = 0$. Thus, due to $D_A = 3$, the classical color charges are three-dimensional vectors constrained to a $2$-sphere with radius $q_2$. The manifold of SU(2) color charges is therefore two-dimensional.
For SU(3), the manifold becomes more complicated: Firstly, due to $D_A = 8$ for SU(3), the quadratic invariant constrains the possible choices of color charges to a $7$-sphere with radius $q_2$. Secondly, the cubic Casimir invariant further constrains the color charge manifold to a six-dimensional submanifold of $\mathbb{R}^8$. It turns out that not all choices of $q_2$ and $q_3$ are admissible.
In contrast to SU(2), where the color charge manifold exists for any value $q_2 > 0$, the SU(3) color charge manifold only exists for certain values of $q_2$ and $q_3$. We address this at the end of Appendix \ref{appen:relnptfctcasimirs}.

\subsection{Integration measure and $n$-point functions}
\label{appen:relnptfctcasimirs}

The generators of SU($N_c$) satisfy the following trace relations
\begin{subequations}
    \label{eq:npointfctg}
    \begin{align}
        \tr{T^a}&=0,\label{eq:trga}\\
        \tr{T^aT^b}&=T_R \delta^{ab},\label{eq:trgagb}\\
        \tr{T^a T^b T^c}&=\frac{A_R}{4}(d_{abc}+i f_{abc}),\label{eq:trgagbgc}
    \end{align}
\end{subequations}
where the anomaly coefficient $A_R$ is given by
\begin{align}
    \label{eq:dynkinanomaly}
    A_R = \left.
      \begin{cases}
        1, & R=F \\
        0, & R=A  
      \end{cases}\right..
\end{align}
Similarly to Eq.~\eqref{eq:npointfctg}, one may impose that the averages performed over classical color charge configurations must satisfy 
\begin{subequations}
    \label{eq:qnpointfct}
    \begin{align}
        \langle Q^a \rangle&\equiv\int\d Q \, Q^a=0, 
        \label{eq:qa_app}
        \\ 
        \langle Q^a Q^b \rangle &\equiv \int\d Q \, Q^aQ^b=T_R \delta^{ab}, 
        \label{eq:qaqb_app}
        \\
        \langle Q^a Q^b Q^c \rangle &\equiv \int\d Q \, Q^aQ^bQ^c=\frac{A_R}{4}d^{abc},
        \label{eq:qaqbqc_app}
    \end{align}
\end{subequations}
where, in the three-point function, the imaginary part was discarded since the classical color charges are real valued and are also symmetric under color indices, while $f^{abc}$ is anti-symmetric. 

The integration measure $dQ$ used in the definition of the $n$-point functions is constrained by the Casimir invariants. For SU(2) it reads
\begin{align}
    \d Q=c_R\,\d^3 Q\, \delta(Q^aQ^a-q_2),
\end{align}
and similarly for SU(3)
\begin{align}
    \d Q=c_R\,\d^8 Q\, \delta(Q^aQ^a-q_2)\delta(d_{abc}Q^aQ^bQ^c-q_3).
\end{align}
The normalization constant $c_R$ is chosen such that the color charge distributions are normalized to unity
\begin{align}
    \label{eq:dq}
    \int \d Q=1.
\end{align}
Once a normalization for the integration measure is chosen, the values for the classical Casimirs $q_2$ and $q_3$ are fixed according to Eqs.~\eqref{eq:qnpointfct}. Contracting the two-point function in Eq.~\eqref{eq:qaqb_app} with $\delta^{ab}$ immediately yields $q_2 = T_R D_A$, which, by virtue of $T_R D_A = D_R C_2(R)$, yields 
\begin{align}
    \label{eq:q2c2}
    q_2(R)=D_RC_2(R).
\end{align}
In an analogous manner, contracting the three-point function of Eq.~\eqref{eq:qaqbqc_app} with $d_{abc}$ yields
\begin{align}
    \label{eq:q3c3}
    q_3(R)=D_RC_3(R),
\end{align}
which is analogous to Eq.~\eqref{eq:q2c2}. Thus, according to Eqs.~\eqref{eq:dr} and~\eqref{eq:c23}, the classical Casimirs are given by
\begin{subequations}
    \label{eq:q23}
    \begin{align}
    &q_2(R) = \left.
          \begin{cases}
            \dfrac{N_c^2-1}{2}, & R=F \\
            N_c(N_c^2-1), & R=A  
          \end{cases}
        \right.,\\[0.5em]
    &q_3(R) = \left.
          \begin{cases}
            \dfrac{(N_c^2-4)(N_c^2-1)}{4N_c}, & R=F \\
            0, & R=A  
          \end{cases}\right..
    \end{align}
\end{subequations}

Similar relationships for the Casimirs of the classical color charges were established in \cite{Kelly:1994dh, Litim:1999id, Litim:1999ns, Litim:2001db}. In \cite{Heinz:1984yq,Heinz:1985qe} within a transport theory, it is argued that a classical description of the color charges holds only for large representations. From this perspective, it is not suitable to assign Casimirs only for a single classical quark or gluon, but rather to work with higher-order representations and introduce them for the whole ensemble. It is within this context that the choices in Eqs.~\eqref{eq:q2c2} and~\eqref{eq:q3c3} for the classical Casimirs $q_2$ and $q_3$ are the product of the dimension of the representation $D_R$ and the group-theoretical Casimirs $C_2$ and $C_3$.

Different choices for the two- and three-point functions would yield other classical quadratic and cubic Casimirs. For example, an elegant choice would be to set $q_2 = C_2$ and $q_3 = C_3$, i.e.~match the classical and the group-theoretical Casimirs directly. This would allows us to avoid the matching procedure of Section  \ref{sec:divisionbydr}, where we showed that observables such as momentum broadening $\langle \delta p^2 \rangle$ must be divided by a factor of $D_R$ in order to reproduce a similar calculation in perturbative QCD. However, we found that this choice is generally not admissible. Using numerical solution methods, we were not able to find a single color charge vector $Q^a$ for SU(3) quarks that satisfies both $q_2 = C_2$ and $q_3 = C_3$ in the fundamental representation. On the other hand, we found possible solutions in the case of SU(3) gluons and for both SU(2) quarks and gluons. It appears that in the case of SU(3) quarks, the color charge manifold embedded in $\mathbb{R}^8$ and constrained by the two Casimir invariants does not exist. On the other hand, the choices in Eqs.~\eqref{eq:q2c2} and \eqref{eq:q3c3} are admissible in the sense that there are valid solution vectors $Q^a$ for both quarks and gluons in SU(2) and SU(3). We require these solution vectors because our color charge sampling method (see Section \ref{subsec:su3charges}) is based on randomly rotating initial color charge vectors. 
For SU(3) gluons, which reside in the adjoint representation, the initial color vector is fixed by $q_2(A)=24$ and $q_3(A)=0$ while SU(3) quarks in fundamental representation are labeled by $q_2(F)=4$ and $q_3(F)=10/3$. In our numerical simulations, we use the initial color vectors
\begin{equation}
    \label{eq:initq0}
    \begin{aligned}
        &\vv{Q}(A)=(4.89898, 0, 0, 0, 0, 0, 0, 0),\\
        &\vv{Q}(F)=(0, 0, 0, 0, -1.69469, 0, 0, -1.06209),
    \end{aligned}
\end{equation}
where we use the notation $\vv{Q}\equiv (Q_1,\dots,Q_8)$. The case of SU(2) is much simpler, because only the quadratic Casimir invariant applies. Possible color charges $Q^a$ are vectors on $2$-sphere with radius $q_2$.


\section{Sampling classical color charges via the Haar measure}
\label{appen:samplecolors}

This Appendix contains an analytical proof for the matching from Eq.~\eqref{eq:matchqs} of the one-, two- and three-point functions of the classical color charges computed via the Haar measure, as given in Eq.~\eqref{eq:nptfcthaar}, to the desired values provided in Eq.~\eqref{eq:qnpointfct_main}. For example, the aim is to derive that the one-point function extracted by integrating over the manifold of SU($N_c$)
\begin{align*}
    \langle Q^a\rangle_U=Q_0^{a^\prime} \int\d U\, U^{a a^\prime}
\end{align*}
where $Q_0^{a^\prime}$ is the initial color vector fixed by the quadratic and cubic Casimirs from Eqs.~\eqref{eq:quad_casimir} and~\eqref{eq:cubic_casimir}, does indeed give the expected value
 \begin{align*}
     \langle Q^a\rangle_U\mapsto \langle Q^a\rangle=0.
 \end{align*}
To this end, we need to perform the integral over matrix elements of $U\in\mathrm{SU}(N_c)$ matrices, as can be seen by using Eq.~\eqref{eq:uadj} to further write 
\begin{align*}
    \int\d U\, U^{a a^\prime}&=\int \d U \, \frac{1}{T_R}\, \tr{T^a U T^{a^\prime} U^\dagger} \\
    &=\frac{1}{T_R}\,T^a_{li}T^{a^\prime}_{jk}\int \d U \, U_{ij}U^\dagger_{kl}
\end{align*}
and similarly for the two- and three-point functions. This matching requires the evaluation of certain integrals over SU($N_c$). In what follows, we calculate them symbolically\footnote{A Python notebook using SymPy \cite{SymPy} where these calculations are carried out explicitly is publicly available at \href{https://github.com/avramescudana/sun\_integrals}{https://github.com/avramescudana/sun\_integrals}.}, in the fundamental representation, and quote the main steps in the derivation.

\subsection{Revising some relevant SU($N_c$) integrals}
\label{appen:sunintegrals}
According to \cite{Zuber:2016xme}, integrals over $U\in\mathrm{SU(N_c)}$ of the type
\begin{align}
    \int \d U \, U_{i_1j_1}\dots U_{i_pj_p}U^\dagger_{k_1l_1}\dots U^\dagger_{k_n l_n},
\end{align}
where $U_{ij}$ denotes the matrix elements of $U$ (in the fundamental representation), may be evaluated from the generating function
\begin{align}
    Z_{p,n}(J,K)=\int \d U \, \left(\tr{KU}\right)^p\left(\tr{JU^\dagger}\right)^n.
\end{align}
We are particularly interested in $Z_{1, 1}$, $Z_{2, 2}$ and $Z_{3,3}$ and their derivatives with respect to $J$ and $K$. For each of these cases, let us see how we may generate our integrals of interest and how to evaluate them, using already computed expressions from \cite{Creutz:1978ub,Carlsson:2008dh,Zuber:2016xme}, for $N_c=3$.

~

\subsubsection{Integral involving one pair of conjugate matrix elements}
For $p=n=1$, the generating function
\begin{align}
    \label{eq:genfct1}
    Z_{1,1}(J,K)=\int \d U \, \tr{KU}\tr{JU^\dagger},
\end{align}
generates, via differentiation, the integral
\begin{align}
    \label{eq:genfct1eq1}
    \frac{\partial^2 }{\partial K_{ji}\partial J_{lk}}Z_{1,1}(J,K)=\int \d U \, U_{ij}U^\dagger_{kl}.
\end{align}
On the other hand, following \cite{Zuber:2016xme}, the generating function Eq.~\eqref{eq:genfct1} is given by
\begin{align}
    Z_{1,1}(J, K) = \frac{1}{N_c}\tr{JK},
\end{align}
which yields
\begin{align}
    \label{eq:genfct1eq2}
    \frac{\partial^2 }{\partial K_{ji}\partial J_{lk}}Z_{1,1}(J,K)=\frac{1}{N_c}\delta_{il}\delta_{jk}.
\end{align}
Collecting Eqs.~\eqref{eq:genfct1eq1} and~\eqref{eq:genfct1eq2} immediately gives
\begin{align}
    \label{eq:intdu1point}
    \int \d U \, U_{ij}U^\dagger_{kl}=\frac{1}{N_c}\delta_{il}\delta_{jk}.
\end{align}
    
\subsubsection{Integral involving two pairs of conjugate matrix elements}
In the case $p=n=2$, the generating function reads
\begin{align}
    Z_{1,1}(J,K)=\int \d U \, \left(\tr{KU}\right)^2\left(\tr{JU^\dagger}\right)^2.
\end{align}
Let us introduce the following simplifying notation
\begin{align}
    \partial^4_{ijklmnop}\equiv \frac{\partial^4}{\partial K_{ji}\partial J_{lk}\partial K_{nm}\partial J_{po}}.
\end{align}
Using similar computations as before, this leads to the following integral over SU($N_c$)
\begin{align}
    \label{eq:genfct2eq1}
    \partial^4_{ijklmnop}Z_{2,2}(J, K) = 4 \int \d U \, U_{ij}U^\dagger_{kl} U_{mn}U^\dagger_{op}.
\end{align}
Alternatively, the generating function is given by \cite{Zuber:2016xme}
\begin{align}
    \begin{split}
    &Z_{2,2}(J, K)\\
    &=2\,\Bigg\{\frac{1}{N_c^2-1}\tr{(JK)^2}-\frac{1}{N_c(N_c^2-1)}\left(\tr{JK}\right)^2\Bigg\}.
    \end{split}
\end{align}
Differentiating the terms gives
\begin{align}
    \begin{split}    \partial^4_{ijklmnop}\left(\tr{JK}\right)^2&=2(\delta_{il}\delta_{jk}\delta_{mp}\delta_{no}+\delta_{ip}\delta_{jo}\delta_{lm}\delta_{kn}),\\    \partial^4_{ijklmnop}\tr{(JK)^2}&=2(\delta_{il}\delta_{jo}\delta_{kn}\delta_{mp}+\delta_{ip}\delta_{jk}\delta_{lm}\delta_{no}).
    \end{split}
\end{align}
Collecting all these results yields the desired integral
\begin{equation}
    \label{eq:intdu2point}
    \begin{aligned}
         &\int \d U \, U_{ij}U^\dagger_{kl} U_{mn}U^\dagger_{op}\\
         &=\frac{1}{N_c^2-1}(\delta_{il}\delta_{jk}\delta_{mp}\delta_{no}+\delta_{ip}\delta_{jo}\delta_{lm}\delta_{kn})-\\
         &\phantom{=}\frac{1}{N_c(N_c^2-1)}(\delta_{il}\delta_{jo}\delta_{kn}\delta_{mp}+\delta_{ip}\delta_{jk}\delta_{lm}\delta_{no}).
    \end{aligned}
\end{equation}


\subsubsection{Integral involving three pairs of conjugate matrix elements}
Taking $p=n=3$ gives the generating function
\begin{align}
    Z_{3,3}(J,K)=\int \d U \, \left(\tr{KU}\right)^3\left(\tr{JU^\dagger}\right)^3.
\end{align}
As before, we use the shorthand
\begin{align}
    \partial^6_{ijklmnopqrst}\equiv \frac{\partial^6}{\partial K_{ji}\partial J_{lk}\partial K_{nm}\partial J_{po}\partial K_{rq}\partial J_{ts}}.
\end{align}
This then leads to the following integral over SU($N_c$):
\begin{align}
    \partial^6_{ijklmnopqrst}Z_{3,3}(J, K) = 36 \int \d U \, U_{ij}U^\dagger_{kl} U_{mn}U^\dagger_{op}U_{qr}U^\dagger_{st}.
\end{align}
According to \cite{Zuber:2016xme}, this generating functional is given by

\begin{widetext}
\begin{align}   
    \begin{split}
    &Z_{3,3}(J,K)\\
    &=6\Bigg\{\frac{N_c^2-2}{(N_c^2-4)(N_c^2-1)N_c}(\tr{JK})^3-\frac{3}{(N_c^2-4)(N_c^2-1)}\tr{JK}\tr{(JK)^2}+\frac{4}{(N_c^2-4)(N_c^2-1)N_c}\tr{(JK)^3}\Bigg\}.
    \end{split}
\end{align}

Performing differentiation for each of these terms yields
\begin{align}
        \begin{split}
        &\partial^6_{ijklmnopqrst}(\tr{JK})^3\\
        &=6 \Big(\delta_{i l} \delta_{j k} \delta_{m p} \delta_{n o} \delta_{q t} \delta_{r s} + \delta_{i l} \delta_{j k} \delta_{m t} \delta_{n s} \delta_{o r} \delta_{p q} + \delta_{i p} \delta_{j o} \delta_{k n} \delta_{l m} \delta_{q t} \delta_{r s} + \delta_{i p} \delta_{j o} \delta_{k r} \delta_{l q} \delta_{m t} \delta_{n s} + \delta_{i t} \delta_{j s} \delta_{k n} \delta_{l m} \delta_{o r} \delta_{p q} \\
        &\phantom{=}+ \ \delta_{i t} \delta_{j s} \delta_{k r} \delta_{l q} \delta_{m p} \delta_{n o}\Big),
        \end{split}
\end{align}
\begin{align}
        \begin{split}
            &\partial^6_{ijklmnopqrst}\tr{JK}\tr{(JK)^2}\\  
        &=2 \Big(\delta_{i l} \delta_{j k} \delta_{m p} \delta_{n s} \delta_{o r} \delta_{q t} + \delta_{i l} \delta_{j k} \delta_{m t} \delta_{n o} \delta_{p q} \delta_{r s} + \delta_{i l} \delta_{j o} \delta_{k n} \delta_{m p} \delta_{q t} \delta_{r s} + \delta_{i l} \delta_{j o} \delta_{k r} \delta_{m t} \delta_{n s} \delta_{p q} + \delta_{i l} \delta_{j s} \delta_{k n} \delta_{m t} \delta_{o r} \delta_{p q} \\
        &\phantom{=} + \delta_{i l} \delta_{j s} \delta_{k r} \delta_{m p} \delta_{n o} \delta_{q t} + \delta_{i p} \delta_{j k} \delta_{l m} \delta_{n o} \delta_{q t} \delta_{r s} + \delta_{i p} \delta_{j k} \delta_{l q} \delta_{m t} \delta_{n s} \delta_{o r} + \delta_{i p} \delta_{j o} \delta_{k n} \delta_{l q} \delta_{m t} \delta_{r s} + \delta_{i p} \delta_{j o} \delta_{k r} \delta_{l m} \delta_{n s} \delta_{q t} \\
        &\phantom{=} +\delta_{i p} \delta_{j s} \delta_{k n} \delta_{l m} \delta_{o r} \delta_{q t} + \delta_{i p} \delta_{j s} \delta_{k r} \delta_{l q} \delta_{m t} \delta_{n o}+\delta_{i t} \delta_{j k} \delta_{l m} \delta_{n s} \delta_{o r} \delta_{p q} +\delta_{i t} \delta_{j k} \delta_{l q} \delta_{m p} \delta_{n o} \delta_{r s} + \delta_{i t} \delta_{j o} \delta_{k n} \delta_{l m} \delta_{p q} \delta_{r s} \\
        &\phantom{=} + \delta_{i t} \delta_{j o} \delta_{k r} \delta_{l q} \delta_{m p} \delta_{n s} + \delta_{i t} \delta_{j s} \delta_{k n} \delta_{l q} \delta_{m p} \delta_{o r} + \delta_{i t} \delta_{j s} \delta_{k r} \delta_{l m} \delta_{n o} \delta_{p q}\Big),
    \end{split}
\end{align}
\begin{align}
        \begin{split}
    &\partial^6_{ijklmnopqrst}\tr{(JK)^3}\\
        &=3 \Big(\delta_{i l} \delta_{j o} \delta_{k n} \delta_{m t} \delta_{p q} \delta_{r s} + \delta_{i l} \delta_{j o} \delta_{k r} \delta_{m p} \delta_{n s} \delta_{q t} + \delta_{i l} \delta_{j s} \delta_{k n} \delta_{m p} \delta_{o r} \delta_{q t} + \delta_{i l} \delta_{j s} \delta_{k r} \delta_{m t} \delta_{n o} \delta_{p q} + \delta_{i p} \delta_{j k} \delta_{l m} \delta_{n s} \delta_{o r} \delta_{q t} \\
        &\phantom{=}  + \delta_{i p} \delta_{j k} \delta_{l q} \delta_{m t} \delta_{n o} \delta_{r s} + \delta_{i p} \delta_{j s} \delta_{k n} \delta_{l q} \delta_{m t} \delta_{o r} + \delta_{i p} \delta_{j s} \delta_{k r} \delta_{l m} \delta_{n o} \delta_{q t} + \delta_{i t} \delta_{j k} \delta_{l m} \delta_{n o} \delta_{p q} \delta_{r s} + \delta_{i t} \delta_{j k} \delta_{l q} \delta_{m p} \delta_{n s} \delta_{o r}\\
        &\phantom{=} +  \delta_{i t} \delta_{j o} \delta_{k n} \delta_{l q} \delta_{m p} \delta_{r s} + \delta_{i t} \delta_{j o} \delta_{k r} \delta_{l m} \delta_{n s} \delta_{p q}\Big).
        \end{split}
\end{align}

Combining all of these results gives us
\begin{align}
    \label{eq:int3u1}
    \begin{split}
        &\int \d U \, U_{ij}U^\dagger_{kl} U_{mn}U^\dagger_{op}U_{qr}U^\dagger_{st}=\frac{1}{6}\Bigg\{\frac{N_c^2-2}{(N_c^2-4)(N_c^2-1)N_c}\partial^6_{ijklmnopqrst}(\tr{JK})^3\\
    &-\frac{3}{(N_c^2-4)(N_c^2-1)}\partial^6_{ijklmnopqrst}\tr{JK}\tr{(JK)^2}+\frac{4}{(N_c^2-4)(N_c^2-1)N_c}\partial^6_{ijklmnopqrst}\tr{(JK)^3}\Bigg\}.
    \end{split}
\end{align}
\end{widetext}
    
\subsection{Matching the one-, two- and three-point functions}
\label{appen:matchnpointfct}
Equipped with the SU($N_c$) integrals from Eqs.~\eqref{eq:intdu1point}, ~\eqref{eq:intdu2point} and~\eqref{eq:int3u1}, we now show that Eq.~\eqref{eq:matchqs} holds for the one-, two- and three-point functions. 

\subsubsection{Matching the one-point functions}
The LHS of Eq.~\eqref{eq:match1pointfct} should be fixed by Eq.~\eqref{eq:qa} as $\langle Q^a\rangle=0$. On the other hand, the RHS contains the term
\begin{align}
    \int \d U \, \tr{T^a U T^{a^\prime} U^\dagger} =T^a_{li}T^{a^\prime}_{jk}\int \d U \, U_{ij}U^\dagger_{kl}.
\end{align}
multiplied by $1/T_F$, see Eq.~\eqref{eq:uadj}. Employing the integral from Eq.~\eqref{eq:intdu1point} immediately gives
\begin{align}
    \begin{split}
    &\int \d U \, \tr {T^a U T^{a^\prime} U^\dagger}=T^a_{li}T^{a^\prime}_{jk}\frac{1}{N_c}\delta_{il}\delta_{jk}\\
    &=\frac{1}{N_c}T^a_{ii}T^{a^\prime}_{jj}=\frac{1}{N_c}\tr{T^a}\tr{T^{a^\prime}}=0,
    \end{split}
\end{align}
since $\tr{T^a}=0$, see Eq.~\eqref{eq:trga}, thus $\langle Q^a\rangle_U=0$.
    
\subsubsection{Matching the two-point functions}
\label{appendixclassiccharges}
Similarly, we want the LHS of Eq.~\eqref{eq:match2pointfct} to be expressible as in Eq.~\eqref{eq:qaqb}. Besides a factor of $1/T_F^2$ from Eq.~\eqref{eq:uadj}, the RHS is proportional to

\begin{equation}
    \begin{aligned}
        &\int \d U \, \tr{T^a U T^{a^\prime} U^\dagger}  \tr{T^b U T^{b^\prime} U^\dagger}\\
        &=T^a_{li}T^{a^\prime}_{jk}T^b_{pm}T^{b^\prime}_{no}\int \d U \, U_{ij}U^\dagger_{kl}U_{mn}U^\dagger_{op}\label{eq:2pointfcteq1},
    \end{aligned}
\end{equation}
\begin{widetext}
in which the integral involving SU($N_c$) matrices may be evaluated from Eq.~\eqref{eq:intdu2point}. Inserting this relation back into Eq.~\eqref{eq:2pointfcteq1} yields
\begin{equation}
    \begin{aligned}
        &\int \d U \, \tr{T^a U T^{a^\prime} U^\dagger}  \tr{T^b U T^{b^\prime} U^\dagger}\\
        &=\frac{1}{N_c^2-1}\Big(T^a_{ii}T^{a^\prime}_{kk}T^b_{mm}T^{b^\prime}_{oo}+T^a_{li}T^b_{il}T^{a^\prime}_{jk}T^{b^\prime}_{kj}\Big)-\frac{1}{N_c(N_c^2-1)}\Big(T^a_{ii}T^{a^\prime}_{jk}T^{b}_{mm}T^{b^\prime}_{kj}+T^a_{li}T^{a^\prime}_{kk}T^b_{il}T^{b^\prime}_{oo}\Big)\\
        &=\frac{1}{N_c^2-1}\Big(\tr{T^a}\tr{T^{a^\prime}}\tr{T^b}\tr{T^{b^\prime}}+\tr{T^aT^b}\tr{T^{a^\prime}T^{b^\prime}}\Big)\\
        &\phantom{=}-\frac{1}{N_c(N_c^2-1)}\Big(\tr{T^a}\tr{T^b}\tr{T^{a^\prime}T^{b^\prime}}+\tr{T^aT^b}\tr{T^{a^\prime}}\tr{T^{b^\prime}}\Big)\\
        &=\frac{1}{N_c^2-1}T_F^2\delta^{ab}\delta^{a^\prime b^\prime},
    \end{aligned}
\end{equation}
\end{widetext}

where in the last step Eqs.~\eqref{eq:trga} and~\eqref{eq:trgagb} were used. This finally gives the RHS of Eq.~\eqref{eq:match2pointfct} as
\begin{align}
    \langle Q^aQ^b\rangle_U=\frac{1}{T_F^2}Q_0^{a^\prime}Q_0^{b^\prime}\frac{1}{N_c^2-1}{T_F^2}\delta^{ab}\delta^{a^\prime b^\prime}.
\end{align}
Using the classical Casimir of Eq.~\eqref{eq:q2} gives $Q_0^{a^\prime}Q_0^{a^\prime}=D_FC_2(F)=(N_c^2-1)T_F$, which finally yields the desired two-point function
\begin{align}
    \langle Q^aQ^b\rangle_U=T_F\delta^{ab},
\end{align}
exactly the two-point function from Eq.~\eqref{eq:qaqb}.

\subsubsection{Matching the three-point functions}
The LHS of Eq.~\eqref{eq:match3pointfct} should be given by Eq.~\eqref{eq:qaqbqc}, while the RHS contains the term

\begin{widetext}
    \begin{align}
        \int \d U \, \tr {T^a U T^{a^\prime} U^\dagger}  \tr{T^b U T^{b^\prime} U^\dagger} \tr{T^c U T^{c^\prime} U^\dagger}&= T^a_{li}T^{a^\prime}_{jk}T^b_{pm}T^{b^\prime}_{no}T^c_{tq}T^{c^\prime}_{rs}\int \d U \, U_{ij}U^\dagger_{kl}U_{mn}U^\dagger_{op}U_{qr}U^\dagger_{st}
    \end{align}    
multiplied by $1/T_F^3=8$ coming from Eq.~\eqref{eq:uadj}. When the integral over SU($N_c$) is evaluated according to Eq.~\eqref{eq:int3u1} and using Eq.~\eqref{eq:trga} along with Eq.~\eqref{eq:trgagbgc}, one may show that the only non-vanishing terms are given by
\begin{equation}
    \begin{aligned}
        &\int \d U \, \tr {T^a U T^{a^\prime} U^\dagger}  \tr{T^b U T^{b^\prime} U^\dagger} \tr{T^c U T^{c^\prime} U^\dagger}\\
        &=\Bigg[\frac{N_c^2-2}{(N_c^2-4)(N_c^2-1)N_c}+\frac{2}{(N_c^2-4)(N_c^2-1)N_c}\Bigg]\Bigg(\tr{T^aT^bT^c}\tr{T^{a^\prime}T^{b^\prime}T^{c^\prime}}+\tr{T^aT^cT^b}\tr{T^{a^\prime}T^{c^\prime}T^{b^\prime}}\Bigg)\\
        &=\frac{N_c}{(N_c^2-4)(N_c^2-1)}\frac{1}{16}\Big[(d_{abc}+\mathrm{i}f_{abc})(d_{a^\prime b^\prime c^\prime}+\mathrm{i}f_{a^\prime b^\prime c^\prime})+(d_{abc}-\mathrm{i}f_{abc})(d_{a^\prime b^\prime c^\prime}-\mathrm{i}f_{a^\prime b^\prime c^\prime})\Big]\\
        &=\frac{1}{8}\frac{N_c}{(N_c^2-4)(N_c^2-1)}(d_{abc}d_{a^\prime b^\prime c^\prime}-f_{abc}f_{a^\prime b^\prime c^\prime}),
    \end{aligned}    
\end{equation}

where we used the fact that $d_{abc}$ is symmetric with respect to $a,b,c$, whereas $f_{abc}$ is anti-symmetric. This then yields
\begin{equation}
    \begin{aligned}
        \langle Q^aQ^bQ^c\rangle_U&=8\sum_{a^\prime b^\prime c^\prime}Q_0^{a^\prime}Q_0^{b^\prime}Q_0^{c^\prime}\int dU \, \tr {T^a U T^{a^\prime} U^\dagger}  \tr {T^b U T^{b^\prime} U^\dagger} \tr{T^c U T^{c^\prime} U^\dagger}\\
        &=\frac{N_c}{(N_c^2-4)(N_c^2-1)}\Big(d_{abc}\underbrace{d_{a^\prime b^\prime c^\prime}Q_0^{a^\prime}Q_0^{b^\prime}Q_0^{c^\prime}}_{ q_3(F)}-f_{abc}\underbrace{f_{a^\prime b^\prime c^\prime}Q_0^{a^\prime}Q_0^{b^\prime}Q_0^{c^\prime}}_{0}\Big),
    \end{aligned}  
\end{equation}
\end{widetext}

in which the antisymmetry of $f_{a^\prime b^\prime c^\prime}$ was used. The further replacement of $q_3(F)=D_F C_3(F)=(N_c^2-4)(N_c^2-1)/(4N_c)$ gives exactly the three-point function from Eq.~\eqref{eq:qaqbqc}
\begin{align}
    \langle Q^aQ^bQ^c\rangle_U=\frac{A_R}{4}d_{abc}.
\end{align}   


\section{Numerical checks}
\label{appen:numchecks}

\subsection{Casimir scaling and SU(2) versus SU(3) comparison}
Our publicly available simulation code \cite{curraun} was subject to various numerical checks. All of them were done for very massive quarks with vanishing transverse momentum, but hold irrespective of particle type or initial momentum.

Firstly, we verified that the Casimirs of the classical color charges are perfectly conserved throughout the evolution, up to numerical precision. This requirement is automatically satisfied by the numerical method used to solve the evolution of the color charge, namely the color rotation with Wilson lines constructed on the lattice, as in Eq.~\eqref{eq:colorrotnum}.

\begin{figure}[!t]
\includegraphics[width=0.85\columnwidth]{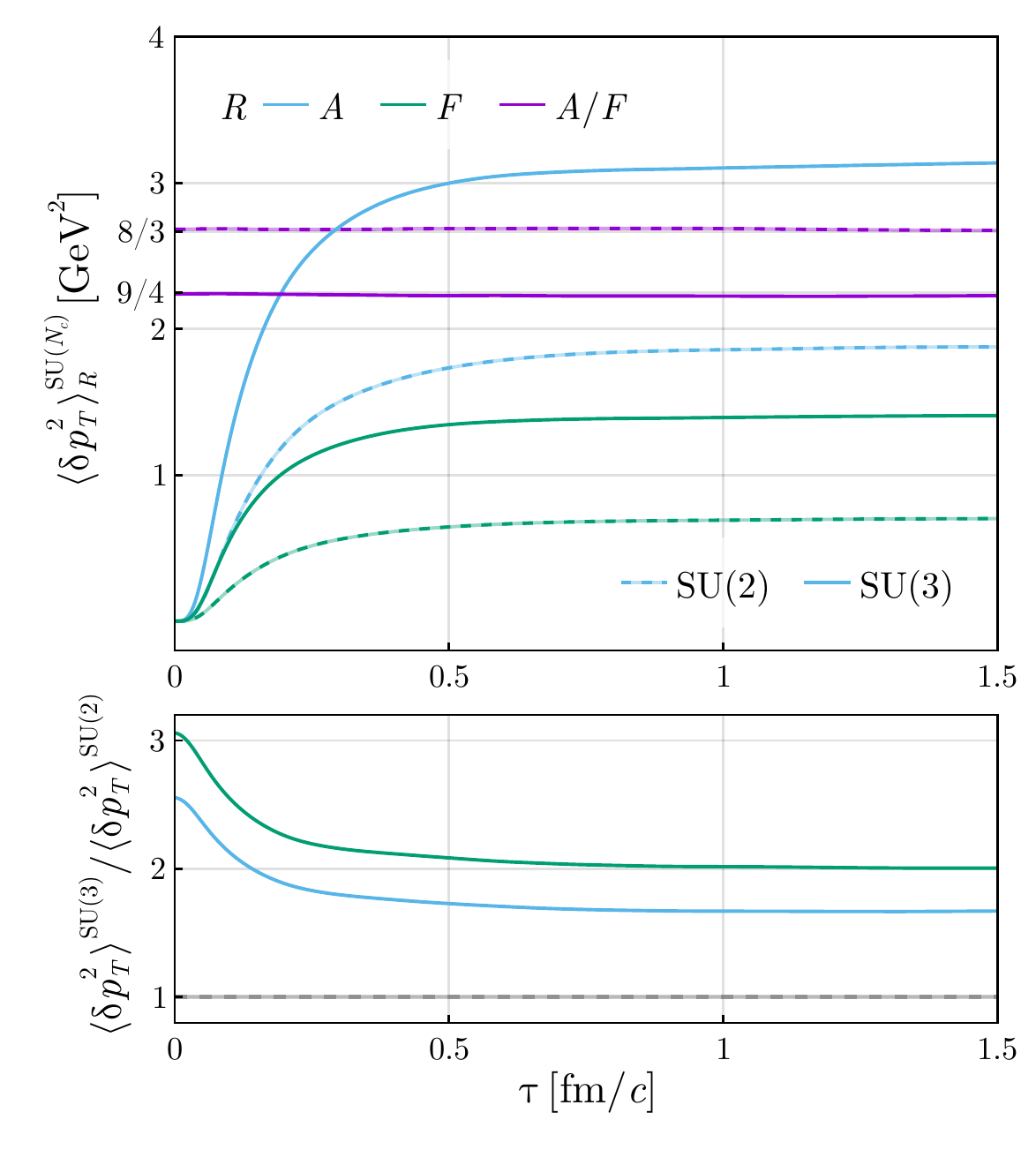}
 \caption{\label{fig:casimirscaling} \textit{(Upper)} Transverse momentum broadening components of very massive quarks, for SU(2) \textit{(dashed line)} and SU(3) \textit{(full line)}, adjoint \textit{(blue)} and fundamental \textit{(green)} representations. Their ratio \textit{(violet)} obeys the Casimir scaling property, as expressed in Eq.~\eqref{eq:casimirscaling}. The band emphasizes the difference between the SU(2) and SU(3) results. \textit{(Lower)} Ratio between SU(2) and SU(3) accumulated transverse momenta, as a function of proper time. There is no trivial scaling at very early times, namely $\delta\tau \lesssim0.5\,\mathrm{fm/}c$. 
 }
\end{figure}

Secondly, we checked that the Casimir scaling expressed in Eq.~\eqref{eq:diffcasscaling} is satisfied at all proper times. For this purpose, we initialized classical color charges in both fundamental and adjoint representations, for SU(2) and SU(3), by appropriately choosing the corresponding Casimir invariants from Eq.~\eqref{eq:q23_main}. We emphasize that classical momentum broadening components are mapped to quantum ones according to Eq.~\eqref{eq:match}. Using Eq.~\eqref{eq:casimirscaling} along with Eq.~\eqref{eq:c23_main}, namely $\left\langle \delta p_{\mu}^{2}\right\rangle_{A}/\left\langle \delta p_{\mu}^{2}\right\rangle_{F}=2N_c^2/(N_c^2-1)$, yields
\begin{align}
    \dfrac{\left\langle \delta p_{\mu}^{2}\right\rangle_{A}}{\left\langle \delta p_{\mu}^{2}\right\rangle_{F}}\Bigg|^\mathrm{SU(2)}=\frac{8}{3},\quad\dfrac{\left\langle \delta p_{\mu}^{2}\right\rangle_{A}}{\left\langle \delta p_{\mu}^{2}\right\rangle_{F}}\Bigg|^\mathrm{SU(3)}=\frac{9}{4}.
\end{align}
These are exactly the results obtained from the numerical simulation, see Fig.~\ref{fig:casimirscaling} (\emph{upper panel}). 

Additionally, we compared SU(2) to SU(3) momentum broadening components, see Fig.~\ref{fig:casimirscaling}  (\emph{lower panel}). We find that, in the very early stages,  there is no simple way to scale the results of SU(2) to SU(3). At later proper times, when the Glasma enters the free-streaming regime and the fields are dilute, there exists a simple scaling from SU(2) to SU(3) momentum broadening, but the scaling factor differs for adjoint and fundamental representations. 

\subsection{Temporal constraint}

\begin{figure}[t]
\includegraphics[width=0.85\columnwidth]{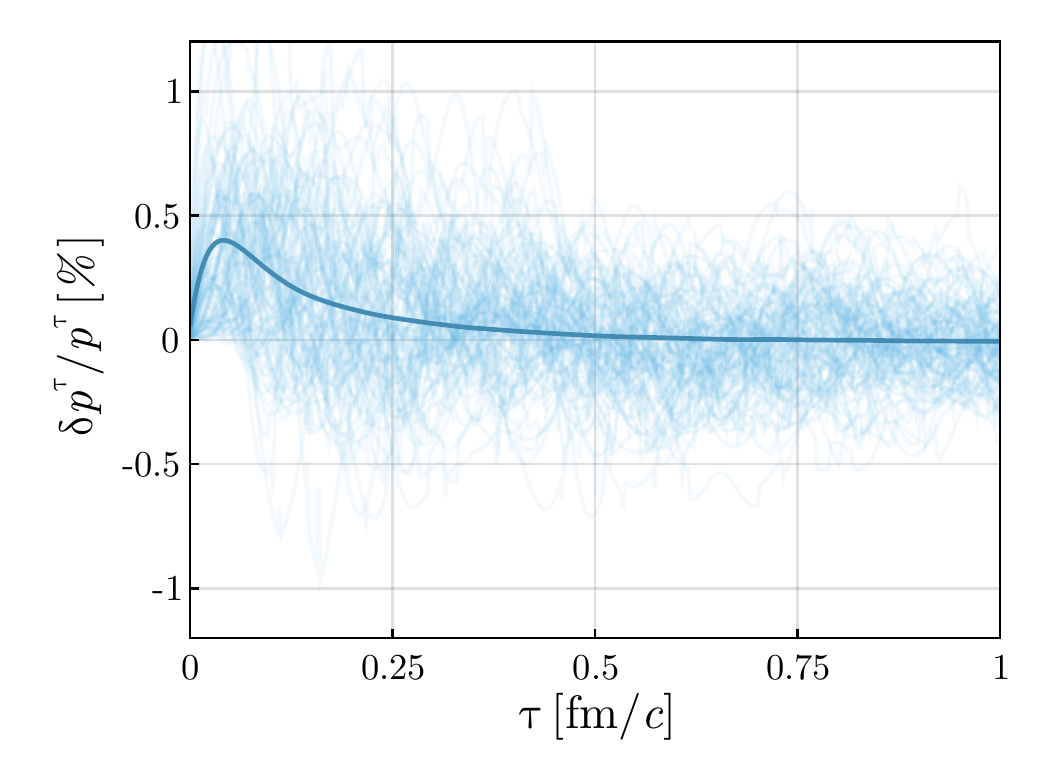}
\caption{\label{fig:ptauconstraint} The relative difference, in percentage, between the temporal component of the momentum $p^\tau$ numerically extracted from Eq.~\eqref{eq:ptau} versus Eq.~\eqref{eq:ptaueq}. The plot contains $N_\mathrm{tp}=100$ test particles \textit{(thin light blue lines)} and their average \textit{(full blue line)}.
}
\end{figure}

Numerically, the temporal component of the momentum is computed from Eq.~\eqref{eq:ptau} but $p^\tau$ may also be extracted using Eq.~\eqref{eq:ptaueq}. We denote by $\delta p^\tau$ the difference between these two distinct ways of computing the temporal component and represent it in Fig.~\ref{fig:ptauconstraint}. The difference is on the percent level and can be understood as an artifact of the lattice discretization.

\nocite{}
\bibliography{references}

\begin{thebibliography}{90}%
\makeatletter
\providecommand \@ifxundefined [1]{%
 \@ifx{#1\undefined}
}%
\providecommand \@ifnum [1]{%
 \ifnum #1\expandafter \@firstoftwo
 \else \expandafter \@secondoftwo
 \fi
}%
\providecommand \@ifx [1]{%
 \ifx #1\expandafter \@firstoftwo
 \else \expandafter \@secondoftwo
 \fi
}%
\providecommand \natexlab [1]{#1}%
\providecommand \enquote  [1]{``#1''}%
\providecommand \bibnamefont  [1]{#1}%
\providecommand \bibfnamefont [1]{#1}%
\providecommand \citenamefont [1]{#1}%
\providecommand \href@noop [0]{\@secondoftwo}%
\providecommand \href [0]{\begingroup \@sanitize@url \@href}%
\providecommand \@href[1]{\@@startlink{#1}\@@href}%
\providecommand \@@href[1]{\endgroup#1\@@endlink}%
\providecommand \@sanitize@url [0]{\catcode `\\12\catcode `\$12\catcode
  `\&12\catcode `\#12\catcode `\^12\catcode `\_12\catcode `\%12\relax}%
\providecommand \@@startlink[1]{}%
\providecommand \@@endlink[0]{}%
\providecommand \url  [0]{\begingroup\@sanitize@url \@url }%
\providecommand \@url [1]{\endgroup\@href {#1}{\urlprefix }}%
\providecommand \urlprefix  [0]{URL }%
\providecommand \Eprint [0]{\href }%
\providecommand \doibase [0]{http://dx.doi.org/}%
\providecommand \selectlanguage [0]{\@gobble}%
\providecommand \bibinfo  [0]{\@secondoftwo}%
\providecommand \bibfield  [0]{\@secondoftwo}%
\providecommand \translation [1]{[#1]}%
\providecommand \BibitemOpen [0]{}%
\providecommand \bibitemStop [0]{}%
\providecommand \bibitemNoStop [0]{.\EOS\space}%
\providecommand \EOS [0]{\spacefactor3000\relax}%
\providecommand \BibitemShut  [1]{\csname bibitem#1\endcsname}%
\let\auto@bib@innerbib\@empty
\bibitem [{\citenamefont {Lappi}\ and\ \citenamefont
  {McLerran}(2006)}]{Lappi:2006fp}%
  \BibitemOpen
  \bibfield  {author} {\bibinfo {author} {\bibfnamefont {T.}~\bibnamefont
  {Lappi}}\ and\ \bibinfo {author} {\bibfnamefont {L.}~\bibnamefont
  {McLerran}},\ }\href {\doibase 10.1016/j.nuclphysa.2006.04.001} {\bibfield
  {journal} {\bibinfo  {journal} {Nucl. Phys. A}\ }\textbf {\bibinfo {volume}
  {772}},\ \bibinfo {pages} {200} (\bibinfo {year} {2006})},\ \Eprint
  {http://arxiv.org/abs/hep-ph/0602189} {arXiv:hep-ph/0602189} \BibitemShut
  {NoStop}%
\bibitem [{\citenamefont {Lappi}(2008{\natexlab{a}})}]{Lappi:2008eq}%
  \BibitemOpen
  \bibfield  {author} {\bibinfo {author} {\bibfnamefont {T.}~\bibnamefont
  {Lappi}},\ }\href {\doibase 10.1088/0954-3899/35/10/104052} {\bibfield
  {journal} {\bibinfo  {journal} {J. Phys. G}\ }\textbf {\bibinfo {volume}
  {35}},\ \bibinfo {pages} {104052} (\bibinfo {year} {2008}{\natexlab{a}})},\
  \Eprint {http://arxiv.org/abs/0804.2338} {arXiv:0804.2338 [hep-ph]}
  \BibitemShut {NoStop}%
\bibitem [{\citenamefont {Fujii}\ \emph {et~al.}(2009)\citenamefont {Fujii},
  \citenamefont {Fukushima},\ and\ \citenamefont {Hidaka}}]{Fujii:2008km}%
  \BibitemOpen
  \bibfield  {author} {\bibinfo {author} {\bibfnamefont {H.}~\bibnamefont
  {Fujii}}, \bibinfo {author} {\bibfnamefont {K.}~\bibnamefont {Fukushima}}, \
  and\ \bibinfo {author} {\bibfnamefont {Y.}~\bibnamefont {Hidaka}},\ }\href
  {\doibase 10.1103/PhysRevC.79.024909} {\bibfield  {journal} {\bibinfo
  {journal} {Phys. Rev. C}\ }\textbf {\bibinfo {volume} {79}},\ \bibinfo
  {pages} {024909} (\bibinfo {year} {2009})},\ \Eprint
  {http://arxiv.org/abs/0811.0437} {arXiv:0811.0437 [hep-ph]} \BibitemShut
  {NoStop}%
\bibitem [{\citenamefont {Fukushima}\ and\ \citenamefont
  {Gelis}(2012)}]{Fukushima:2011nq}%
  \BibitemOpen
  \bibfield  {author} {\bibinfo {author} {\bibfnamefont {K.}~\bibnamefont
  {Fukushima}}\ and\ \bibinfo {author} {\bibfnamefont {F.}~\bibnamefont
  {Gelis}},\ }\href {\doibase 10.1016/j.nuclphysa.2011.11.003} {\bibfield
  {journal} {\bibinfo  {journal} {Nucl. Phys. A}\ }\textbf {\bibinfo {volume}
  {874}},\ \bibinfo {pages} {108} (\bibinfo {year} {2012})},\ \Eprint
  {http://arxiv.org/abs/1106.1396} {arXiv:1106.1396 [hep-ph]} \BibitemShut
  {NoStop}%
\bibitem [{\citenamefont {Iancu}\ and\ \citenamefont
  {Venugopalan}(2003)}]{Iancu:2003xm}%
  \BibitemOpen
  \bibfield  {author} {\bibinfo {author} {\bibfnamefont {E.}~\bibnamefont
  {Iancu}}\ and\ \bibinfo {author} {\bibfnamefont {R.}~\bibnamefont
  {Venugopalan}},\ }\enquote {\bibinfo {title} {{The Color Glass Condensate and
  high-energy scattering in QCD}},}\ in\ \href {\doibase
  10.1142/9789812795533_0005} {\emph {\bibinfo {booktitle} {{Quark-gluon plasma
  4}}}},\ \bibinfo {editor} {edited by\ \bibinfo {editor} {\bibfnamefont
  {R.~C.}\ \bibnamefont {Hwa}}\ and\ \bibinfo {editor} {\bibfnamefont {X.-N.}\
  \bibnamefont {Wang}}}\ (\bibinfo {year} {2003})\ pp.\ \bibinfo {pages}
  {249--3363},\ \Eprint {http://arxiv.org/abs/hep-ph/0303204}
  {arXiv:hep-ph/0303204} \BibitemShut {NoStop}%
\bibitem [{\citenamefont {Gelis}\ \emph {et~al.}(2010)\citenamefont {Gelis},
  \citenamefont {Iancu}, \citenamefont {Jalilian-Marian},\ and\ \citenamefont
  {Venugopalan}}]{Gelis:2010nm}%
  \BibitemOpen
  \bibfield  {author} {\bibinfo {author} {\bibfnamefont {F.}~\bibnamefont
  {Gelis}}, \bibinfo {author} {\bibfnamefont {E.}~\bibnamefont {Iancu}},
  \bibinfo {author} {\bibfnamefont {J.}~\bibnamefont {Jalilian-Marian}}, \ and\
  \bibinfo {author} {\bibfnamefont {R.}~\bibnamefont {Venugopalan}},\ }\href
  {\doibase 10.1146/annurev.nucl.010909.083629} {\bibfield  {journal} {\bibinfo
   {journal} {Ann. Rev. Nucl. Part. Sci.}\ }\textbf {\bibinfo {volume} {60}},\
  \bibinfo {pages} {463} (\bibinfo {year} {2010})},\ \Eprint
  {http://arxiv.org/abs/1002.0333} {arXiv:1002.0333 [hep-ph]} \BibitemShut
  {NoStop}%
\bibitem [{\citenamefont {Gelis}(2013)}]{Gelis:2012ri}%
  \BibitemOpen
  \bibfield  {author} {\bibinfo {author} {\bibfnamefont {F.}~\bibnamefont
  {Gelis}},\ }\href {\doibase 10.1142/S0217751X13300019} {\bibfield  {journal}
  {\bibinfo  {journal} {Int. J. Mod. Phys. A}\ }\textbf {\bibinfo {volume}
  {28}},\ \bibinfo {pages} {1330001} (\bibinfo {year} {2013})},\ \Eprint
  {http://arxiv.org/abs/1211.3327} {arXiv:1211.3327 [hep-ph]} \BibitemShut
  {NoStop}%
\bibitem [{\citenamefont {Lappi}(2003)}]{Lappi:2003bi}%
  \BibitemOpen
  \bibfield  {author} {\bibinfo {author} {\bibfnamefont {T.}~\bibnamefont
  {Lappi}},\ }\href {\doibase 10.1103/PhysRevC.67.054903} {\bibfield  {journal}
  {\bibinfo  {journal} {Phys. Rev. C}\ }\textbf {\bibinfo {volume} {67}},\
  \bibinfo {pages} {054903} (\bibinfo {year} {2003})},\ \Eprint
  {http://arxiv.org/abs/hep-ph/0303076} {arXiv:hep-ph/0303076} \BibitemShut
  {NoStop}%
\bibitem [{\citenamefont {Lappi}(2006)}]{Lappi:2006hq}%
  \BibitemOpen
  \bibfield  {author} {\bibinfo {author} {\bibfnamefont {T.}~\bibnamefont
  {Lappi}},\ }\href {\doibase 10.1016/j.physletb.2006.10.017} {\bibfield
  {journal} {\bibinfo  {journal} {Phys. Lett. B}\ }\textbf {\bibinfo {volume}
  {643}},\ \bibinfo {pages} {11} (\bibinfo {year} {2006})},\ \Eprint
  {http://arxiv.org/abs/hep-ph/0606207} {arXiv:hep-ph/0606207} \BibitemShut
  {NoStop}%
\bibitem [{\citenamefont {M\"uller}(2019)}]{Muller:2019bwd}%
  \BibitemOpen
  \bibfield  {author} {\bibinfo {author} {\bibfnamefont {D.}~\bibnamefont
  {M\"uller}},\ }\emph {\bibinfo {title} {{Simulations of the Glasma in
  3+1D}}},\ \href@noop {} {Ph.D. thesis},\ \bibinfo  {school} {Vienna, Tech.
  U.} (\bibinfo {year} {2019}),\ \Eprint {http://arxiv.org/abs/1904.04267}
  {arXiv:1904.04267 [hep-ph]} \BibitemShut {NoStop}%
\bibitem [{\citenamefont {Liu}\ \emph {et~al.}(2006)\citenamefont {Liu},
  \citenamefont {Rajagopal},\ and\ \citenamefont {Wiedemann}}]{Liu:2006ug}%
  \BibitemOpen
  \bibfield  {author} {\bibinfo {author} {\bibfnamefont {H.}~\bibnamefont
  {Liu}}, \bibinfo {author} {\bibfnamefont {K.}~\bibnamefont {Rajagopal}}, \
  and\ \bibinfo {author} {\bibfnamefont {U.~A.}\ \bibnamefont {Wiedemann}},\
  }\href {\doibase 10.1103/PhysRevLett.97.182301} {\bibfield  {journal}
  {\bibinfo  {journal} {Phys. Rev. Lett.}\ }\textbf {\bibinfo {volume} {97}},\
  \bibinfo {pages} {182301} (\bibinfo {year} {2006})},\ \Eprint
  {http://arxiv.org/abs/hep-ph/0605178} {arXiv:hep-ph/0605178} \BibitemShut
  {NoStop}%
\bibitem [{\citenamefont {Arnold}\ and\ \citenamefont
  {Xiao}(2008)}]{Arnold:2008vd}%
  \BibitemOpen
  \bibfield  {author} {\bibinfo {author} {\bibfnamefont {P.~B.}\ \bibnamefont
  {Arnold}}\ and\ \bibinfo {author} {\bibfnamefont {W.}~\bibnamefont {Xiao}},\
  }\href {\doibase 10.1103/PhysRevD.78.125008} {\bibfield  {journal} {\bibinfo
  {journal} {Phys. Rev. D}\ }\textbf {\bibinfo {volume} {78}},\ \bibinfo
  {pages} {125008} (\bibinfo {year} {2008})},\ \Eprint
  {http://arxiv.org/abs/0810.1026} {arXiv:0810.1026 [hep-ph]} \BibitemShut
  {NoStop}%
\bibitem [{\citenamefont {Caron-Huot}(2009)}]{Caron-Huot:2008zna}%
  \BibitemOpen
  \bibfield  {author} {\bibinfo {author} {\bibfnamefont {S.}~\bibnamefont
  {Caron-Huot}},\ }\href {\doibase 10.1103/PhysRevD.79.065039} {\bibfield
  {journal} {\bibinfo  {journal} {Phys. Rev. D}\ }\textbf {\bibinfo {volume}
  {79}},\ \bibinfo {pages} {065039} (\bibinfo {year} {2009})},\ \Eprint
  {http://arxiv.org/abs/0811.1603} {arXiv:0811.1603 [hep-ph]} \BibitemShut
  {NoStop}%
\bibitem [{\citenamefont {Majumder}\ \emph {et~al.}(2009)\citenamefont
  {Majumder}, \citenamefont {M{\"u}ller},\ and\ \citenamefont
  {Mr{\'o}wczy{\'n}ski}}]{Majumder:2009cf}%
  \BibitemOpen
  \bibfield  {author} {\bibinfo {author} {\bibfnamefont {A.}~\bibnamefont
  {Majumder}}, \bibinfo {author} {\bibfnamefont {B.}~\bibnamefont
  {M{\"u}ller}}, \ and\ \bibinfo {author} {\bibfnamefont {S.}~\bibnamefont
  {Mr{\'o}wczy{\'n}ski}},\ }\href {\doibase 10.1103/PhysRevD.80.125020}
  {\bibfield  {journal} {\bibinfo  {journal} {Phys. Rev. D}\ }\textbf {\bibinfo
  {volume} {80}},\ \bibinfo {pages} {125020} (\bibinfo {year} {2009})},\
  \Eprint {http://arxiv.org/abs/0903.3683} {arXiv:0903.3683 [hep-ph]}
  \BibitemShut {NoStop}%
\bibitem [{\citenamefont {Schenke}\ \emph {et~al.}(2009)\citenamefont
  {Schenke}, \citenamefont {Strickland}, \citenamefont {Dumitru}, \citenamefont
  {Nara},\ and\ \citenamefont {Greiner}}]{Schenke:2008gg}%
  \BibitemOpen
  \bibfield  {author} {\bibinfo {author} {\bibfnamefont {B.}~\bibnamefont
  {Schenke}}, \bibinfo {author} {\bibfnamefont {M.}~\bibnamefont {Strickland}},
  \bibinfo {author} {\bibfnamefont {A.}~\bibnamefont {Dumitru}}, \bibinfo
  {author} {\bibfnamefont {Y.}~\bibnamefont {Nara}}, \ and\ \bibinfo {author}
  {\bibfnamefont {C.}~\bibnamefont {Greiner}},\ }\href {\doibase
  10.1103/PhysRevC.79.034903} {\bibfield  {journal} {\bibinfo  {journal} {Phys.
  Rev. C}\ }\textbf {\bibinfo {volume} {79}},\ \bibinfo {pages} {034903}
  (\bibinfo {year} {2009})},\ \Eprint {http://arxiv.org/abs/0810.1314}
  {arXiv:0810.1314 [hep-ph]} \BibitemShut {NoStop}%
\bibitem [{\citenamefont {Banerjee}\ \emph {et~al.}(2012)\citenamefont
  {Banerjee}, \citenamefont {Datta}, \citenamefont {Gavai},\ and\ \citenamefont
  {Majumdar}}]{Banerjee:2011ra}%
  \BibitemOpen
  \bibfield  {author} {\bibinfo {author} {\bibfnamefont {D.}~\bibnamefont
  {Banerjee}}, \bibinfo {author} {\bibfnamefont {S.}~\bibnamefont {Datta}},
  \bibinfo {author} {\bibfnamefont {R.}~\bibnamefont {Gavai}}, \ and\ \bibinfo
  {author} {\bibfnamefont {P.}~\bibnamefont {Majumdar}},\ }\href {\doibase
  10.1103/PhysRevD.85.014510} {\bibfield  {journal} {\bibinfo  {journal} {Phys.
  Rev. D}\ }\textbf {\bibinfo {volume} {85}},\ \bibinfo {pages} {014510}
  (\bibinfo {year} {2012})},\ \Eprint {http://arxiv.org/abs/1109.5738}
  {arXiv:1109.5738 [hep-lat]} \BibitemShut {NoStop}%
\bibitem [{\citenamefont {Panero}\ \emph {et~al.}(2014)\citenamefont {Panero},
  \citenamefont {Rummukainen},\ and\ \citenamefont
  {Sch\"afer}}]{Panero:2013pla}%
  \BibitemOpen
  \bibfield  {author} {\bibinfo {author} {\bibfnamefont {M.}~\bibnamefont
  {Panero}}, \bibinfo {author} {\bibfnamefont {K.}~\bibnamefont {Rummukainen}},
  \ and\ \bibinfo {author} {\bibfnamefont {A.}~\bibnamefont {Sch\"afer}},\
  }\href {\doibase 10.1103/PhysRevLett.112.162001} {\bibfield  {journal}
  {\bibinfo  {journal} {Phys. Rev. Lett.}\ }\textbf {\bibinfo {volume} {112}},\
  \bibinfo {pages} {162001} (\bibinfo {year} {2014})},\ \Eprint
  {http://arxiv.org/abs/1307.5850} {arXiv:1307.5850 [hep-ph]} \BibitemShut
  {NoStop}%
\bibitem [{\citenamefont {Boguslavski}\ \emph {et~al.}(2018)\citenamefont
  {Boguslavski}, \citenamefont {Kurkela}, \citenamefont {Lappi},\ and\
  \citenamefont {Peuron}}]{Boguslavski:2018beu}%
  \BibitemOpen
  \bibfield  {author} {\bibinfo {author} {\bibfnamefont {K.}~\bibnamefont
  {Boguslavski}}, \bibinfo {author} {\bibfnamefont {A.}~\bibnamefont
  {Kurkela}}, \bibinfo {author} {\bibfnamefont {T.}~\bibnamefont {Lappi}}, \
  and\ \bibinfo {author} {\bibfnamefont {J.}~\bibnamefont {Peuron}},\ }\href
  {\doibase 10.1103/PhysRevD.98.014006} {\bibfield  {journal} {\bibinfo
  {journal} {Phys. Rev. D}\ }\textbf {\bibinfo {volume} {98}},\ \bibinfo
  {pages} {014006} (\bibinfo {year} {2018})},\ \Eprint
  {http://arxiv.org/abs/1804.01966} {arXiv:1804.01966 [hep-ph]} \BibitemShut
  {NoStop}%
\bibitem [{\citenamefont {Altenkort}\ \emph {et~al.}(2021)\citenamefont
  {Altenkort}, \citenamefont {Eller}, \citenamefont {Kaczmarek}, \citenamefont
  {Mazur}, \citenamefont {Moore},\ and\ \citenamefont
  {Shu}}]{Altenkort:2020fgs}%
  \BibitemOpen
  \bibfield  {author} {\bibinfo {author} {\bibfnamefont {L.}~\bibnamefont
  {Altenkort}}, \bibinfo {author} {\bibfnamefont {A.~M.}\ \bibnamefont
  {Eller}}, \bibinfo {author} {\bibfnamefont {O.}~\bibnamefont {Kaczmarek}},
  \bibinfo {author} {\bibfnamefont {L.}~\bibnamefont {Mazur}}, \bibinfo
  {author} {\bibfnamefont {G.~D.}\ \bibnamefont {Moore}}, \ and\ \bibinfo
  {author} {\bibfnamefont {H.-T.}\ \bibnamefont {Shu}},\ }\href {\doibase
  10.1103/PhysRevD.103.014511} {\bibfield  {journal} {\bibinfo  {journal}
  {Phys. Rev. D}\ }\textbf {\bibinfo {volume} {103}},\ \bibinfo {pages}
  {014511} (\bibinfo {year} {2021})},\ \Eprint
  {http://arxiv.org/abs/2009.13553} {arXiv:2009.13553 [hep-lat]} \BibitemShut
  {NoStop}%
\bibitem [{\citenamefont {Litim}\ and\ \citenamefont
  {Manuel}(1999{\natexlab{a}})}]{Litim:1999id}%
  \BibitemOpen
  \bibfield  {author} {\bibinfo {author} {\bibfnamefont {D.~F.}\ \bibnamefont
  {Litim}}\ and\ \bibinfo {author} {\bibfnamefont {C.}~\bibnamefont {Manuel}},\
  }\href {\doibase 10.1016/S0550-3213(99)00531-3} {\bibfield  {journal}
  {\bibinfo  {journal} {Nucl. Phys. B}\ }\textbf {\bibinfo {volume} {562}},\
  \bibinfo {pages} {237} (\bibinfo {year} {1999}{\natexlab{a}})},\ \Eprint
  {http://arxiv.org/abs/hep-ph/9906210} {arXiv:hep-ph/9906210} \BibitemShut
  {NoStop}%
\bibitem [{\citenamefont {Litim}\ and\ \citenamefont
  {Manuel}(1999{\natexlab{b}})}]{Litim:1999ns}%
  \BibitemOpen
  \bibfield  {author} {\bibinfo {author} {\bibfnamefont {D.~F.}\ \bibnamefont
  {Litim}}\ and\ \bibinfo {author} {\bibfnamefont {C.}~\bibnamefont {Manuel}},\
  }\href {\doibase 10.1103/PhysRevLett.82.4981} {\bibfield  {journal} {\bibinfo
   {journal} {Phys. Rev. Lett.}\ }\textbf {\bibinfo {volume} {82}},\ \bibinfo
  {pages} {4981} (\bibinfo {year} {1999}{\natexlab{b}})},\ \Eprint
  {http://arxiv.org/abs/hep-ph/9902430} {arXiv:hep-ph/9902430} \BibitemShut
  {NoStop}%
\bibitem [{\citenamefont {Litim}\ and\ \citenamefont
  {Manuel}(2002)}]{Litim:2001db}%
  \BibitemOpen
  \bibfield  {author} {\bibinfo {author} {\bibfnamefont {D.~F.}\ \bibnamefont
  {Litim}}\ and\ \bibinfo {author} {\bibfnamefont {C.}~\bibnamefont {Manuel}},\
  }\href {\doibase 10.1016/S0370-1573(02)00015-7} {\bibfield  {journal}
  {\bibinfo  {journal} {Phys. Rept.}\ }\textbf {\bibinfo {volume} {364}},\
  \bibinfo {pages} {451} (\bibinfo {year} {2002})},\ \Eprint
  {http://arxiv.org/abs/hep-ph/0110104} {arXiv:hep-ph/0110104} \BibitemShut
  {NoStop}%
\bibitem [{\citenamefont {Ipp}\ \emph {et~al.}(2020{\natexlab{a}})\citenamefont
  {Ipp}, \citenamefont {M\"uller},\ and\ \citenamefont {Schuh}}]{Ipp:2020mjc}%
  \BibitemOpen
  \bibfield  {author} {\bibinfo {author} {\bibfnamefont {A.}~\bibnamefont
  {Ipp}}, \bibinfo {author} {\bibfnamefont {D.~I.}\ \bibnamefont {M\"uller}}, \
  and\ \bibinfo {author} {\bibfnamefont {D.}~\bibnamefont {Schuh}},\ }\href
  {\doibase 10.1103/PhysRevD.102.074001} {\bibfield  {journal} {\bibinfo
  {journal} {Phys. Rev. D}\ }\textbf {\bibinfo {volume} {102}},\ \bibinfo
  {pages} {074001} (\bibinfo {year} {2020}{\natexlab{a}})},\ \Eprint
  {http://arxiv.org/abs/2001.10001} {arXiv:2001.10001 [hep-ph]} \BibitemShut
  {NoStop}%
\bibitem [{\citenamefont {Ipp}\ \emph {et~al.}(2020{\natexlab{b}})\citenamefont
  {Ipp}, \citenamefont {M\"uller},\ and\ \citenamefont {Schuh}}]{Ipp:2020nfu}%
  \BibitemOpen
  \bibfield  {author} {\bibinfo {author} {\bibfnamefont {A.}~\bibnamefont
  {Ipp}}, \bibinfo {author} {\bibfnamefont {D.~I.}\ \bibnamefont {M\"uller}}, \
  and\ \bibinfo {author} {\bibfnamefont {D.}~\bibnamefont {Schuh}},\ }\href
  {\doibase 10.1016/j.physletb.2020.135810} {\bibfield  {journal} {\bibinfo
  {journal} {Phys. Lett. B}\ }\textbf {\bibinfo {volume} {810}},\ \bibinfo
  {pages} {135810} (\bibinfo {year} {2020}{\natexlab{b}})},\ \Eprint
  {http://arxiv.org/abs/2009.14206} {arXiv:2009.14206 [hep-ph]} \BibitemShut
  {NoStop}%
\bibitem [{\citenamefont {Boguslavski}\ \emph {et~al.}(2020)\citenamefont
  {Boguslavski}, \citenamefont {Kurkela}, \citenamefont {Lappi},\ and\
  \citenamefont {Peuron}}]{Boguslavski:2020tqz}%
  \BibitemOpen
  \bibfield  {author} {\bibinfo {author} {\bibfnamefont {K.}~\bibnamefont
  {Boguslavski}}, \bibinfo {author} {\bibfnamefont {A.}~\bibnamefont
  {Kurkela}}, \bibinfo {author} {\bibfnamefont {T.}~\bibnamefont {Lappi}}, \
  and\ \bibinfo {author} {\bibfnamefont {J.}~\bibnamefont {Peuron}},\ }\href
  {\doibase 10.1007/JHEP09(2020)077} {\bibfield  {journal} {\bibinfo  {journal}
  {JHEP}\ }\textbf {\bibinfo {volume} {09}},\ \bibinfo {pages} {077} (\bibinfo
  {year} {2020})},\ \Eprint {http://arxiv.org/abs/2005.02418} {arXiv:2005.02418
  [hep-ph]} \BibitemShut {NoStop}%
\bibitem [{\citenamefont {Boguslavski}\ \emph {et~al.}(2021)\citenamefont
  {Boguslavski}, \citenamefont {Kurkela}, \citenamefont {Lappi},\ and\
  \citenamefont {Peuron}}]{Boguslavski:2020mzh}%
  \BibitemOpen
  \bibfield  {author} {\bibinfo {author} {\bibfnamefont {K.}~\bibnamefont
  {Boguslavski}}, \bibinfo {author} {\bibfnamefont {A.}~\bibnamefont
  {Kurkela}}, \bibinfo {author} {\bibfnamefont {T.}~\bibnamefont {Lappi}}, \
  and\ \bibinfo {author} {\bibfnamefont {J.}~\bibnamefont {Peuron}},\ }\href
  {\doibase 10.1016/j.nuclphysa.2020.121970} {\bibfield  {journal} {\bibinfo
  {journal} {Nucl. Phys. A}\ }\textbf {\bibinfo {volume} {1005}},\ \bibinfo
  {pages} {121970} (\bibinfo {year} {2021})},\ \Eprint
  {http://arxiv.org/abs/2001.11863} {arXiv:2001.11863 [hep-ph]} \BibitemShut
  {NoStop}%
\bibitem [{\citenamefont {Das}\ \emph {et~al.}(2015)\citenamefont {Das},
  \citenamefont {Ruggieri}, \citenamefont {Mazumder}, \citenamefont {Greco},\
  and\ \citenamefont {Alam}}]{Das:2015aga}%
  \BibitemOpen
  \bibfield  {author} {\bibinfo {author} {\bibfnamefont {S.~K.}\ \bibnamefont
  {Das}}, \bibinfo {author} {\bibfnamefont {M.}~\bibnamefont {Ruggieri}},
  \bibinfo {author} {\bibfnamefont {S.}~\bibnamefont {Mazumder}}, \bibinfo
  {author} {\bibfnamefont {V.}~\bibnamefont {Greco}}, \ and\ \bibinfo {author}
  {\bibfnamefont {J.}~\bibnamefont {Alam}},\ }\href {\doibase
  10.1088/0954-3899/42/9/095108} {\bibfield  {journal} {\bibinfo  {journal} {J.
  Phys. G}\ }\textbf {\bibinfo {volume} {42}},\ \bibinfo {pages} {095108}
  (\bibinfo {year} {2015})},\ \Eprint {http://arxiv.org/abs/1501.07521}
  {arXiv:1501.07521 [nucl-th]} \BibitemShut {NoStop}%
\bibitem [{\citenamefont {Das}\ \emph {et~al.}(2017)\citenamefont {Das},
  \citenamefont {Ruggieri}, \citenamefont {Scardina}, \citenamefont {Plumari},\
  and\ \citenamefont {Greco}}]{Das:2017dsh}%
  \BibitemOpen
  \bibfield  {author} {\bibinfo {author} {\bibfnamefont {S.~K.}\ \bibnamefont
  {Das}}, \bibinfo {author} {\bibfnamefont {M.}~\bibnamefont {Ruggieri}},
  \bibinfo {author} {\bibfnamefont {F.}~\bibnamefont {Scardina}}, \bibinfo
  {author} {\bibfnamefont {S.}~\bibnamefont {Plumari}}, \ and\ \bibinfo
  {author} {\bibfnamefont {V.}~\bibnamefont {Greco}},\ }\href {\doibase
  10.1088/1361-6471/aa815a} {\bibfield  {journal} {\bibinfo  {journal} {J.
  Phys. G}\ }\textbf {\bibinfo {volume} {44}},\ \bibinfo {pages} {095102}
  (\bibinfo {year} {2017})},\ \Eprint {http://arxiv.org/abs/1701.05123}
  {arXiv:1701.05123 [nucl-th]} \BibitemShut {NoStop}%
\bibitem [{\citenamefont {Ruggieri}\ and\ \citenamefont
  {Das}(2018)}]{Ruggieri:2018rzi}%
  \BibitemOpen
  \bibfield  {author} {\bibinfo {author} {\bibfnamefont {M.}~\bibnamefont
  {Ruggieri}}\ and\ \bibinfo {author} {\bibfnamefont {S.~K.}\ \bibnamefont
  {Das}},\ }\href {\doibase 10.1103/PhysRevD.98.094024} {\bibfield  {journal}
  {\bibinfo  {journal} {Phys. Rev. D}\ }\textbf {\bibinfo {volume} {98}},\
  \bibinfo {pages} {094024} (\bibinfo {year} {2018})},\ \Eprint
  {http://arxiv.org/abs/1805.09617} {arXiv:1805.09617 [nucl-th]} \BibitemShut
  {NoStop}%
\bibitem [{\citenamefont {Sun}\ \emph {et~al.}(2019)\citenamefont {Sun},
  \citenamefont {Coci}, \citenamefont {Das}, \citenamefont {Plumari},
  \citenamefont {Ruggieri},\ and\ \citenamefont {Greco}}]{Sun:2019fud}%
  \BibitemOpen
  \bibfield  {author} {\bibinfo {author} {\bibfnamefont {Y.}~\bibnamefont
  {Sun}}, \bibinfo {author} {\bibfnamefont {G.}~\bibnamefont {Coci}}, \bibinfo
  {author} {\bibfnamefont {S.~K.}\ \bibnamefont {Das}}, \bibinfo {author}
  {\bibfnamefont {S.}~\bibnamefont {Plumari}}, \bibinfo {author} {\bibfnamefont
  {M.}~\bibnamefont {Ruggieri}}, \ and\ \bibinfo {author} {\bibfnamefont
  {V.}~\bibnamefont {Greco}},\ }\href {\doibase 10.1016/j.physletb.2019.134933}
  {\bibfield  {journal} {\bibinfo  {journal} {Phys. Lett. B}\ }\textbf
  {\bibinfo {volume} {798}},\ \bibinfo {pages} {134933} (\bibinfo {year}
  {2019})},\ \Eprint {http://arxiv.org/abs/1902.06254} {arXiv:1902.06254
  [nucl-th]} \BibitemShut {NoStop}%
\bibitem [{\citenamefont {Liu}\ \emph {et~al.}(2020)\citenamefont {Liu},
  \citenamefont {Plumari}, \citenamefont {Das}, \citenamefont {Greco},\ and\
  \citenamefont {Ruggieri}}]{Liu:2019lac}%
  \BibitemOpen
  \bibfield  {author} {\bibinfo {author} {\bibfnamefont {J.~H.}\ \bibnamefont
  {Liu}}, \bibinfo {author} {\bibfnamefont {S.}~\bibnamefont {Plumari}},
  \bibinfo {author} {\bibfnamefont {S.~K.}\ \bibnamefont {Das}}, \bibinfo
  {author} {\bibfnamefont {V.}~\bibnamefont {Greco}}, \ and\ \bibinfo {author}
  {\bibfnamefont {M.}~\bibnamefont {Ruggieri}},\ }\href {\doibase
  10.1103/PhysRevC.102.044902} {\bibfield  {journal} {\bibinfo  {journal}
  {Phys. Rev. C}\ }\textbf {\bibinfo {volume} {102}},\ \bibinfo {pages}
  {044902} (\bibinfo {year} {2020})},\ \Eprint
  {http://arxiv.org/abs/1911.02480} {arXiv:1911.02480 [nucl-th]} \BibitemShut
  {NoStop}%
\bibitem [{\citenamefont {Liu}\ \emph {et~al.}(2021)\citenamefont {Liu},
  \citenamefont {Das}, \citenamefont {Greco},\ and\ \citenamefont
  {Ruggieri}}]{Liu:2020cpj}%
  \BibitemOpen
  \bibfield  {author} {\bibinfo {author} {\bibfnamefont {J.-H.}\ \bibnamefont
  {Liu}}, \bibinfo {author} {\bibfnamefont {S.~K.}\ \bibnamefont {Das}},
  \bibinfo {author} {\bibfnamefont {V.}~\bibnamefont {Greco}}, \ and\ \bibinfo
  {author} {\bibfnamefont {M.}~\bibnamefont {Ruggieri}},\ }\href {\doibase
  10.1103/PhysRevD.103.034029} {\bibfield  {journal} {\bibinfo  {journal}
  {Phys. Rev. D}\ }\textbf {\bibinfo {volume} {103}},\ \bibinfo {pages}
  {034029} (\bibinfo {year} {2021})},\ \Eprint
  {http://arxiv.org/abs/2011.05818} {arXiv:2011.05818 [hep-ph]} \BibitemShut
  {NoStop}%
\bibitem [{\citenamefont {Khowal}\ \emph {et~al.}(2022)\citenamefont {Khowal},
  \citenamefont {Das}, \citenamefont {Oliva},\ and\ \citenamefont
  {Ruggieri}}]{Khowal:2021zoo}%
  \BibitemOpen
  \bibfield  {author} {\bibinfo {author} {\bibfnamefont {P.}~\bibnamefont
  {Khowal}}, \bibinfo {author} {\bibfnamefont {S.~K.}\ \bibnamefont {Das}},
  \bibinfo {author} {\bibfnamefont {L.}~\bibnamefont {Oliva}}, \ and\ \bibinfo
  {author} {\bibfnamefont {M.}~\bibnamefont {Ruggieri}},\ }\href {\doibase
  10.1140/epjp/s13360-022-02517-w} {\bibfield  {journal} {\bibinfo  {journal}
  {Eur. Phys. J. Plus}\ }\textbf {\bibinfo {volume} {137}},\ \bibinfo {pages}
  {307} (\bibinfo {year} {2022})},\ \Eprint {http://arxiv.org/abs/2110.14610}
  {arXiv:2110.14610 [hep-ph]} \BibitemShut {NoStop}%
\bibitem [{\citenamefont {Ruggieri}\ \emph {et~al.}(2022)\citenamefont
  {Ruggieri}, \citenamefont {Pooja}, \citenamefont {Prakash},\ and\
  \citenamefont {Das}}]{Ruggieri:2022kxv}%
  \BibitemOpen
  \bibfield  {author} {\bibinfo {author} {\bibfnamefont {M.}~\bibnamefont
  {Ruggieri}}, \bibinfo {author} {\bibnamefont {Pooja}}, \bibinfo {author}
  {\bibfnamefont {J.}~\bibnamefont {Prakash}}, \ and\ \bibinfo {author}
  {\bibfnamefont {S.~K.}\ \bibnamefont {Das}},\ }\href {\doibase
  10.1103/PhysRevD.106.034032} {\bibfield  {journal} {\bibinfo  {journal}
  {Phys. Rev. D}\ }\textbf {\bibinfo {volume} {106}},\ \bibinfo {pages}
  {034032} (\bibinfo {year} {2022})},\ \Eprint
  {http://arxiv.org/abs/2203.06712} {arXiv:2203.06712 [hep-ph]} \BibitemShut
  {NoStop}%
\bibitem [{\citenamefont {Carrington}\ \emph {et~al.}(2020)\citenamefont
  {Carrington}, \citenamefont {Czajka},\ and\ \citenamefont
  {Mr{\'o}wczy{\'n}ski}}]{Carrington:2020sww}%
  \BibitemOpen
  \bibfield  {author} {\bibinfo {author} {\bibfnamefont {M.~E.}\ \bibnamefont
  {Carrington}}, \bibinfo {author} {\bibfnamefont {A.}~\bibnamefont {Czajka}},
  \ and\ \bibinfo {author} {\bibfnamefont {S.}~\bibnamefont
  {Mr{\'o}wczy{\'n}ski}},\ }\href {\doibase 10.1016/j.nuclphysa.2020.121914}
  {\bibfield  {journal} {\bibinfo  {journal} {Nucl. Phys. A}\ }\textbf
  {\bibinfo {volume} {1001}},\ \bibinfo {pages} {121914} (\bibinfo {year}
  {2020})},\ \Eprint {http://arxiv.org/abs/2001.05074} {arXiv:2001.05074
  [nucl-th]} \BibitemShut {NoStop}%
\bibitem [{\citenamefont {Carrington}\ \emph
  {et~al.}(2022{\natexlab{a}})\citenamefont {Carrington}, \citenamefont
  {Czajka},\ and\ \citenamefont {Mr{\'o}wczy{\'n}ski}}]{Carrington:2021dvw}%
  \BibitemOpen
  \bibfield  {author} {\bibinfo {author} {\bibfnamefont {M.~E.}\ \bibnamefont
  {Carrington}}, \bibinfo {author} {\bibfnamefont {A.}~\bibnamefont {Czajka}},
  \ and\ \bibinfo {author} {\bibfnamefont {S.}~\bibnamefont
  {Mr{\'o}wczy{\'n}ski}},\ }\href {\doibase 10.1016/j.physletb.2022.137464}
  {\bibfield  {journal} {\bibinfo  {journal} {Phys. Lett. B}\ }\textbf
  {\bibinfo {volume} {834}},\ \bibinfo {pages} {137464} (\bibinfo {year}
  {2022}{\natexlab{a}})},\ \Eprint {http://arxiv.org/abs/2112.06812}
  {arXiv:2112.06812 [hep-ph]} \BibitemShut {NoStop}%
\bibitem [{\citenamefont {Carrington}\ \emph
  {et~al.}(2022{\natexlab{b}})\citenamefont {Carrington}, \citenamefont
  {Czajka},\ and\ \citenamefont {Mr{\'o}wczy{\'n}ski}}]{Carrington:2022bnv}%
  \BibitemOpen
  \bibfield  {author} {\bibinfo {author} {\bibfnamefont {M.~E.}\ \bibnamefont
  {Carrington}}, \bibinfo {author} {\bibfnamefont {A.}~\bibnamefont {Czajka}},
  \ and\ \bibinfo {author} {\bibfnamefont {S.}~\bibnamefont
  {Mr{\'o}wczy{\'n}ski}},\ }\href {\doibase 10.1103/PhysRevC.105.064910}
  {\bibfield  {journal} {\bibinfo  {journal} {Phys. Rev. C}\ }\textbf {\bibinfo
  {volume} {105}},\ \bibinfo {pages} {064910} (\bibinfo {year}
  {2022}{\natexlab{b}})},\ \Eprint {http://arxiv.org/abs/2202.00357}
  {arXiv:2202.00357 [nucl-th]} \BibitemShut {NoStop}%
\bibitem [{\citenamefont {Andres}\ \emph {et~al.}(2020)\citenamefont {Andres},
  \citenamefont {Armesto}, \citenamefont {Niemi}, \citenamefont {Paatelainen},\
  and\ \citenamefont {Salgado}}]{Andres:2019eus}%
  \BibitemOpen
  \bibfield  {author} {\bibinfo {author} {\bibfnamefont {C.}~\bibnamefont
  {Andres}}, \bibinfo {author} {\bibfnamefont {N.}~\bibnamefont {Armesto}},
  \bibinfo {author} {\bibfnamefont {H.}~\bibnamefont {Niemi}}, \bibinfo
  {author} {\bibfnamefont {R.}~\bibnamefont {Paatelainen}}, \ and\ \bibinfo
  {author} {\bibfnamefont {C.~A.}\ \bibnamefont {Salgado}},\ }\href {\doibase
  10.1016/j.physletb.2020.135318} {\bibfield  {journal} {\bibinfo  {journal}
  {Phys. Lett. B}\ }\textbf {\bibinfo {volume} {803}},\ \bibinfo {pages}
  {135318} (\bibinfo {year} {2020})},\ \Eprint
  {http://arxiv.org/abs/1902.03231} {arXiv:1902.03231 [hep-ph]} \BibitemShut
  {NoStop}%
\bibitem [{\citenamefont {Andres}\ \emph {et~al.}(2023)\citenamefont {Andres},
  \citenamefont {Apolin\'ario}, \citenamefont {Dominguez}, \citenamefont
  {Martinez},\ and\ \citenamefont {Salgado}}]{Andres:2022bql}%
  \BibitemOpen
  \bibfield  {author} {\bibinfo {author} {\bibfnamefont {C.}~\bibnamefont
  {Andres}}, \bibinfo {author} {\bibfnamefont {L.}~\bibnamefont
  {Apolin\'ario}}, \bibinfo {author} {\bibfnamefont {F.}~\bibnamefont
  {Dominguez}}, \bibinfo {author} {\bibfnamefont {M.~G.}\ \bibnamefont
  {Martinez}}, \ and\ \bibinfo {author} {\bibfnamefont {C.~A.}\ \bibnamefont
  {Salgado}},\ }\href {\doibase 10.1007/JHEP03(2023)189} {\bibfield  {journal}
  {\bibinfo  {journal} {JHEP}\ }\textbf {\bibinfo {volume} {03}},\ \bibinfo
  {pages} {189} (\bibinfo {year} {2023})},\ \Eprint
  {http://arxiv.org/abs/2211.10161} {arXiv:2211.10161 [hep-ph]} \BibitemShut
  {NoStop}%
\bibitem [{\citenamefont {Avramescu}\ \emph {et~al.}(2022)\citenamefont
  {Avramescu}, \citenamefont {B\u{a}ran}, \citenamefont {Greco}, \citenamefont
  {Ipp}, \citenamefont {M\"uller},\ and\ \citenamefont
  {Ruggieri}}]{Avramescu:2022vkd}%
  \BibitemOpen
  \bibfield  {author} {\bibinfo {author} {\bibfnamefont {D.}~\bibnamefont
  {Avramescu}}, \bibinfo {author} {\bibfnamefont {V.}~\bibnamefont
  {B\u{a}ran}}, \bibinfo {author} {\bibfnamefont {V.}~\bibnamefont {Greco}},
  \bibinfo {author} {\bibfnamefont {A.}~\bibnamefont {Ipp}}, \bibinfo {author}
  {\bibfnamefont {D.~I.}\ \bibnamefont {M\"uller}}, \ and\ \bibinfo {author}
  {\bibfnamefont {M.}~\bibnamefont {Ruggieri}},\ }in\ \href@noop {} {\emph
  {\bibinfo {booktitle} {{29th International Conference on Ultra-relativistic
  Nucleus-Nucleus Collisions}}}}\ (\bibinfo {year} {2022})\ \Eprint
  {http://arxiv.org/abs/2208.04781} {arXiv:2208.04781 [hep-ph]} \BibitemShut
  {NoStop}%
\bibitem [{\citenamefont {McLerran}\ and\ \citenamefont
  {Venugopalan}(1994{\natexlab{a}})}]{McLerran:1993ni}%
  \BibitemOpen
  \bibfield  {author} {\bibinfo {author} {\bibfnamefont {L.~D.}\ \bibnamefont
  {McLerran}}\ and\ \bibinfo {author} {\bibfnamefont {R.}~\bibnamefont
  {Venugopalan}},\ }\href {\doibase 10.1103/PhysRevD.49.2233} {\bibfield
  {journal} {\bibinfo  {journal} {Phys. Rev. D}\ }\textbf {\bibinfo {volume}
  {49}},\ \bibinfo {pages} {2233} (\bibinfo {year} {1994}{\natexlab{a}})},\
  \Eprint {http://arxiv.org/abs/hep-ph/9309289} {arXiv:hep-ph/9309289}
  \BibitemShut {NoStop}%
\bibitem [{\citenamefont {McLerran}\ and\ \citenamefont
  {Venugopalan}(1994{\natexlab{b}})}]{McLerran:1993ka}%
  \BibitemOpen
  \bibfield  {author} {\bibinfo {author} {\bibfnamefont {L.~D.}\ \bibnamefont
  {McLerran}}\ and\ \bibinfo {author} {\bibfnamefont {R.}~\bibnamefont
  {Venugopalan}},\ }\href {\doibase 10.1103/PhysRevD.49.3352} {\bibfield
  {journal} {\bibinfo  {journal} {Phys. Rev. D}\ }\textbf {\bibinfo {volume}
  {49}},\ \bibinfo {pages} {3352} (\bibinfo {year} {1994}{\natexlab{b}})},\
  \Eprint {http://arxiv.org/abs/hep-ph/9311205} {arXiv:hep-ph/9311205}
  \BibitemShut {NoStop}%
\bibitem [{\citenamefont {McLerran}\ and\ \citenamefont
  {Venugopalan}(1994{\natexlab{c}})}]{McLerran:1994vd}%
  \BibitemOpen
  \bibfield  {author} {\bibinfo {author} {\bibfnamefont {L.~D.}\ \bibnamefont
  {McLerran}}\ and\ \bibinfo {author} {\bibfnamefont {R.}~\bibnamefont
  {Venugopalan}},\ }\href {\doibase 10.1103/PhysRevD.50.2225} {\bibfield
  {journal} {\bibinfo  {journal} {Phys. Rev. D}\ }\textbf {\bibinfo {volume}
  {50}},\ \bibinfo {pages} {2225} (\bibinfo {year} {1994}{\natexlab{c}})},\
  \Eprint {http://arxiv.org/abs/hep-ph/9402335} {arXiv:hep-ph/9402335}
  \BibitemShut {NoStop}%
\bibitem [{\citenamefont {Kovner}\ \emph {et~al.}(1995)\citenamefont {Kovner},
  \citenamefont {McLerran},\ and\ \citenamefont {Weigert}}]{Kovner:1995ja}%
  \BibitemOpen
  \bibfield  {author} {\bibinfo {author} {\bibfnamefont {A.}~\bibnamefont
  {Kovner}}, \bibinfo {author} {\bibfnamefont {L.~D.}\ \bibnamefont
  {McLerran}}, \ and\ \bibinfo {author} {\bibfnamefont {H.}~\bibnamefont
  {Weigert}},\ }\href {\doibase 10.1103/PhysRevD.52.6231} {\bibfield  {journal}
  {\bibinfo  {journal} {Phys. Rev. D}\ }\textbf {\bibinfo {volume} {52}},\
  \bibinfo {pages} {6231} (\bibinfo {year} {1995})},\ \Eprint
  {http://arxiv.org/abs/hep-ph/9502289} {arXiv:hep-ph/9502289} \BibitemShut
  {NoStop}%
\bibitem [{\citenamefont {Krasnitz}\ and\ \citenamefont
  {Venugopalan}(1999)}]{Krasnitz:1998ns}%
  \BibitemOpen
  \bibfield  {author} {\bibinfo {author} {\bibfnamefont {A.}~\bibnamefont
  {Krasnitz}}\ and\ \bibinfo {author} {\bibfnamefont {R.}~\bibnamefont
  {Venugopalan}},\ }\href {\doibase 10.1016/S0550-3213(99)00366-1} {\bibfield
  {journal} {\bibinfo  {journal} {Nucl. Phys. B}\ }\textbf {\bibinfo {volume}
  {557}},\ \bibinfo {pages} {237} (\bibinfo {year} {1999})},\ \Eprint
  {http://arxiv.org/abs/hep-ph/9809433} {arXiv:hep-ph/9809433} \BibitemShut
  {NoStop}%
\bibitem [{\citenamefont {Fukushima}(2008)}]{Fukushima:2007ki}%
  \BibitemOpen
  \bibfield  {author} {\bibinfo {author} {\bibfnamefont {K.}~\bibnamefont
  {Fukushima}},\ }\href {\doibase 10.1103/PhysRevD.77.074005} {\bibfield
  {journal} {\bibinfo  {journal} {Phys. Rev. D}\ }\textbf {\bibinfo {volume}
  {77}},\ \bibinfo {pages} {074005} (\bibinfo {year} {2008})},\ \Eprint
  {http://arxiv.org/abs/0711.2364} {arXiv:0711.2364 [hep-ph]} \BibitemShut
  {NoStop}%
\bibitem [{\citenamefont {Wong}(1970)}]{Wong:1970fu}%
  \BibitemOpen
  \bibfield  {author} {\bibinfo {author} {\bibfnamefont {S.~K.}\ \bibnamefont
  {Wong}},\ }\href {\doibase 10.1007/BF02892134} {\bibfield  {journal}
  {\bibinfo  {journal} {Nuovo Cim. A}\ }\textbf {\bibinfo {volume} {65}},\
  \bibinfo {pages} {689} (\bibinfo {year} {1970})}\BibitemShut {NoStop}%
\bibitem [{\citenamefont {Boozer}(2011)}]{Boozer_2011}%
  \BibitemOpen
  \bibfield  {author} {\bibinfo {author} {\bibfnamefont {A.~D.}\ \bibnamefont
  {Boozer}},\ }\href {\doibase 10.1119/1.3606478} {\bibfield  {journal}
  {\bibinfo  {journal} {American Journal of Physics}\ }\textbf {\bibinfo
  {volume} {79}},\ \bibinfo {pages} {925} (\bibinfo {year} {2011})}\BibitemShut
  {NoStop}%
\bibitem [{\citenamefont {Hu}\ and\ \citenamefont
  {M{\"u}ller}(1997)}]{Hu:1996sf}%
  \BibitemOpen
  \bibfield  {author} {\bibinfo {author} {\bibfnamefont {C.~R.}\ \bibnamefont
  {Hu}}\ and\ \bibinfo {author} {\bibfnamefont {B.}~\bibnamefont
  {M{\"u}ller}},\ }\href {\doibase 10.1016/S0370-2693(97)00851-4} {\bibfield
  {journal} {\bibinfo  {journal} {Phys. Lett. B}\ }\textbf {\bibinfo {volume}
  {409}},\ \bibinfo {pages} {377} (\bibinfo {year} {1997})},\ \Eprint
  {http://arxiv.org/abs/hep-ph/9611292} {arXiv:hep-ph/9611292} \BibitemShut
  {NoStop}%
\bibitem [{\citenamefont {Moore}\ \emph {et~al.}(1998)\citenamefont {Moore},
  \citenamefont {Hu},\ and\ \citenamefont {M{\"u}ller}}]{Moore:1997sn}%
  \BibitemOpen
  \bibfield  {author} {\bibinfo {author} {\bibfnamefont {G.~D.}\ \bibnamefont
  {Moore}}, \bibinfo {author} {\bibfnamefont {C.~R.}\ \bibnamefont {Hu}}, \
  and\ \bibinfo {author} {\bibfnamefont {B.}~\bibnamefont {M{\"u}ller}},\
  }\href {\doibase 10.1103/PhysRevD.58.045001} {\bibfield  {journal} {\bibinfo
  {journal} {Phys. Rev. D}\ }\textbf {\bibinfo {volume} {58}},\ \bibinfo
  {pages} {045001} (\bibinfo {year} {1998})},\ \Eprint
  {http://arxiv.org/abs/hep-ph/9710436} {arXiv:hep-ph/9710436} \BibitemShut
  {NoStop}%
\bibitem [{\citenamefont {Dumitru}\ \emph {et~al.}(2007)\citenamefont
  {Dumitru}, \citenamefont {Nara},\ and\ \citenamefont
  {Strickland}}]{Dumitru:2006pz}%
  \BibitemOpen
  \bibfield  {author} {\bibinfo {author} {\bibfnamefont {A.}~\bibnamefont
  {Dumitru}}, \bibinfo {author} {\bibfnamefont {Y.}~\bibnamefont {Nara}}, \
  and\ \bibinfo {author} {\bibfnamefont {M.}~\bibnamefont {Strickland}},\
  }\href {\doibase 10.1103/PhysRevD.75.025016} {\bibfield  {journal} {\bibinfo
  {journal} {Phys. Rev. D}\ }\textbf {\bibinfo {volume} {75}},\ \bibinfo
  {pages} {025016} (\bibinfo {year} {2007})},\ \Eprint
  {http://arxiv.org/abs/hep-ph/0604149} {arXiv:hep-ph/0604149} \BibitemShut
  {NoStop}%
\bibitem [{\citenamefont {Schenke}(2008)}]{Schenke:2008hw}%
  \BibitemOpen
  \bibfield  {author} {\bibinfo {author} {\bibfnamefont {B.}~\bibnamefont
  {Schenke}},\ }\emph {\bibinfo {title} {{Collective Phenomena in the
  Non-Equilibrium Quark-Gluon Plasma}}},\ \href@noop {} {Ph.D. thesis},\
  \bibinfo  {school} {Frankfurt U.} (\bibinfo {year} {2008}),\ \Eprint
  {http://arxiv.org/abs/0810.4306} {arXiv:0810.4306 [hep-ph]} \BibitemShut
  {NoStop}%
\bibitem [{\citenamefont {Kelly}\ \emph {et~al.}(1994)\citenamefont {Kelly},
  \citenamefont {Liu}, \citenamefont {Lucchesi},\ and\ \citenamefont
  {Manuel}}]{Kelly:1994dh}%
  \BibitemOpen
  \bibfield  {author} {\bibinfo {author} {\bibfnamefont {P.~F.}\ \bibnamefont
  {Kelly}}, \bibinfo {author} {\bibfnamefont {Q.}~\bibnamefont {Liu}}, \bibinfo
  {author} {\bibfnamefont {C.}~\bibnamefont {Lucchesi}}, \ and\ \bibinfo
  {author} {\bibfnamefont {C.}~\bibnamefont {Manuel}},\ }\href {\doibase
  10.1103/PhysRevD.50.4209} {\bibfield  {journal} {\bibinfo  {journal} {Phys.
  Rev. D}\ }\textbf {\bibinfo {volume} {50}},\ \bibinfo {pages} {4209}
  (\bibinfo {year} {1994})},\ \Eprint {http://arxiv.org/abs/hep-ph/9406285}
  {arXiv:hep-ph/9406285} \BibitemShut {NoStop}%
\bibitem [{\citenamefont {Johnson}(1989)}]{Johnson:1988qm}%
  \BibitemOpen
  \bibfield  {author} {\bibinfo {author} {\bibfnamefont {K.}~\bibnamefont
  {Johnson}},\ }\href {\doibase 10.1016/0003-4916(89)90120-6} {\bibfield
  {journal} {\bibinfo  {journal} {Annals Phys.}\ }\textbf {\bibinfo {volume}
  {192}},\ \bibinfo {pages} {104} (\bibinfo {year} {1989})}\BibitemShut
  {NoStop}%
\bibitem [{\citenamefont {Bulgac}\ and\ \citenamefont
  {Kusnezov}(1990)}]{Bulgac_1990}%
  \BibitemOpen
  \bibfield  {author} {\bibinfo {author} {\bibfnamefont {A.}~\bibnamefont
  {Bulgac}}\ and\ \bibinfo {author} {\bibfnamefont {D.}~\bibnamefont
  {Kusnezov}},\ }\href {\doibase https://doi.org/10.1016/0003-4916(90)90373-V}
  {\bibfield  {journal} {\bibinfo  {journal} {Annals of Physics}\ }\textbf
  {\bibinfo {volume} {199}},\ \bibinfo {pages} {187} (\bibinfo {year}
  {1990})}\BibitemShut {NoStop}%
\bibitem [{\citenamefont {Brambilla}\ \emph {et~al.}(2020)\citenamefont
  {Brambilla}, \citenamefont {Leino}, \citenamefont {Petreczky},\ and\
  \citenamefont {Vairo}}]{Brambilla:2020siz}%
  \BibitemOpen
  \bibfield  {author} {\bibinfo {author} {\bibfnamefont {N.}~\bibnamefont
  {Brambilla}}, \bibinfo {author} {\bibfnamefont {V.}~\bibnamefont {Leino}},
  \bibinfo {author} {\bibfnamefont {P.}~\bibnamefont {Petreczky}}, \ and\
  \bibinfo {author} {\bibfnamefont {A.}~\bibnamefont {Vairo}},\ }\href
  {\doibase 10.1103/PhysRevD.102.074503} {\bibfield  {journal} {\bibinfo
  {journal} {Phys. Rev. D}\ }\textbf {\bibinfo {volume} {102}},\ \bibinfo
  {pages} {074503} (\bibinfo {year} {2020})},\ \Eprint
  {http://arxiv.org/abs/2007.10078} {arXiv:2007.10078 [hep-lat]} \BibitemShut
  {NoStop}%
\bibitem [{\citenamefont {Casalderrey-Solana}\ and\ \citenamefont
  {Teaney}(2007)}]{Casalderrey-Solana:2007ahi}%
  \BibitemOpen
  \bibfield  {author} {\bibinfo {author} {\bibfnamefont {J.}~\bibnamefont
  {Casalderrey-Solana}}\ and\ \bibinfo {author} {\bibfnamefont
  {D.}~\bibnamefont {Teaney}},\ }\href {\doibase 10.1088/1126-6708/2007/04/039}
  {\bibfield  {journal} {\bibinfo  {journal} {JHEP}\ }\textbf {\bibinfo
  {volume} {04}},\ \bibinfo {pages} {039} (\bibinfo {year} {2007})},\ \Eprint
  {http://arxiv.org/abs/hep-th/0701123} {arXiv:hep-th/0701123} \BibitemShut
  {NoStop}%
\bibitem [{\citenamefont {D'Eramo}\ \emph {et~al.}(2011)\citenamefont
  {D'Eramo}, \citenamefont {Liu},\ and\ \citenamefont
  {Rajagopal}}]{DEramo:2010wup}%
  \BibitemOpen
  \bibfield  {author} {\bibinfo {author} {\bibfnamefont {F.}~\bibnamefont
  {D'Eramo}}, \bibinfo {author} {\bibfnamefont {H.}~\bibnamefont {Liu}}, \ and\
  \bibinfo {author} {\bibfnamefont {K.}~\bibnamefont {Rajagopal}},\ }\href
  {\doibase 10.1103/PhysRevD.84.065015} {\bibfield  {journal} {\bibinfo
  {journal} {Phys. Rev. D}\ }\textbf {\bibinfo {volume} {84}},\ \bibinfo
  {pages} {065015} (\bibinfo {year} {2011})},\ \Eprint
  {http://arxiv.org/abs/1006.1367} {arXiv:1006.1367 [hep-ph]} \BibitemShut
  {NoStop}%
\bibitem [{\citenamefont {Carrington}\ \emph {et~al.}(2017)\citenamefont
  {Carrington}, \citenamefont {Mr{\'o}wczy{\'n}ski},\ and\ \citenamefont
  {Schenke}}]{Carrington:2016mhd}%
  \BibitemOpen
  \bibfield  {author} {\bibinfo {author} {\bibfnamefont {M.~E.}\ \bibnamefont
  {Carrington}}, \bibinfo {author} {\bibfnamefont {S.}~\bibnamefont
  {Mr{\'o}wczy{\'n}ski}}, \ and\ \bibinfo {author} {\bibfnamefont
  {B.}~\bibnamefont {Schenke}},\ }\href {\doibase 10.1103/PhysRevC.95.024906}
  {\bibfield  {journal} {\bibinfo  {journal} {Phys. Rev. C}\ }\textbf {\bibinfo
  {volume} {95}},\ \bibinfo {pages} {024906} (\bibinfo {year} {2017})},\
  \Eprint {http://arxiv.org/abs/1607.02359} {arXiv:1607.02359 [hep-ph]}
  \BibitemShut {NoStop}%
\bibitem [{\citenamefont {Laine}\ and\ \citenamefont
  {Manuel}(2002)}]{Laine:2001my}%
  \BibitemOpen
  \bibfield  {author} {\bibinfo {author} {\bibfnamefont {M.}~\bibnamefont
  {Laine}}\ and\ \bibinfo {author} {\bibfnamefont {C.}~\bibnamefont {Manuel}},\
  }\href {\doibase 10.1103/PhysRevD.65.077902} {\bibfield  {journal} {\bibinfo
  {journal} {Phys. Rev. D}\ }\textbf {\bibinfo {volume} {65}},\ \bibinfo
  {pages} {077902} (\bibinfo {year} {2002})},\ \Eprint
  {http://arxiv.org/abs/hep-ph/0111113} {arXiv:hep-ph/0111113} \BibitemShut
  {NoStop}%
\bibitem [{\citenamefont {Ghiglieri}\ and\ \citenamefont
  {Kim}(2018)}]{Ghiglieri:2018ltw}%
  \BibitemOpen
  \bibfield  {author} {\bibinfo {author} {\bibfnamefont {J.}~\bibnamefont
  {Ghiglieri}}\ and\ \bibinfo {author} {\bibfnamefont {H.}~\bibnamefont
  {Kim}},\ }\href {\doibase 10.1007/JHEP12(2018)049} {\bibfield  {journal}
  {\bibinfo  {journal} {JHEP}\ }\textbf {\bibinfo {volume} {12}},\ \bibinfo
  {pages} {049} (\bibinfo {year} {2018})},\ \Eprint
  {http://arxiv.org/abs/1809.01349} {arXiv:1809.01349 [hep-ph]} \BibitemShut
  {NoStop}%
\bibitem [{\citenamefont {Lappi}(2008{\natexlab{b}})}]{Lappi:2007ku}%
  \BibitemOpen
  \bibfield  {author} {\bibinfo {author} {\bibfnamefont {T.}~\bibnamefont
  {Lappi}},\ }\href {\doibase 10.1140/epjc/s10052-008-0588-4} {\bibfield
  {journal} {\bibinfo  {journal} {Eur. Phys. J. C}\ }\textbf {\bibinfo {volume}
  {55}},\ \bibinfo {pages} {285} (\bibinfo {year} {2008}{\natexlab{b}})},\
  \Eprint {http://arxiv.org/abs/0711.3039} {arXiv:0711.3039 [hep-ph]}
  \BibitemShut {NoStop}%
\bibitem [{cur()}]{curraun}%
  \BibitemOpen
  \href@noop {} {}\bibinfo {note}
  {\url{https://gitlab.com/openpixi/curraun}}\BibitemShut {NoStop}%
\bibitem [{\citenamefont {Workman}\ \emph {et~al.}(2022)\citenamefont {Workman}
  \emph {et~al.}}]{ParticleDataGroup:2022pth}%
  \BibitemOpen
  \bibfield  {author} {\bibinfo {author} {\bibfnamefont {R.~L.}\ \bibnamefont
  {Workman}} \emph {et~al.} (\bibinfo {collaboration} {Particle Data Group}),\
  }\href {\doibase 10.1093/ptep/ptac097} {\bibfield  {journal} {\bibinfo
  {journal} {PTEP}\ }\textbf {\bibinfo {volume} {2022}},\ \bibinfo {pages}
  {083C01} (\bibinfo {year} {2022})}\BibitemShut {NoStop}%
\bibitem [{\citenamefont {Krasnitz}\ and\ \citenamefont
  {Venugopalan}(2001)}]{Krasnitz:2000gz}%
  \BibitemOpen
  \bibfield  {author} {\bibinfo {author} {\bibfnamefont {A.}~\bibnamefont
  {Krasnitz}}\ and\ \bibinfo {author} {\bibfnamefont {R.}~\bibnamefont
  {Venugopalan}},\ }\href {\doibase 10.1103/PhysRevLett.86.1717} {\bibfield
  {journal} {\bibinfo  {journal} {Phys. Rev. Lett.}\ }\textbf {\bibinfo
  {volume} {86}},\ \bibinfo {pages} {1717} (\bibinfo {year} {2001})},\ \Eprint
  {http://arxiv.org/abs/hep-ph/0007108} {arXiv:hep-ph/0007108} \BibitemShut
  {NoStop}%
\bibitem [{\citenamefont {Lappi}\ and\ \citenamefont
  {Peuron}(2017)}]{Lappi:2016ato}%
  \BibitemOpen
  \bibfield  {author} {\bibinfo {author} {\bibfnamefont {T.}~\bibnamefont
  {Lappi}}\ and\ \bibinfo {author} {\bibfnamefont {J.}~\bibnamefont {Peuron}},\
  }\href {\doibase 10.1103/PhysRevD.95.014025} {\bibfield  {journal} {\bibinfo
  {journal} {Phys. Rev. D}\ }\textbf {\bibinfo {volume} {95}},\ \bibinfo
  {pages} {014025} (\bibinfo {year} {2017})},\ \Eprint
  {http://arxiv.org/abs/1610.03711} {arXiv:1610.03711 [hep-ph]} \BibitemShut
  {NoStop}%
\bibitem [{\citenamefont {Lappi}\ and\ \citenamefont
  {Peuron}(2018)}]{Lappi:2017ckt}%
  \BibitemOpen
  \bibfield  {author} {\bibinfo {author} {\bibfnamefont {T.}~\bibnamefont
  {Lappi}}\ and\ \bibinfo {author} {\bibfnamefont {J.}~\bibnamefont {Peuron}},\
  }\href {\doibase 10.1103/PhysRevD.97.034017} {\bibfield  {journal} {\bibinfo
  {journal} {Phys. Rev. D}\ }\textbf {\bibinfo {volume} {97}},\ \bibinfo
  {pages} {034017} (\bibinfo {year} {2018})},\ \Eprint
  {http://arxiv.org/abs/1712.02194} {arXiv:1712.02194 [hep-lat]} \BibitemShut
  {NoStop}%
\bibitem [{\citenamefont {Schenke}\ \emph
  {et~al.}(2012{\natexlab{a}})\citenamefont {Schenke}, \citenamefont
  {Tribedy},\ and\ \citenamefont {Venugopalan}}]{Schenke:2012wb}%
  \BibitemOpen
  \bibfield  {author} {\bibinfo {author} {\bibfnamefont {B.}~\bibnamefont
  {Schenke}}, \bibinfo {author} {\bibfnamefont {P.}~\bibnamefont {Tribedy}}, \
  and\ \bibinfo {author} {\bibfnamefont {R.}~\bibnamefont {Venugopalan}},\
  }\href {\doibase 10.1103/PhysRevLett.108.252301} {\bibfield  {journal}
  {\bibinfo  {journal} {Phys. Rev. Lett.}\ }\textbf {\bibinfo {volume} {108}},\
  \bibinfo {pages} {252301} (\bibinfo {year} {2012}{\natexlab{a}})},\ \Eprint
  {http://arxiv.org/abs/1202.6646} {arXiv:1202.6646 [nucl-th]} \BibitemShut
  {NoStop}%
\bibitem [{\citenamefont {Schenke}\ \emph
  {et~al.}(2012{\natexlab{b}})\citenamefont {Schenke}, \citenamefont
  {Tribedy},\ and\ \citenamefont {Venugopalan}}]{Schenke:2012hg}%
  \BibitemOpen
  \bibfield  {author} {\bibinfo {author} {\bibfnamefont {B.}~\bibnamefont
  {Schenke}}, \bibinfo {author} {\bibfnamefont {P.}~\bibnamefont {Tribedy}}, \
  and\ \bibinfo {author} {\bibfnamefont {R.}~\bibnamefont {Venugopalan}},\
  }\href {\doibase 10.1103/PhysRevC.86.034908} {\bibfield  {journal} {\bibinfo
  {journal} {Phys. Rev. C}\ }\textbf {\bibinfo {volume} {86}},\ \bibinfo
  {pages} {034908} (\bibinfo {year} {2012}{\natexlab{b}})},\ \Eprint
  {http://arxiv.org/abs/1206.6805} {arXiv:1206.6805 [hep-ph]} \BibitemShut
  {NoStop}%
\bibitem [{\citenamefont {M\"antysaari}\ and\ \citenamefont
  {Schenke}(2016)}]{Mantysaari:2016jaz}%
  \BibitemOpen
  \bibfield  {author} {\bibinfo {author} {\bibfnamefont {H.}~\bibnamefont
  {M\"antysaari}}\ and\ \bibinfo {author} {\bibfnamefont {B.}~\bibnamefont
  {Schenke}},\ }\href {\doibase 10.1103/PhysRevD.94.034042} {\bibfield
  {journal} {\bibinfo  {journal} {Phys. Rev. D}\ }\textbf {\bibinfo {volume}
  {94}},\ \bibinfo {pages} {034042} (\bibinfo {year} {2016})},\ \Eprint
  {http://arxiv.org/abs/1607.01711} {arXiv:1607.01711 [hep-ph]} \BibitemShut
  {NoStop}%
\bibitem [{\citenamefont {Mäntysaari}\ \emph {et~al.}(2022)\citenamefont
  {Mäntysaari}, \citenamefont {Schenke}, \citenamefont {Shen},\ and\
  \citenamefont {Zhao}}]{Mantysaari_2022}%
  \BibitemOpen
  \bibfield  {author} {\bibinfo {author} {\bibfnamefont {H.}~\bibnamefont
  {Mäntysaari}}, \bibinfo {author} {\bibfnamefont {B.}~\bibnamefont
  {Schenke}}, \bibinfo {author} {\bibfnamefont {C.}~\bibnamefont {Shen}}, \
  and\ \bibinfo {author} {\bibfnamefont {W.}~\bibnamefont {Zhao}},\ }\href
  {\doibase 10.48550/ARXIV.2208.00396} {\enquote {\bibinfo {title} {Bayesian
  inference of the fluctuating proton shape in dis and hadronic collisions},}\
  } (\bibinfo {year} {2022})\BibitemShut {NoStop}%
\bibitem [{\citenamefont {Demirci}\ \emph {et~al.}(2022)\citenamefont
  {Demirci}, \citenamefont {Lappi},\ and\ \citenamefont
  {Schlichting}}]{Demirci:2022wuy}%
  \BibitemOpen
  \bibfield  {author} {\bibinfo {author} {\bibfnamefont {S.}~\bibnamefont
  {Demirci}}, \bibinfo {author} {\bibfnamefont {T.}~\bibnamefont {Lappi}}, \
  and\ \bibinfo {author} {\bibfnamefont {S.}~\bibnamefont {Schlichting}},\
  }\href {\doibase 10.1103/PhysRevD.106.074025} {\bibfield  {journal} {\bibinfo
   {journal} {Phys. Rev. D}\ }\textbf {\bibinfo {volume} {106}},\ \bibinfo
  {pages} {074025} (\bibinfo {year} {2022})},\ \Eprint
  {http://arxiv.org/abs/2206.05207} {arXiv:2206.05207 [hep-ph]} \BibitemShut
  {NoStop}%
\bibitem [{\citenamefont {Gelfand}\ \emph {et~al.}(2016)\citenamefont
  {Gelfand}, \citenamefont {Ipp},\ and\ \citenamefont
  {M\"uller}}]{Gelfand:2016yho}%
  \BibitemOpen
  \bibfield  {author} {\bibinfo {author} {\bibfnamefont {D.}~\bibnamefont
  {Gelfand}}, \bibinfo {author} {\bibfnamefont {A.}~\bibnamefont {Ipp}}, \ and\
  \bibinfo {author} {\bibfnamefont {D.}~\bibnamefont {M\"uller}},\ }\href
  {\doibase 10.1103/PhysRevD.94.014020} {\bibfield  {journal} {\bibinfo
  {journal} {Phys. Rev. D}\ }\textbf {\bibinfo {volume} {94}},\ \bibinfo
  {pages} {014020} (\bibinfo {year} {2016})},\ \Eprint
  {http://arxiv.org/abs/1605.07184} {arXiv:1605.07184 [hep-ph]} \BibitemShut
  {NoStop}%
\bibitem [{\citenamefont {Ipp}\ and\ \citenamefont
  {M\"uller}(2017)}]{Ipp:2017lho}%
  \BibitemOpen
  \bibfield  {author} {\bibinfo {author} {\bibfnamefont {A.}~\bibnamefont
  {Ipp}}\ and\ \bibinfo {author} {\bibfnamefont {D.}~\bibnamefont {M\"uller}},\
  }\href {\doibase 10.1016/j.physletb.2017.05.032} {\bibfield  {journal}
  {\bibinfo  {journal} {Phys. Lett. B}\ }\textbf {\bibinfo {volume} {771}},\
  \bibinfo {pages} {74} (\bibinfo {year} {2017})},\ \Eprint
  {http://arxiv.org/abs/1703.00017} {arXiv:1703.00017 [hep-ph]} \BibitemShut
  {NoStop}%
\bibitem [{\citenamefont {Ipp}\ and\ \citenamefont
  {M\"uller}(2020)}]{Ipp:2020igo}%
  \BibitemOpen
  \bibfield  {author} {\bibinfo {author} {\bibfnamefont {A.}~\bibnamefont
  {Ipp}}\ and\ \bibinfo {author} {\bibfnamefont {D.~I.}\ \bibnamefont
  {M\"uller}},\ }\href {\doibase 10.1140/epja/s10050-020-00241-6} {\bibfield
  {journal} {\bibinfo  {journal} {Eur. Phys. J. A}\ }\textbf {\bibinfo {volume}
  {56}},\ \bibinfo {pages} {243} (\bibinfo {year} {2020})},\ \Eprint
  {http://arxiv.org/abs/2009.02044} {arXiv:2009.02044 [hep-ph]} \BibitemShut
  {NoStop}%
\bibitem [{\citenamefont {Schlichting}\ and\ \citenamefont
  {Singh}(2021)}]{Schlichting:2020wrv}%
  \BibitemOpen
  \bibfield  {author} {\bibinfo {author} {\bibfnamefont {S.}~\bibnamefont
  {Schlichting}}\ and\ \bibinfo {author} {\bibfnamefont {P.}~\bibnamefont
  {Singh}},\ }\href {\doibase 10.1103/PhysRevD.103.014003} {\bibfield
  {journal} {\bibinfo  {journal} {Phys. Rev. D}\ }\textbf {\bibinfo {volume}
  {103}},\ \bibinfo {pages} {014003} (\bibinfo {year} {2021})},\ \Eprint
  {http://arxiv.org/abs/2010.11172} {arXiv:2010.11172 [hep-ph]} \BibitemShut
  {NoStop}%
\bibitem [{\citenamefont {Schenke}\ and\ \citenamefont
  {Schlichting}(2016)}]{Schenke:2016ksl}%
  \BibitemOpen
  \bibfield  {author} {\bibinfo {author} {\bibfnamefont {B.}~\bibnamefont
  {Schenke}}\ and\ \bibinfo {author} {\bibfnamefont {S.}~\bibnamefont
  {Schlichting}},\ }\href {\doibase 10.1103/PhysRevC.94.044907} {\bibfield
  {journal} {\bibinfo  {journal} {Phys. Rev. C}\ }\textbf {\bibinfo {volume}
  {94}},\ \bibinfo {pages} {044907} (\bibinfo {year} {2016})},\ \Eprint
  {http://arxiv.org/abs/1605.07158} {arXiv:1605.07158 [hep-ph]} \BibitemShut
  {NoStop}%
\bibitem [{\citenamefont {McDonald}\ \emph {et~al.}(2019)\citenamefont
  {McDonald}, \citenamefont {Jeon},\ and\ \citenamefont
  {Gale}}]{McDonald:2018wql}%
  \BibitemOpen
  \bibfield  {author} {\bibinfo {author} {\bibfnamefont {S.}~\bibnamefont
  {McDonald}}, \bibinfo {author} {\bibfnamefont {S.}~\bibnamefont {Jeon}}, \
  and\ \bibinfo {author} {\bibfnamefont {C.}~\bibnamefont {Gale}},\ }\href
  {\doibase 10.1016/j.nuclphysa.2018.08.014} {\bibfield  {journal} {\bibinfo
  {journal} {Nucl. Phys. A}\ }\textbf {\bibinfo {volume} {982}},\ \bibinfo
  {pages} {239} (\bibinfo {year} {2019})},\ \Eprint
  {http://arxiv.org/abs/1807.05409} {arXiv:1807.05409 [nucl-th]} \BibitemShut
  {NoStop}%
\bibitem [{\citenamefont {McDonald}\ \emph {et~al.}(2021)\citenamefont
  {McDonald}, \citenamefont {Jeon},\ and\ \citenamefont
  {Gale}}]{McDonald:2020oyf}%
  \BibitemOpen
  \bibfield  {author} {\bibinfo {author} {\bibfnamefont {S.}~\bibnamefont
  {McDonald}}, \bibinfo {author} {\bibfnamefont {S.}~\bibnamefont {Jeon}}, \
  and\ \bibinfo {author} {\bibfnamefont {C.}~\bibnamefont {Gale}},\ }\href
  {\doibase 10.1016/j.nuclphysa.2020.121771} {\bibfield  {journal} {\bibinfo
  {journal} {Nucl. Phys. A}\ }\textbf {\bibinfo {volume} {1005}},\ \bibinfo
  {pages} {121771} (\bibinfo {year} {2021})},\ \Eprint
  {http://arxiv.org/abs/2001.08636} {arXiv:2001.08636 [nucl-th]} \BibitemShut
  {NoStop}%
\bibitem [{\citenamefont {Ipp}\ \emph {et~al.}(2021)\citenamefont {Ipp},
  \citenamefont {M\"uller}, \citenamefont {Schlichting},\ and\ \citenamefont
  {Singh}}]{Ipp:2021lwz}%
  \BibitemOpen
  \bibfield  {author} {\bibinfo {author} {\bibfnamefont {A.}~\bibnamefont
  {Ipp}}, \bibinfo {author} {\bibfnamefont {D.~I.}\ \bibnamefont {M\"uller}},
  \bibinfo {author} {\bibfnamefont {S.}~\bibnamefont {Schlichting}}, \ and\
  \bibinfo {author} {\bibfnamefont {P.}~\bibnamefont {Singh}},\ }\href
  {\doibase 10.1103/PhysRevD.104.114040} {\bibfield  {journal} {\bibinfo
  {journal} {Phys. Rev. D}\ }\textbf {\bibinfo {volume} {104}},\ \bibinfo
  {pages} {114040} (\bibinfo {year} {2021})},\ \Eprint
  {http://arxiv.org/abs/2109.05028} {arXiv:2109.05028 [hep-ph]} \BibitemShut
  {NoStop}%
\bibitem [{\citenamefont {Ipp}\ and\ \citenamefont
  {M\"uller}(2018)}]{Ipp:2018hai}%
  \BibitemOpen
  \bibfield  {author} {\bibinfo {author} {\bibfnamefont {A.}~\bibnamefont
  {Ipp}}\ and\ \bibinfo {author} {\bibfnamefont {D.}~\bibnamefont {M\"uller}},\
  }\href {\doibase 10.1140/epjc/s10052-018-6323-x} {\bibfield  {journal}
  {\bibinfo  {journal} {Eur. Phys. J. C}\ }\textbf {\bibinfo {volume} {78}},\
  \bibinfo {pages} {884} (\bibinfo {year} {2018})},\ \Eprint
  {http://arxiv.org/abs/1804.01995} {arXiv:1804.01995 [hep-lat]} \BibitemShut
  {NoStop}%
\bibitem [{\citenamefont {Hauksson}\ \emph {et~al.}(2022)\citenamefont
  {Hauksson}, \citenamefont {Jeon},\ and\ \citenamefont
  {Gale}}]{Hauksson:2021okc}%
  \BibitemOpen
  \bibfield  {author} {\bibinfo {author} {\bibfnamefont {S.}~\bibnamefont
  {Hauksson}}, \bibinfo {author} {\bibfnamefont {S.}~\bibnamefont {Jeon}}, \
  and\ \bibinfo {author} {\bibfnamefont {C.}~\bibnamefont {Gale}},\ }\href
  {\doibase 10.1103/PhysRevC.105.014914} {\bibfield  {journal} {\bibinfo
  {journal} {Phys. Rev. C}\ }\textbf {\bibinfo {volume} {105}},\ \bibinfo
  {pages} {014914} (\bibinfo {year} {2022})},\ \Eprint
  {http://arxiv.org/abs/2109.04575} {arXiv:2109.04575 [hep-ph]} \BibitemShut
  {NoStop}%
\bibitem [{\citenamefont {Moore}\ \emph {et~al.}(2021)\citenamefont {Moore},
  \citenamefont {Schlichting}, \citenamefont {Schlusser},\ and\ \citenamefont
  {Soudi}}]{Moore:2021jwe}%
  \BibitemOpen
  \bibfield  {author} {\bibinfo {author} {\bibfnamefont {G.~D.}\ \bibnamefont
  {Moore}}, \bibinfo {author} {\bibfnamefont {S.}~\bibnamefont {Schlichting}},
  \bibinfo {author} {\bibfnamefont {N.}~\bibnamefont {Schlusser}}, \ and\
  \bibinfo {author} {\bibfnamefont {I.}~\bibnamefont {Soudi}},\ }\href
  {\doibase 10.1007/JHEP10(2021)059} {\bibfield  {journal} {\bibinfo  {journal}
  {JHEP}\ }\textbf {\bibinfo {volume} {10}},\ \bibinfo {pages} {059} (\bibinfo
  {year} {2021})},\ \Eprint {http://arxiv.org/abs/2105.01679} {arXiv:2105.01679
  [hep-ph]} \BibitemShut {NoStop}%
\bibitem [{\citenamefont {Haber}(2021)}]{Haber:2019sgz}%
  \BibitemOpen
  \bibfield  {author} {\bibinfo {author} {\bibfnamefont {H.~E.}\ \bibnamefont
  {Haber}},\ }\href {\doibase 10.21468/SciPostPhysLectNotes.21} {\bibfield
  {journal} {\bibinfo  {journal} {SciPost Phys. Lect. Notes}\ }\textbf
  {\bibinfo {volume} {21}},\ \bibinfo {pages} {1} (\bibinfo {year} {2021})},\
  \Eprint {http://arxiv.org/abs/1912.13302} {arXiv:1912.13302 [math-ph]}
  \BibitemShut {NoStop}%
\bibitem [{\citenamefont {Heinz}(1985)}]{Heinz:1984yq}%
  \BibitemOpen
  \bibfield  {author} {\bibinfo {author} {\bibfnamefont {U.~W.}\ \bibnamefont
  {Heinz}},\ }\href {\doibase 10.1016/0003-4916(85)90336-7} {\bibfield
  {journal} {\bibinfo  {journal} {Annals Phys.}\ }\textbf {\bibinfo {volume}
  {161}},\ \bibinfo {pages} {48} (\bibinfo {year} {1985})}\BibitemShut
  {NoStop}%
\bibitem [{\citenamefont {Heinz}(1986)}]{Heinz:1985qe}%
  \BibitemOpen
  \bibfield  {author} {\bibinfo {author} {\bibfnamefont {U.~W.}\ \bibnamefont
  {Heinz}},\ }\href {\doibase 10.1016/0003-4916(86)90114-4} {\bibfield
  {journal} {\bibinfo  {journal} {Annals Phys.}\ }\textbf {\bibinfo {volume}
  {168}},\ \bibinfo {pages} {148} (\bibinfo {year} {1986})}\BibitemShut
  {NoStop}%
\bibitem [{\citenamefont {Meurer}\ \emph {et~al.}(2017)\citenamefont {Meurer},
  \citenamefont {Smith}, \citenamefont {Paprocki}, \citenamefont
  {\v{C}ert\'{i}k}, \citenamefont {Kirpichev}, \citenamefont {Rocklin},
  \citenamefont {Kumar}, \citenamefont {Ivanov}, \citenamefont {Moore},
  \citenamefont {Singh}, \citenamefont {Rathnayake}, \citenamefont {Vig},
  \citenamefont {Granger}, \citenamefont {Muller}, \citenamefont {Bonazzi},
  \citenamefont {Gupta}, \citenamefont {Vats}, \citenamefont {Johansson},
  \citenamefont {Pedregosa}, \citenamefont {Curry}, \citenamefont {Terrel},
  \citenamefont {Rou\v{c}ka}, \citenamefont {Saboo}, \citenamefont {Fernando},
  \citenamefont {Kulal}, \citenamefont {Cimrman},\ and\ \citenamefont
  {Scopatz}}]{SymPy}%
  \BibitemOpen
  \bibfield  {author} {\bibinfo {author} {\bibfnamefont {A.}~\bibnamefont
  {Meurer}}, \bibinfo {author} {\bibfnamefont {C.~P.}\ \bibnamefont {Smith}},
  \bibinfo {author} {\bibfnamefont {M.}~\bibnamefont {Paprocki}}, \bibinfo
  {author} {\bibfnamefont {O.}~\bibnamefont {\v{C}ert\'{i}k}}, \bibinfo
  {author} {\bibfnamefont {S.~B.}\ \bibnamefont {Kirpichev}}, \bibinfo {author}
  {\bibfnamefont {M.}~\bibnamefont {Rocklin}}, \bibinfo {author} {\bibfnamefont
  {A.}~\bibnamefont {Kumar}}, \bibinfo {author} {\bibfnamefont
  {S.}~\bibnamefont {Ivanov}}, \bibinfo {author} {\bibfnamefont {J.~K.}\
  \bibnamefont {Moore}}, \bibinfo {author} {\bibfnamefont {S.}~\bibnamefont
  {Singh}}, \bibinfo {author} {\bibfnamefont {T.}~\bibnamefont {Rathnayake}},
  \bibinfo {author} {\bibfnamefont {S.}~\bibnamefont {Vig}}, \bibinfo {author}
  {\bibfnamefont {B.~E.}\ \bibnamefont {Granger}}, \bibinfo {author}
  {\bibfnamefont {R.~P.}\ \bibnamefont {Muller}}, \bibinfo {author}
  {\bibfnamefont {F.}~\bibnamefont {Bonazzi}}, \bibinfo {author} {\bibfnamefont
  {H.}~\bibnamefont {Gupta}}, \bibinfo {author} {\bibfnamefont
  {S.}~\bibnamefont {Vats}}, \bibinfo {author} {\bibfnamefont {F.}~\bibnamefont
  {Johansson}}, \bibinfo {author} {\bibfnamefont {F.}~\bibnamefont
  {Pedregosa}}, \bibinfo {author} {\bibfnamefont {M.~J.}\ \bibnamefont
  {Curry}}, \bibinfo {author} {\bibfnamefont {A.~R.}\ \bibnamefont {Terrel}},
  \bibinfo {author} {\bibfnamefont {v.}~\bibnamefont {Rou\v{c}ka}}, \bibinfo
  {author} {\bibfnamefont {A.}~\bibnamefont {Saboo}}, \bibinfo {author}
  {\bibfnamefont {I.}~\bibnamefont {Fernando}}, \bibinfo {author}
  {\bibfnamefont {S.}~\bibnamefont {Kulal}}, \bibinfo {author} {\bibfnamefont
  {R.}~\bibnamefont {Cimrman}}, \ and\ \bibinfo {author} {\bibfnamefont
  {A.}~\bibnamefont {Scopatz}},\ }\href {\doibase 10.7717/peerj-cs.103}
  {\bibfield  {journal} {\bibinfo  {journal} {PeerJ Computer Science}\ }\textbf
  {\bibinfo {volume} {3}},\ \bibinfo {pages} {e103} (\bibinfo {year}
  {2017})}\BibitemShut {NoStop}%
\bibitem [{\citenamefont {Zuber}(2017)}]{Zuber:2016xme}%
  \BibitemOpen
  \bibfield  {author} {\bibinfo {author} {\bibfnamefont {J.-B.}\ \bibnamefont
  {Zuber}},\ }\href {\doibase 10.1088/1751-8113/50/1/015203} {\bibfield
  {journal} {\bibinfo  {journal} {J. Phys. A}\ }\textbf {\bibinfo {volume}
  {50}},\ \bibinfo {pages} {015203} (\bibinfo {year} {2017})},\ \Eprint
  {http://arxiv.org/abs/1611.00236} {arXiv:1611.00236 [math-ph]} \BibitemShut
  {NoStop}%
\bibitem [{\citenamefont {Creutz}(1978)}]{Creutz:1978ub}%
  \BibitemOpen
  \bibfield  {author} {\bibinfo {author} {\bibfnamefont {M.}~\bibnamefont
  {Creutz}},\ }\href {\doibase 10.1063/1.523581} {\bibfield  {journal}
  {\bibinfo  {journal} {J. Math. Phys.}\ }\textbf {\bibinfo {volume} {19}},\
  \bibinfo {pages} {2043} (\bibinfo {year} {1978})}\BibitemShut {NoStop}%
\bibitem [{\citenamefont {Carlsson}(2008)}]{Carlsson:2008dh}%
  \BibitemOpen
  \bibfield  {author} {\bibinfo {author} {\bibfnamefont {J.}~\bibnamefont
  {Carlsson}},\ }\href@noop {} {\  (\bibinfo {year} {2008})},\ \Eprint
  {http://arxiv.org/abs/0802.3409} {arXiv:0802.3409 [hep-lat]} \BibitemShut
  {NoStop}%
\end{thebibliography}%

\end{document}